\documentclass[final,5p,times, twocolumn, authoryear]{elsarticle}

\usepackage{amssymb}
\usepackage{physics}
\usepackage{aas_macros}
\usepackage{xcolor}
\usepackage{xspace}
\usepackage{algorithm}
\usepackage{algpseudocode}

\usepackage{hyperref} 
\hypersetup{
    colorlinks=true,
    citecolor=blue,
    linkcolor=blue,
    urlcolor=blue,
    }

\definecolor{cadmiumred}{rgb}{0.89, 0.0, .138}

\definecolor{scc}{rgb}{0.6, 0.2, .8}

\definecolor{gsc}{rgb}{0.0, 0.48, 0.28}

\definecolor{gsc}{rgb}{1.0, 0.0, 0.0}

\journal{Astronomy and computing}
\graphicspath{ {./images/} }

\begin{document}

\begin{frontmatter}




\title{A Quantum Genetic Algorithm with application to Cosmological Parameters Estimation}


\author[inst1]{Giuseppe Sarracino}

\author[inst2,inst_INFN_Roma]{Vincenzo Fabrizio Cardone}

\author[inst2,inst_INFN_Roma]{Roberto Scaramella}

\author[inst1]{Giuseppe Riccio}

\author[inst_Bol_OAS]{Andrea Bulgarelli}

\author[inst_Bol_INAF,inst_Bol_INFN]{Carlo Burigana}

\author[inst_Tri_INAF,inst_Tri_Univ]{Luca Cappelli}

\author[inst1]{Stefano Cavuoti}

\author[inst_Cat_INAF]{Farida Farsian}

\author[inst1]{Irene Graziotti}

\author[inst_Bol_OAS]{Massimo Meneghetti}

\author[inst_Tri_INAF]{Giuseppe Murante}

\author[inst_Bol_OAS]{Nicolò Parmiggiani}

\author[inst_Cat_INAF]{Alessandro Rizzo}

\author[inst_Cat_INAF]{Francesco Schillirò}

\author[inst2]{Vincenzo Testa}

\author[inst_Bol_INAF,inst_Bol_INFN]{Tiziana Trombetti}

\affiliation[inst1]{organization={INAF-Osservatorio Astronomico di Capodimonte},
            addressline={Via Moiariello 16}, 
            city={Napoli},
            postcode={80131}, 
            state={Italy},
            }

\affiliation[inst2]{organization={INAF-Osservatorio Astronomico di Roma},
            addressline={Via Frascati 33}, 
            city={Monteporzio Catone, Roma},
            postcode={00078}, 
            state={Italy}
            }

\affiliation[inst_INFN_Roma]{organization={INFN Sezione di Roma 1},
            addressline={Piazzale Aldo Moro 2, Edificio G. Marconi}, 
            city={Roma},
            postcode={00185}, 
            state={Italy},
            }

\affiliation[inst_Bol_OAS]{organization={INAF-Osservatorio di astrofisica e scienza dello spazio},
            addressline={Via Pietro Gobetti 93/3}, 
            city={ Bologna},
            postcode={ 40129}, 
            state={Italy},
            }

\affiliation[inst_Bol_INAF]{organization={INAF-Istituto di Radioastronomia},
            addressline={Via Piero Gobetti 101}, 
            city={ Bologna},
            postcode={ 40129}, 
            state={Italy},
            }

\affiliation[inst_Bol_INFN]{organization={INFN Sezione di Bologna},
            addressline={Via Irnerio 46}, 
            city={ Bologna},
            postcode={ 40127}, 
            state={Italy},
            }

\affiliation[inst_Tri_INAF]{organization={INAF-Osservatorio Astronomico di Trieste},
            addressline={Via Tiepolo 11}, 
            city={ Trieste},
            postcode={ I-34131}, 
            state={Italy},
            }

\affiliation[inst_Tri_Univ]{organization={Università degli Studi di Trieste, Dipartimento di Fisica},
            addressline={Via A. Valerio 2}, 
            city={ Trieste},
            postcode={ 34127}, 
            state={Italy}
            }

\affiliation[inst_Cat_INAF]{organization={INAF-Osservatorio Astronomico di Catania},
            addressline={Via S. Sofia 78}, 
            city={ Catania},
            postcode={ 95123}, 
            state={Italy},
            }

\begin{abstract}
An Amplitude-Encoded Quantum Genetic Algorithm (AEQGA) has been developed to minimize $\chi^2$ functions of different cosmological probes (Supernovae Type Ia, Baryon Acoustic Oscillations, Cosmic Microwave Background Radiation), to find the best-fit value for two cosmological parameters, namely the Hubble Constant and the density matter content of the Universe today. Our main aim is to pave the way to testing the adoption of quantum optimization in the inference of the cosmological parameters that describe the universe evolution. AEQGA computes the merit function classically, and then uses a quantum circuit to entangle the population and perform crossover and mutation operations. The results show consistency with the isocontours of the objective functions. We then tested the general behavior of AEQGA as a function of its hyperparameters and compared it with a second quantum genetic algorithm found in the literature as well as with classical algorithms, finding consistent results. 

\end{abstract}




\end{frontmatter}


\section{Introduction}
\label{sec_Introduction}

Quantum Computing (QC) is a novel framework linking fundamental concepts of Mathematics, Computer Science, and Quantum Mechanics \citep{Yanofsky_2007}.
The interest in QC dates back to the end of the last century \citep{Feynman_1982,  Unruh_1995, DiVincenzo_1995, Narayanan_1996}, although first investigations were mainly devoted to theoretical foundations. In that period, the first quantum algorithms were created, enabling an impressive advantage in terms of computational costs in comparison with the corresponding classical computations, at least theoretically. The Grover's algorithm for quantum search \citep{Grover_1998}, and Shor's one for integer factorization \citep{Shor_1999} were some of the first examples.

Nowadays, QC is a new frontier that has been rapidly evolving in recent years, attracting significant interest from the scientific community \citep{Steane_1998, AHARONOV_1999, ladd_2010, Li_2020}. The key idea behind this framework is a fundamental reinterpretation of numerical computations utilizing Quantum Mechanics concepts, by employing the so-called qubit as the information unit instead of the classical bit. The main difference is that, unlike classical bits, which can exist in one of two states (0 or 1), qubits can exist in a linear combination of basis states because of quantum superposition, which collapses once the state of the qubit is classically measured. This property, along with another key concept called \textit{Entanglement} (for which states of different qubits are interconnected via their wave function regardless of the physical distance between them) allows, on an ideal quantum hardware and for specific tasks fully using these properties, to perform complex operations significantly reducing time and/or computing resources with respect to classical frameworks. This would be particularly useful in specific instances, involving some of the so-called Nondeterministic Polynomial (NP)-hard problems, i.e. operations that can be reduced to NP problems in a polynomial time, NP problems being computations whose solutions can be verified (not solved) in a polynomial time \citep{Cook_1971}.

 More recently, remarkable technological improvements have enabled the creation of proper quantum computers \citep{Massarotti_2023}. This allowed for quantum supremacy to be finally achieved for an ``ad hoc" problem \citep{Arute_2019}, which has been proven to be realistically impossible to solve via classical computations in a reasonable amount of time. Specifically, from a complex quantum circuit composed of gates layered with 53 qubits, the authors sampled a probability distribution that is extremely difficult to sample classically. This task would have been solved in approximately 10.000 years with a classical supercomputer, while {\it the sampling has been completed} in only 200 seconds through quantum computations. These impressive gains, however, cannot be generalized, and each class of problems has to be studied on its own.

One of the main goals for QC is to improve the underlying hardware of quantum computers, i.e. finding new strategies to build qubits, the physical connection between them, and a proper classical interface to effectively interact with other hardware, while also developing proper quantum algorithms using their properties, finding realistic fields of application in science where this framework can be successfully applied, reaching the so-called ``quantum utility". For a recent review of quantum algorithms, see \cite{Montanaro_2016, Dalzell_2023}.


From a hardware point of view, two main paths (among the others) have been followed by the scientific community, one related to the adiabatic quantum computations \citep{Albash_2018} based on quantum annealers, while the other on the so-called \textit{gate-based} quantum computations and hardware \citep{Hagouel_2012}. 

The quantum annealer is a hardware built following the idea of classical simulated annealers, where the system is cooled to adiabatically reach the minimum energy of a specific configuration \citep{Somma_2012, Rajak_2022}. It specializes in combinatorial optimization problems \citep{Djidjev_2018} by evolving a quantum system toward the ground state of a problem-specific Hamiltonian typically mapped to an Ising model \citep{Kadowaki_1998}. This approach has been followed for a variety of applications \citep{Lucas_2014, Orus_2019}, but it has also faced different challenges related to the hardware implementations of algorithms, even if potential improvements on the classical results have been found \citep{Ronnow_2014, Tasseff_2024}. Very recently, evidence for quantum advantage for this kind of hardware has been reached \citep{King_2025}, even if there are some discussions on these claims, especially on the advantage with respect to classical methods \citep{tindall_2025}.

Gate-based QC is instead based on operations applied on qubits via unitary matrices \citep{van_de_Wetering_2021}, building the so-called quantum circuits whose results can then be measured classically. As this approach is followed in our analysis, more details on this are found in $\S$~\ref{sub_sec_brief_introd}. Gate-based QC is generally more flexible, given that it can be used for a plethora of different applications \citep{Sachdeva_2024}. 

Given the many fields in which a strong computational effort is becoming mandatory, employing quantum computers to integrate classical computations is of great interest to the scientific community. Astrophysics and cosmology, in particular, rely more and more on massive datasets and high-dimensional parameter spaces, whereas traditional computational methods often struggle with scalability and efficiency \citep{Piras_2024}.

In the last decades, the astronomical community has moved from an epoch of data paucity to one of data richness \citep{Comparato_2007}, thanks to both space missions like Gaia \citep{Ripepi_2023}, and ground-based ones like the Very Large Telescope (VLT, \citealt{Lilly_2007}). In this sense, a fundamental observational campaign symbolizing this paradigm shift is the Sloan Digital Sky Survey (SDSS, \citealt{York_2000}), and the impressive size of astronomical data it has gathered. 

To give an idea of the size of astrophysical and cosmological datasets, data release 17 of the SDSS has a volume of 245 TB, to be added to the total volume of the previous data releases, which is equal to 407 TB \citep{Abdurro_uf_2022}. This paradigm change will be further exacerbated with the current and next generations of scientific missions, like \textit{Euclid} \citep{Laureijs_2011, Scaramella_2022, Mellier_2024}, and the Vera C. Rubin Observatory \citep{Abell_2009}. Thus, new strategies that can manage enormous databases in a limited amount of time are mandatory. 


In this context, classical machine learning methods are becoming more and more common in these fields \citep{Sen_2022}, for many tasks like classification \citep{Cavuoti_2014, Brescia_2015, Angora_2020}, estimation of features associated with specific sources of a catalog \citep{Laurino_2011, Brescia_2014, Delli_Veneri_2019, Razim_2021}, feature selection \citep{D_Isanto_2018}. For a complete review, see \cite{Ivezic_2020}. Another important point is the fact that astronomical data analysis benefits strongly from the rapid evolution of classical hardware, like the extensive diffusion of Graphics Processing Units (GPUs). From this point of view, fully exploiting these resources requires a general reworking of the astrophysical and cosmological codes \citep{Taffoni_2020}. 

It is important to note that, related to these classical frameworks, quantum machine learning \citep{Schuld_2014} based on gate formalism is currently being extensively studied \citep{farsian_2025}. Also, quantum computations can be intrinsically parallelized, because superposition allows to perform operations on linear combinations of basis states \citep{Markidis_2024}. This is the reason why we aim to explore possible applications of QC in astrophysics, given that it can be naturally associated with two frameworks that are becoming more and more popular in this field.

To understand the possible beneficial effects of QC in this field, here we derive, using QC properties in a genetic algorithm, the Hubble constant $H_0$, and today's matter density content $\Omega_M$ by modeling data from Supernovae Type Ia (SNe Ia), Baryon Acoustic Oscillations (BAO), and the Cosmic Microwave Background (CMB) Radiation. It is worth stressing that the determination of $(\Omega_M, H_0)$ is a standard task with well-known and accurately working classical methods already available. This allows us to directly test the results obtained with our Amplitude-Encoded Quantum Genetic Algorithm (AEQGA) on a simple problem with known solutions. The aim is to find the same solution with this new QC approach, thus showing its potential before moving to more complicated challenges involving a larger number of cosmological parameters and datasets.

In $\S$~\ref{sec_Data_Model}, we present the cosmological data used in our analysis and how we implemented the cosmological objective function that we are considering for our genetic algorithm. In $\S$~\ref{Sec_our_QGA}, we present AEQGA, explaining each section of the code. In $\S$~\ref{Sec_Results} we show the main results. In $\S$~\ref{sec_comparison} we show comparisons between AEQGA and others, both classical and quantum \citep{Acampora_2021}. Finally, in $\S$~\ref{Sec_Conclusions} we draw some conclusions. In $\S$~\ref{sub_sec_brief_introd} an introduction to quantum computing in general is found. In $\S$~\ref{Sec_Introduction_QGA}, we introduce genetic algorithms, both classical and quantum, already analyzed in the literature, while in $\S$~\ref{App_Reiability_tests} we show the reliability tests performed to confirm the goodness of the results of AEQGA.

\section{Data and model}
\label{sec_Data_Model}
As previously mentioned, we consider three cosmological probes for our analysis: SNe Ia, BAO, and CMB. 
Having three probes, one could ideally combine all of them to define a single objective function as the sum of the $\chi^2$ values of each probe. However, caution is needed because of the so-called $H_0$ tension \citep{Abdalla_2022}, i.e., the fact that estimates of the Hubble constant from local distance estimators (such as Cepheids) strongly disagree with those inferred from early universe probes. Since SNe Ia have been calibrated on the Classical Cepheids following the cosmological ladder approach \citep{Brout_2022, Scolnic_2022}, they can be considered late-time cosmological probes. On the other hand, both the CMB spectrum and the BAO measurements are early-time cosmological probes, so we can safely combine them. Since it is not our aim here to address the $H_0$ tension issue, in the following, we will run our algorithm using SNe Ia and CMB\,+\,BAO separately. This will also allow us to check its performance on different datasets. 

\subsection{SNe Ia} \label{sub_sec_SNe_Ia}


SNe Ia are a particular type of Supernova explosion occurring in binary systems after the accretion of mass of a white dwarf from its companion star. These objects are of fundamental importance for cosmology \citep{Riess_1998, Abdalla_2022}, because of their role as standardizable candles \citep{Phillips_1993}, i.e. the possibility of recovering their luminosity distance with intrinsic physical characteristics similar for all the SNe Ia. The luminosity distance is defined as \citep{hogg-1999}

\begin{equation} \label{luminosity distance}
    d_{\rm L}(z)=(1+z)d_{\rm M}(z)~,
\end{equation}
where $d_{\rm M}(z)$ is the transverse comoving distance
\begin{equation} \label{comoving flat}
    d_{\rm M}(z)=\frac{c}{H_0} \int_0^z \frac{dz'}{E(z')}~,
\end{equation}
where $E(z)$ is defined as
\begin{equation} \label{E(z)}
    E(z)=\frac{H(z)}{H_0}=\sqrt{\Omega_r(1+z)^4+\Omega_M(1+z)^3+\Omega_k(1+z)^2+\Omega_{\Lambda}}~.
\end{equation}
Here:
\begin{itemize}
    \item $\Omega_r$ is the radiation density.
    \item $\Omega_k$ is the spatial curvature density.
    \item $\Omega_{\Lambda}$ is the Dark Energy component.
\end{itemize}
 These parameters refer to the current epoch of the Universe. Under the assumption of flat $\Lambda$ Cold Dark Matter (CDM) model, $\Omega_k=0$ and today $\Omega_r$ is negligible. The measured quantity is the distance modulus, defined as 
\begin{equation}
    \mu_{th, SNe Ia}=m-M=5 \log (d_L)+25,
\label{muth}
\end{equation}
where $m$ is the apparent magnitude of the SN Ia, $M$ its absolute magnitude, and the luminosity distance is expressed in Mpc. The degeneracy between $M$ and $H_0$ is the main reason why SNe IA need to be calibrated on other primary distance indicators, like the previously mentioned Classical Cepheids, as is done in the SH0ES program \citep{Riess_2022}.
This quantity is compared with the observed distance modulus, $\mu_{obs}$ of the SN Ia by defining a $\chi^2$ function. We use the SNe Ia Pantheon+ sample \citep{Scolnic_2022, Brout_2022}, which is a compilation of 1701 light curves gathered from 1550 different sources, and one of the most refined  SNe Ia datasets employed for cosmological computations. Here, the $\chi^2$ function is defined as 
\begin{equation} \label{eq_chi2_SNe}
    \chi^2_{SNe Ia}= (\mu_{th}-\mu_{obs})^T \mathcal{C}_{SNe Ia}^{-1} (\mu_{th}-\mu_{obs})~,   
\end{equation}
 where $\mathcal{C}_{SNe Ia}^{-1}$ is the inverse of the covariance matrix for the Pantheon+ data \citep{Scolnic_2022}, for which we have considered both statistical and systematic contributions, as done in \citep{Sarracino_2022}. All these functions are dependent on the cosmological parameters, among which are those we analyze, $H_0$ and $\Omega_M$. \\
 
To decrease the computational cost, we have precomputed the integral in Eq.~(\ref{luminosity distance}) for 100 and 300 evenly spaced $\Omega_M$ values over the range of interest $(0.0, 0.5)$. For each $\Omega_M$ selected by the algorithm, we look for the closest value in the grid and select the corresponding integral value. We have verified that the mean and standard deviation provided by our quantum genetic algorithm, as described in $\S$~\ref{Sec_Results}, do not change if we increase the number of points in the grid. This means that the tiny loss of precision in the approximation of the luminosity distance integral is a reasonable cost for speeding up the computation of the objective function. This is also the reason why we have not performed linear interpolations on these values. We refer to $\S$~\ref{sub_sec_approximations} for the difference between the pre-computed and analytical $\chi^2$ functions.

 \subsection{BAO} \label{sub_sec_BAO}
{\it Baryon Acoustic Oscillations} are fluctuations in the density of the visible baryonic matter of the universe, caused by acoustic density waves in the primordial plasma \citep{Eisenstein_2005}. We have used a set of 16 BAO-related data points, as done in \citet{Dainotti_2022e, Spallicci_2022, Sarracino_2022}, that have been derived from different observations of galaxy correlations, Lyman $\alpha$ forests, and quasars. This dataset has been gathered from the results shown in \citet{Beutler_2011, Blake_2012, Ross_2015, du_Mas_des_Bourboux_2020, Alam_2021}. We note that, while for the SNe Ia each measurement is associated with the luminosity distance, for the BAO each measurement may be associated with a different cosmological quantity, even if they can all be associated to the angular-diameter distance, typically used for the so-called distance reguli, as the BAO are. These cosmological quantities are defined as

\begin{equation}
        d_V(z) \coloneqq \biggr[{d_M}^2(z)\frac{cz}{H(z)} \biggr]^\frac{1}{3},
    \label{eq_dilationscale}
    \end{equation}
    \begin{equation}
        A(z) \coloneqq \frac{100 d_V(z)\sqrt{\Omega_M h^2}}{cz},
        \label{Aparameter}
    \end{equation}
    \begin{equation}
        d_H(z) \coloneqq \frac{c}{H(z)},
        \label{dH}
    \end{equation}
    and the comoving distance defined in Eq.~(\ref{comoving flat}). Here, $h=H_0/100  \ {\rm km/s/Mpc}$. $d_V(z)$ is known as the volume averaged distance \citep{Carter_2018}, $A(z)$ as the acoustic parameter \citep{Blake_2012}, and $d_H(z)$ is a distance directly defined via $H(z)$.
    
    Some of these measurements are rescaled with the sound horizon $r_d$. Thus, in our computations, we have implemented the following numerical approximation \citep{Aubourg_2015, Sharov_2016}:

    \begin{equation}
    r_d=\frac{55.154 \cdot e^{[-72.3(\Omega_{\nu}h^{2}+0.0006)^2]}}{(\Omega_{M}h^{2})^{0.25351}(\Omega_{b}h^{2})^{0.12807}} Mpc,
    \label{eq_rsfiducialtrue}
    \end{equation}
    where $\Omega_{b}$ is the baryonic density of the universe and $\Omega_{\nu}$ is the neutrino density. Since we are interested in deriving only $\Omega_M$ and $H_0$ in our tests, we set $\Omega_{b}\cdot h^2=0.02237$ \citep{Planck2020} and $\Omega_{\nu} \cdot h^2=0.00064$, corresponding to a total of neutrino masses $\Sigma m_{\nu} \approx 0.06$ \citep{Lesgourgues_2006, Planck2020}. 
We define a $\chi^2$ merit function as:

    \begin{equation} \label{eq_chi2_BAO}
    \chi^2_{BAO}= (D_{th, BAO}-D_{obs, BAO})^T  \mathcal{C}_{BAO}^{-1} (D_{th, BAO}-D_{obs, BAO})~,   
    \end{equation}
    where $D_{BAO}$ is the BAO dataset including the measured values of the quantities shown in Eqs. (\ref{eq_dilationscale}, \ref{Aparameter}, \ref{dH}) eventually rescaled with Eq.~(\ref{eq_rsfiducialtrue}). We retrieve the covariance matrix from \citet{Beutler_2011, Blake_2012, Ross_2015, du_Mas_des_Bourboux_2020, Alam_2021}, by combining all the different measurements and hence including off-diagonal terms in the covariance matrix as they have been reported in the datasets. When the measurements are uncorrelated, we compute $\chi^2_{BAO}$ for each one, and add all the contributions. In total, our BAO dataset is composed of 16 measurements. Out of these, 10 present off-diagonal terms in the covariance matrix, all of which are not negligible. 
    
    \subsection{CMB} \label{sub_sec_CMB}
    
    Being here interested in presenting a novel methodology rather than the constraints themselves, we used only a subset of the latest CMB measurements by the \textit{Planck} mission \citep{Planck2020}. In particular, we downloaded the Temperature-Temperature (TT) Power Spectrum and compared it with the theoretical spectrum computed from the Code for Anisotropies in the Microwave Background (\texttt{CAMB}) python interface \citep{Lewis_2000} fixing all the cosmological parameters but $H_0$ and $\Omega_{M}$ to their best-fit values reported in \cite{Planck2020}. The agreement is quantified by a usual $\chi^2$ function between the measured spectrum and the corresponding squared difference obtained with the theoretical fit evaluated independently for each multipole moment. To reduce the computational time, we tested two strategies. In the first approach, we pre-computed the spectra for three different grids of $\Omega_M$ and $H_0$ values before the evaluation of the objective function. In the second, instead, we used an emulator, Parameters for the Impatient Cosmologist (\texttt{PICO}) \citep{Fendt_2007, Fendt_2007b}, which is able to interpolate the TT spectra in a faster way with respect to \texttt{CAMB}. Among these two strategies, we chose to use \texttt{PICO} for our computations. For details, see $\S$~\ref{sub_sec_approximations}. 

\section{Amplitude-Encoded Quantum Genetic Algorithm} \label{Sec_our_QGA}
 In this section, we describe the implementation of AEQGA.
The peculiarities of the problem we want to address have motivated us to develop a hybrid quantum genetic algorithm performing quantum operations only over a part of it. In particular, the merit evaluation of the $\chi^2$ functions associated with each set of cosmological parameters is computed classically. On the contrary, encoding the population, genetic operations, and decoding are quantum operations. Needing to compute the merit function to find the set of parameters that optimize it somewhat hinders the advantage of a quantum algorithm, as it is also the most computationally expensive step. Nevertheless, the quantum implementation of the crossover and mutation operations allows for a comparison with their classical counterparts, hence highlighting the usefulness of the QC approach. 

As mentioned in $\S$~\ref{sec_Introduction}, our goal is finding the best-fit parameters for cosmological functions in different parameter spaces, in particular for $\Omega_M$ and $H_0$. AEQGA has been designed following the general scheme of genetic algorithms, thus considering as hyperparameters, among the others detailed in $\S$~\ref{sub_sec_overview_Experiments}, the population size (whose individuals are pairs of $\Omega_M$ and $H_0$ values), and the number of generations for each iteration of the genetic algorithm. 

As a starting point, we initialize the population by randomly picking values over the ranges $[0.0, 0.5]$ for $\Omega_M$ and $[60, 80] \ {\rm km/s/Mpc}$ for $H_0$ (all the results shown for $H_0$ inside our analysis are in $ \ {\rm km/s/Mpc}$). We then perform the classical evaluation of the objective function for each individual of the population, and the selection of the best-fitting individuals, i.e., those corresponding to the smallest values of the function. Then, the quantum part of the algorithm proceeds as sketched below:

\begin{itemize}
    \item encoding of the population inside the quantum circuit;
    \item quantum crossover and quantum mutation operations;
    \item quantum decoding.
\end{itemize}

For the evaluation of the objective function, we proceed classically by computing the $\chi^2$ as described in $\S$~\ref{sec_Data_Model}. 

For the selection of the individuals inside the population, we consider the $25\%$ of our population that obtains the best values for the objective function, and we store them for the next step, to be sure that for each generation the new population will be at least as good as the previous one when finding the best-fit individuals. We then repopulate the rest of the population by duplicating the best individuals, to obtain another $25\%$ of the total new population, and randomly pick the rest $50\%$ inside the intervals defined for the starting population. This means that $75\%$ of the population effectively enters the quantum circuits. In summary, at each generation:
\begin{itemize}
    \item We select the $25\%$ best individuals, that do not enter the quantum circuits;
    \item duplicate them to produce another 25\%, resulting in 50\% of the new population, that enters into a specific quantum circuit;
    \item generate the remaining 50\% randomly, on which quantum operations are also performed in another circuit.
    
\end{itemize}

These two subsets of the population (the duplicates and the randomly picked) are then encoded in two different quantum circuits. This is because of the different ranges considered for the decoding (see $\S$~\ref{sub_sec_Quantum_Decoding}), and for matching the number of qubits properly with the classical elements.  Each element of our quantum circuit is described in the following subsections, taking as a reference point 3 qubits. A schematic pseudo-code for this algorithm is shown in Alg. \ref{Alg_AEQGA}.

\begin{algorithm}[H]
\caption{Amplitude-Encoded Quantum Genetic Algorithm}
\begin{algorithmic}[1]
\State \textbf{Input}: Population size $n_p$, number of generations $n_g$, crossover probability $p_c$, mutation probability $p_m$
\State Initialize population $P$ with $n_p$ individuals
\For{$t = 1$ to $n_g$}
    \State Evaluate fitness of each individual in $P$
    \State Select top individuals $P_{\text{elite}} \subset P$ with size $n_p/4$.
    \State Duplication of the top individuals $P_{\text{elite copy}}$.
    \State Generate random population $P_{\text{rand}}$ so that $P_{\text{rand}}$+$P_{\text{elite}}$+$P_{\text{elite copy}}$=$P$.
    \State Quantum encode $P_{\text{elite copy}}$ and $P_{\text{random}}$ into two different circuits
    \For{each dimension $d$}
        \State Apply quantum crossover with probability $p_c$
        \State Apply quantum mutation with probability $p_m$
        \State Measure circuit to decode updated individuals (two different decoding schemes for$P_{\text{random}}$ and $P_{\text{elite copy}}$)
    \EndFor
    \State Combine: $P \gets P_{\text{elite}} \cup P_{\text{decoded}}$
\EndFor
\State \textbf{Return:} Best individual in final population
\end{algorithmic} \label{Alg_AEQGA}
\end{algorithm}

\begin{figure*}\centering 
\includegraphics[width=0.9\hsize]{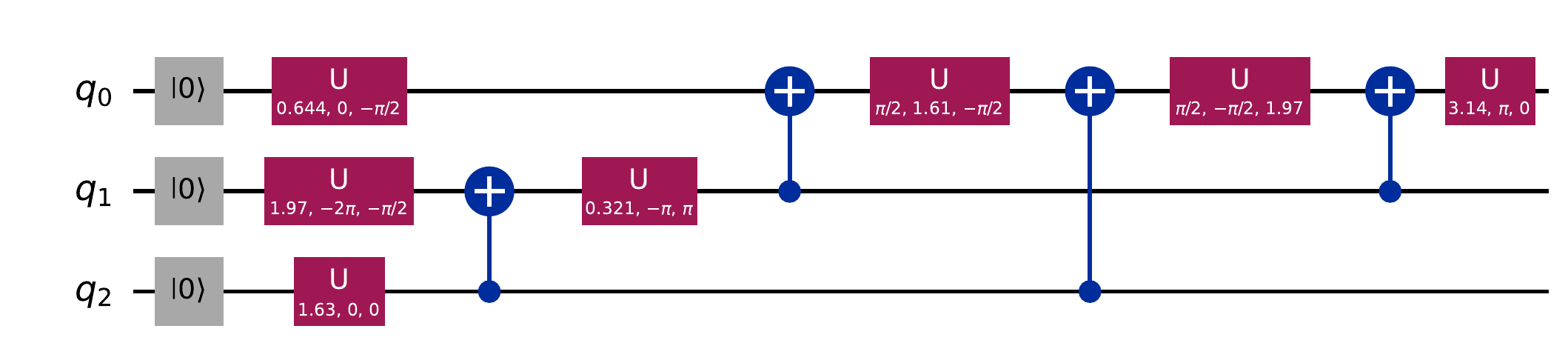}
\caption{The quantum encoding step for a population of 8 individuals and thus in a 3-qubit quantum circuit. The numbers inside the rotation gates depend on the values of the starting classical population. This is the expanded version of the state preparation block shown in green in Fig. \ref{Fig_Quantum_Circuit}. For the insights on the gates used in this circuit, see $\S$~\ref{Sec_our_QGA}.}
\label{Fig_Quantum_Initialization}
\end{figure*}

\begin{figure*} 
\includegraphics[width=1\hsize]{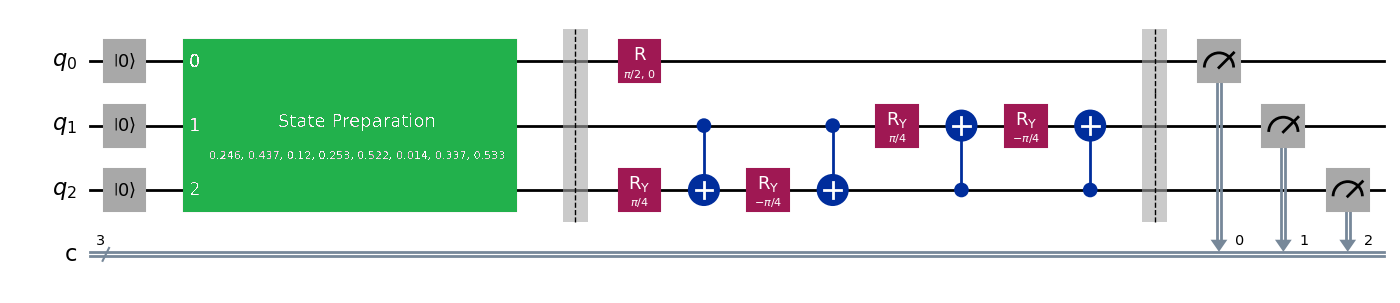}
\caption{The quantum circuit for our algorithm, built for a population of 8 real values encoded in 3 qubits. Apart from the state preparation block previously detailed, the other gates are represent the operations of quantum mutation (on qubit 0) and quantum crossover (on qubits 1 and 2).}
\label{Fig_Quantum_Circuit}
\end{figure*}

\subsection{Amplitude Quantum Encoding} \label{sub_sec_Quantum_Encoding}
 The process of encoding real data in quantum circuits can be performed in different ways. In $\S$~\ref{Sec_Introduction_QGA}, for instance, we describe the binary encoding, which translates a value in its binary notation, and then associates it with the corresponding quantum state using the appropriate number of qubits, which is given by the number of digits in binary notation used to express the number itself. This poses limitations on the classical sets that need to be discretized and limited in depth to perform this encoding. Furthermore, the number of qubits depends on the characteristics of the values rather than on the population size (for instance, on the number of significant digits for a specific number). 

The strategy we followed, instead, is that of amplitude encoding \citep{Gonzalez_Conde_2024}, according to which one normalizes the classical array containing the population and then assigns to each number the amplitude of the quantum states rather than the quantum states themselves. Quantitatively, this is described as follows: Given a classical array of real values $\mathbf{x} = [x_0, x_1, \dots, x_{N-1}]$, this is first normalized so that $\sum_{i=0}^{N-1} |x_i|^2 = 1$. This can be mapped into a quantum state defined as 
\begin{equation} \label{amplitude_encoding}
    |\psi\rangle = \sum_{i=0}^{N-1} x_i |i\rangle,
\end{equation}
where $\{|i\rangle\}$ denotes the computational basis of an $n$-qubit system with $N = 2^n$ states, while $x_i$ are effectively the amplitudes associated to this basis. The main advantage of this procedure is that the number of qubits necessary to encode a given population scales logarithmically with the size of the population itself, and does not depend on the classical values inside the population because it acts on rescaled arrays. The main drawback is that, being the population encoded not by the quantum states but by their amplitude, this does not allow for search routines used in other quantum algorithms, like the ones presented in $\S$~\ref{sub_sec_QGA_lit}. Also, the decoding phase has to be designed accordingly, by considering that the relevant information is not the quantum state itself, but the associated probability.

This encoding, as well as the rest of the circuit, has been implemented using Qiskit\footnote{https://quantum.cloud.ibm.com/docs/en/guides} \citep{Javadi-Abhari_2024}. This package allows us to emulate quantum computers' behaviour in classical ones; indeed, all the results shown in our analysis have been obtained by this emulator. The amplitude encoding can be easily implemented via the initialize function, once the population set, i.e., the vectors with the individual values for $\Omega_M$ and $H_0$, respectively, is normalized accordingly. 

We now describe how we have divided the population in the encoding phase. The elements representing $\Omega_M$ and $H_0$ are inserted in separate but similar and independent quantum circuits, given that the encoding and decoding phases depend on the interval range in which these values are picked. Taking as a reference point a population of 16 individuals, the 4 providing the best values for the objective function do not enter inside any quantum circuit, the 4 duplicates are encoded in a 2-qubit quantum circuit, the remaining 8 in a different 3-qubit quantum circuit. More generally, for a population of $n_p$ individuals, $\log_2 (n_p)-2$ qubits are necessary for the quantum circuit regarding the duplicates of the best subset, and $\log_2 (n_p)-1$ qubits for the circuit associated with the rest of the population. The encoding phase is the same for both circuits. This is also the main reason why we have used the aforementioned percentages to divide our individuals, so that they match exactly the circuit we build according to this encoding strategy.

In Fig.~\ref{Fig_Quantum_Initialization} we present the decomposition of the initialize function of Qiskit for the encoding of a population composed of 8 individuals, corresponding to the first block shown in Fig.~\ref{Fig_Quantum_Circuit}, where the entire circuit is shown. In this plot, the $U$ gates represent both the most general rotation of three Euler Angles, as in Eq.~(\ref{rotation}) shown in the $\S$~\ref{sub_sec_brief_introd}, while the operations linking two qubits are the CNOT gates, defined in Eq.~(\ref{CNOT}). The $R$ gates are instead rotations around specific axes that will be detailed in the next subsection. We note that a certain degree of entanglement is achieved by the combined use of rotation and CNOT gates, because they link all the qubits inside the circuit. 

\subsection{Quantum Crossover and Mutation} \label{sub_sec_Quantum_Crossover_Mutation}
Because of the encoding strategy of the classic populations, what can significantly modify the population after the final measurements of the quantum circuit are operations that modify the probability associated with a quantum state rather than the quantum state itself. This is because, as we have encoded our results in the amplitudes, what we want to focus on is not how many times a particular state is measured, as it would have been the case if we were to use the binary encoding, but the overall distributions of the measurements, which are more strictly linked to their overall distributions of amplitudes. Thus, we have chosen to represent both the genetic operations via rotations around the $y$ and $x$ axes, being the rotation around the $z$-axis a phase shift irrelevant to the final probability. The quantum crossover is defined as a coupled conditioned rotation around the $y$-axis performed on two randomly picked qubits. The rotation has been chosen to be $\pi/2$ to maximize the variation on the probability associated with the involved quantum states due to the crossover. Mathematically, following the formalism of quantum gates, this operation is described as follows, taking the first qubit as the control and the second as the target:
\begin{equation}
    \text{CR}_y\left(\frac{\pi}{2}\right) = |0\rangle\langle 0| \otimes I + |1\rangle\langle 1| \otimes R_y\left(\frac{\pi}{2}\right),
\end{equation}
with
\begin{equation}
    R_y\left(\frac{\pi}{2}\right) = \frac{1}{\sqrt{2}} \begin{pmatrix} 1 & -1 \\ 1 & 1 \end{pmatrix}.
\end{equation}
Then, the quantum crossover is completed by performing the same operation but switching the roles of the qubits. The total matrix $U_{\text{cross}}$ obtained by combining these two operation is thus 
\begin{equation} \label{matrix_crossover}
    U_{\text{cross}} =
\begin{pmatrix}
1 & 0 & 0 & 0 \\
0 & \frac{1}{\sqrt{2}} & -\frac{1}{2} & -\frac{1}{2} \\
0 & 0 & \frac{1}{\sqrt{2}} & -\frac{1}{\sqrt{2}} \\
0 & \frac{1}{\sqrt{2}} & \frac{1}{2} & \frac{1}{2}
\end{pmatrix}
\end{equation}

For the quantum mutation, instead, we have chosen to schematize it as a rotation of $\pi/2$ around the $x$-axis, performed randomly on one of the qubits in the quantum circuit. This operation is expressed by the matrix $U_{\text{mut}}$ defined as
\begin{equation} \label{matrix_mutation}
    U_{\text{mut}}=R_x\left(\frac{\pi}{2}\right) = 
\frac{1}{\sqrt{2}} \begin{pmatrix}
1 & -i \\
-i & 1
\end{pmatrix}
\end{equation}

Both quantum operations are shown in the second block of Fig.~\ref{Fig_Quantum_Circuit}. In this plot, the mutation is expressed by the rotation gate on the first qubit, while the crossover is represented by the combination of rotations and CNOT gates operating on the second and third qubits.

\subsection{Quantum Decoding} \label{sub_sec_Quantum_Decoding}
The last step is the classical measurements. Here, the number of simulations of the quantum circuit has to be provided, as we are emulating it on a classical machine via Qiskit. Then, different information can be extracted, such as the probabilities associated with different quantum states and the final state vector of the system. We stress again that for AEQGA what is important are the associated probabilities to the quantum states rather than the singular measurements on the states themselves, to underline the difference with the binary encoding. Hence, we have to entirely focus on these for our decoding strategy.

We have followed two different paths for the two quantum circuits necessary for the entire set of individuals. 

For the 8 individuals randomly drawn, the decoding normalizes the number of outcomes for each quantum state, mapped in the interval range of the population, so that the highest value is mapped in the upper limit of the population range ($0.5$ for $\Omega_M$ and $80$ for $H_0$) and the lowest number in the lower limit ($0$ for $\Omega_M$ and $60$ for $H_0$)\footnote{We note that, if more than one state presents 0 measurements, these would all correspond to the lower limit of the populations ($0$ for $\Omega_M$ and $60$ for $H_0$), thus not allowing to explore the entire interval range with all the individuals. To avoid this possibility, we have substituted the cases in which the exact lower limit is obtained by the algorithm with a random value in the proper interval range.}. Quantitatively, this operation in general can be summarized as follows
\begin{equation} \label{deconding_1}
    x_i = a + (b - a) \cdot \frac{c_i - c_{\min}}{c_{\max} - c_{\min}},
\end{equation}
where  $[a, b]$ is the mapped interval, $c_i$ is the measurement count for the $i$-th quantum state, $c_{\min} = \min_i c_i$, and $c_{\max} = \max_i c_i$.
\begin{figure*}[ht]
\centering
\includegraphics[width=0.46\hsize]{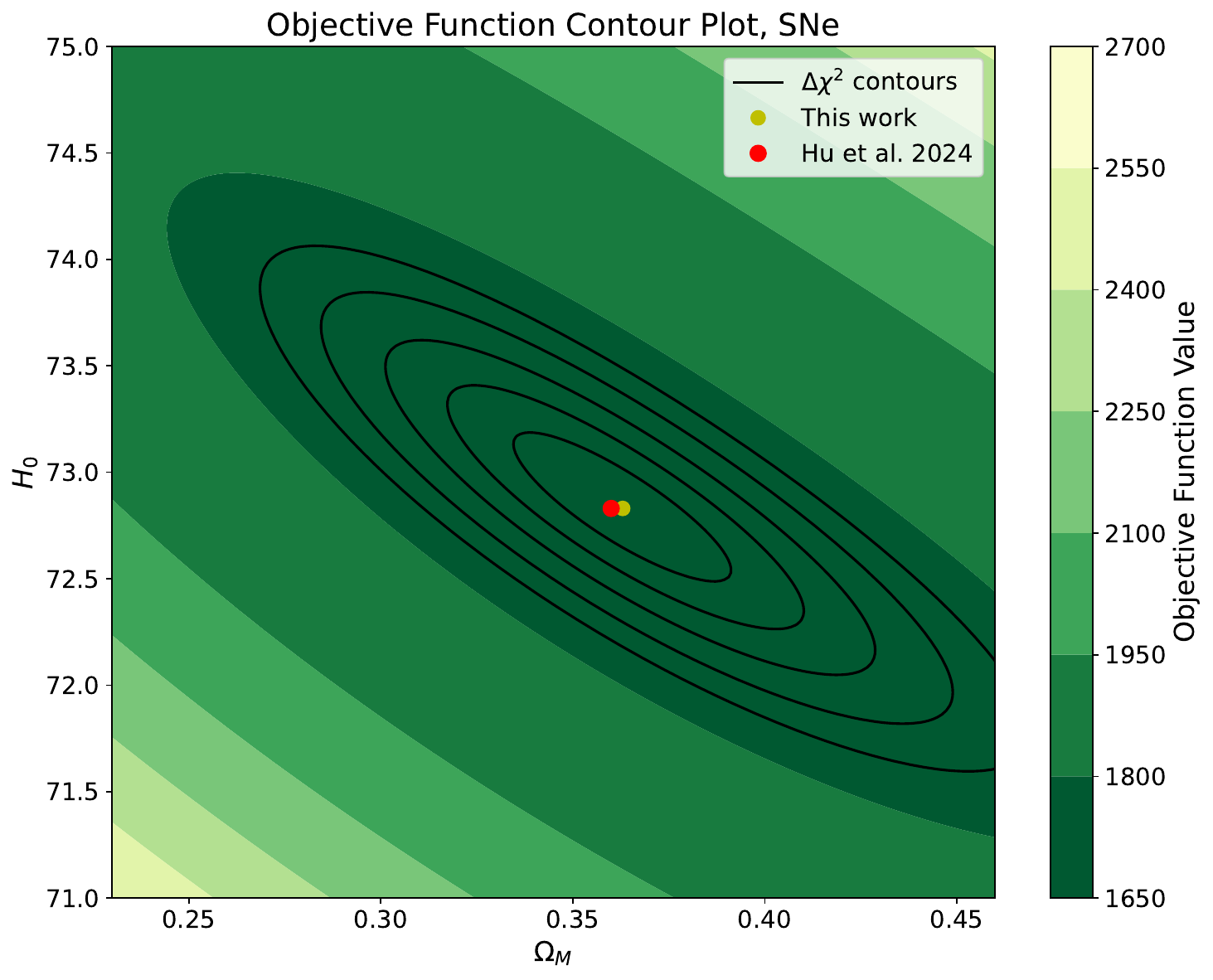}
\includegraphics[width=0.46\hsize]{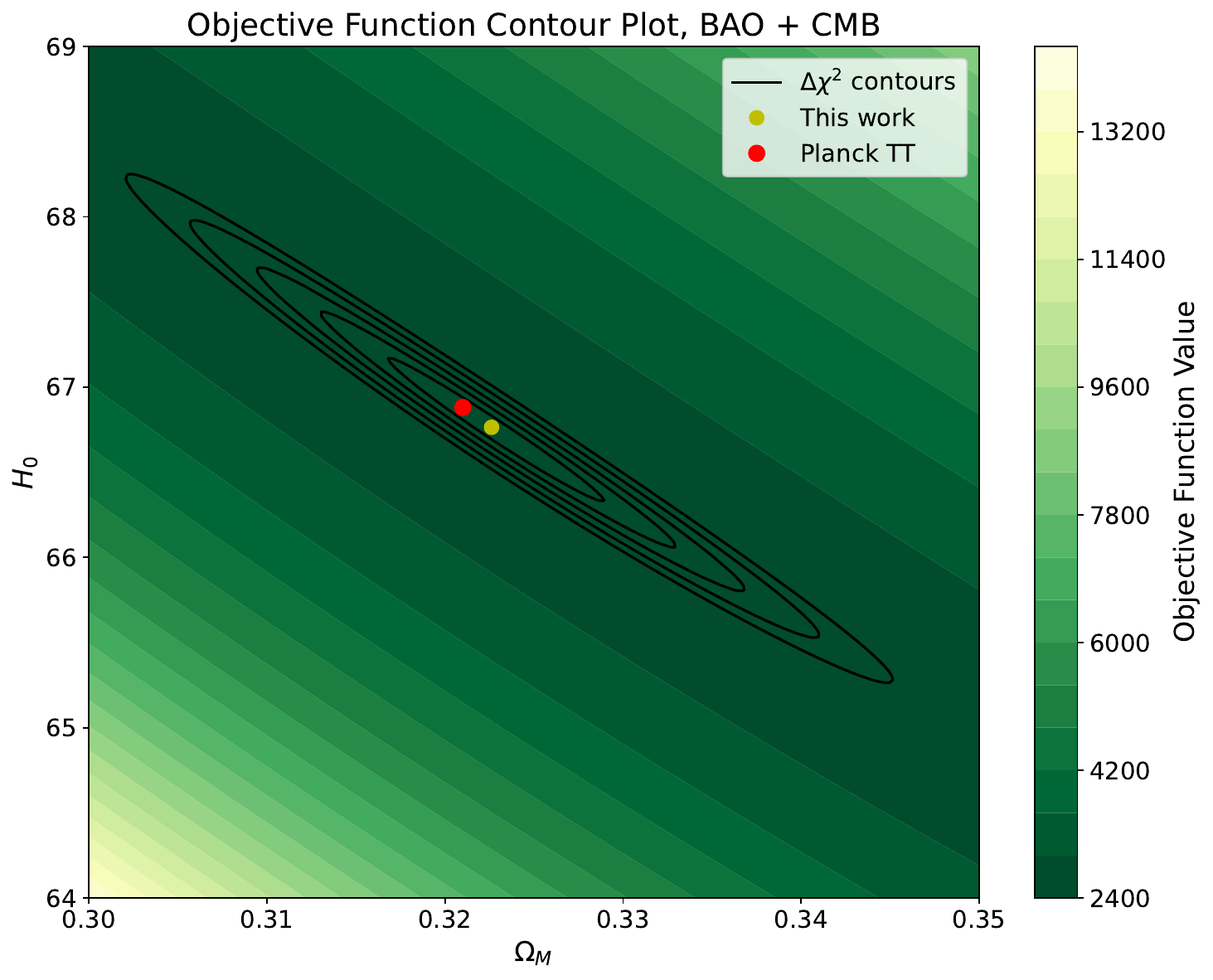}
\caption{Analytical contour map of the objective functions we study in this work. Left panel: for the SNe Ia. Right panel: for CMB\,+\,BAO. The black lines represent the $1,2,3,4$ and $5 \sigma$ analytical contours. With red dots, two results (one for SNe Ia, the other for CMB) found in the literature to compare with our objective functions.}
\label{Fig_Contour_Maps}
\end{figure*}

For the 4 individuals originated by the duplication of the best pairs from the previous generation, instead, a different decoding process has been considered. Instead of spanning the entire population range, a box has been chosen around the region delimited by the best individuals from the previous population, with the idea of allowing this subset to `` remember" the results from the previous generation, without completely disrupting them. Here, the values of the individuals are not mapped directly in the entire interval range, but derived from the percentages of the measurements associated with each quantum state. Mathematically, this operation is described as follows: we select a restricted interval $[n_{\min}, n_{\max}]$ around the region defined by the best individuals of the previous generation, and the new individuals from the decoding are derived from 
\begin{equation} \label{deconding_2}
    x_i = n_{\min} + (n_{\max} - n_{\min}) \cdot \sqrt{p_i},
\end{equation}
where
\begin{equation}
    p_i = \frac{c_i}{\sum_j c_j},
\end{equation}
are the percentages of the given measurements. The $\sqrt{p_i}$ has been used phenomenologically to find values that are more probable in the central region of the box in which we have limited the best individuals. The quantum decoding is summarized in Alg. \ref{Alg_Quantum_Decoding}.

\begin{algorithm} 
\caption{AEQGA Decoding from Quantum Measurements}
\begin{algorithmic}[1]
\State \textbf{Input}: Quantum circuit $\mathcal{C}$ (with measurements), number of qubits $n_{\text{q}}$, decoding flag $\texttt{flag}\in\{\texttt{best},\texttt{random}\}$, interval $[n_{\min},n_{\max}]$, shots $S$ 
\State Initialize simulator with seed $s$ and run $\mathcal{C}$ for $S$ shots to obtain counts
\If{$\texttt{flag} = \texttt{best}$} \Comment{Restricted box around elites; probability–based decoding}
    \State $C \leftarrow \sum_i c_i$ \Comment{total measurements}
    \State $p_i \leftarrow  c_i/C$
    \State $x_i \leftarrow n_{\min} + (n_{\max}-n_{\min})\cdot \sqrt{p_i}$ \Comment{$\sqrt{p_i}$ focuses samples toward box center}
\Else \Comment{$\texttt{random}$: full-range min–max decoding}
    \State $c_{\min} \leftarrow \min_i c_i$, \quad $c_{\max} \leftarrow \max_i c_i$
    \State $x_i \leftarrow n_{\min} + (n_{\max}-n_{\min})\cdot \dfrac{c_i - c_{\min}}{c_{\max}-c_{\min}}$
    \EndIf
    \State \textbf{(Post–process)}: if any $x_i = n_{\min}$ exactly, replace it with a draw from $\mathcal{U}(n_{\min}, n_{\max})$

\State \textbf{Output}: Decoded vector $\mathbf{x}$ of the classical individuals for the populations
\end{algorithmic} \label{Alg_Quantum_Decoding}
\end{algorithm}

\subsection{Overview of Hyperparameters and Experiments} \label{sub_sec_overview_Experiments}

After the quantum decoding, a new generation starts with a novel merit evaluation, repeating the cycle. After the last generation, the best-fit values for $\Omega_M$ and $H_0$ are saved (i.e., the pair corresponding to the lowest value for the objective function in the algorithm). Then, AEQGA is run for different iterations, to find a mean for the obtained values of $\Omega_M$ and $H_0$ and to compute the related uncertainties from the distribution of results. Thus, as a summary, AEQGA contains the following hyperparameters, which can be regulated:
\begin{itemize}
    \item The number of individuals inside the population, $n_p$, directly linked to the number of qubits used for the quantum circuit. For the experiments concerning how the results are influenced by the number of generations and the crossover and mutation probabilities, it has been set to 16. This corresponds to two circuits of 2 and 3 qubits, for the duplicated and randomly picked parts, respectively. We also recall that the $\Omega_M$ and $H_0$ populations go into different independent circuits. Instead, for the plots concerning one run of the algorithm given a certain number of iterations, this number has been set to 32, because of an interesting effect on the results given by $n_p$ which will be detailed in $\S$~\ref{sub_sec_experiments_generation_individuals}. 
    \item The number of generations for a single iteration of the genetic algorithm, $n_g$. For all the experiments but the ones related to the influence of this hyperparameter on the results, it has been set to 50.
    \item The number of iterations needed for a statistical analysis on the precision of the results, $n_i$, set to 300.
    \item Crossover and mutation probabilities inside the quantum circuit. The effects on the precision of the results are extensively studied in $\S$~\ref{sub_sec_experiments_crossover_mutation}.
\end{itemize}

Having only two cosmological parameters, we can compute the objective function over the full 2D plane, draw contours, and directly look for the best fit. Note that this brute force approach can hardly be performed for a much larger number of parameters. This is why we focused on a problem with a small number of dimensions as the first application of AEQGA. Fig.~\ref{Fig_Contour_Maps} shows the analytical contours of equal objective function when using only SNe Ia (left) or CMB\,+\,BAO (right). The global minimum sits in $(\Omega_M, H_0) = (0.363, 72.82)$ for SNe Ia, and in $(\Omega_M, H_0) = (0.323, 66.76)$ for CMB\,+\,BAO with no evidence of further local minima. In these plots are also reported two results found from the literature. On the left panel, the one taken from \citet{Hu_2024} for the Pantheon+, chosen because in their analysis they vary only $(\Omega_M, H_0)$ as in our case. On the right panel, instead, the one related only to the CMB TT spectrum taken from \citet{Planck2020}, with the caveat that their analysis considers directly other cosmological parameters and does not consider BAO for that specific result. Nevertheless, in both cases, they are consistent with our objective functions.

\section{Results and Comparisons} \label{Sec_Results}

For our computations, we have used a machine made up of 1 CPU, model intel(R) Core(TM) i9-10980XE CPU @ 3.00GHz. The CPU has 18 cores and 2 threads per core. The RAM of the hardware is 128 GBs. For our purposes and analyses, we used up to 8 cores, usually divided for different runs. A single run of AEQGA with 4 cores, $n_p=16$, $n_g=50$, and $n_i=300$ takes around 3900 seconds and less than 1 GB of RAM during its run. This computational time scales linearly with $n_p, n_g,$ and $n_i$. As previously mentioned, the most time-consuming part of AEQGA is the evaluation of the merit function, followed by the simulations of the quantum circuit in the measurement phase. 

\begin{figure*}
\centering
\includegraphics[width=0.45\hsize]{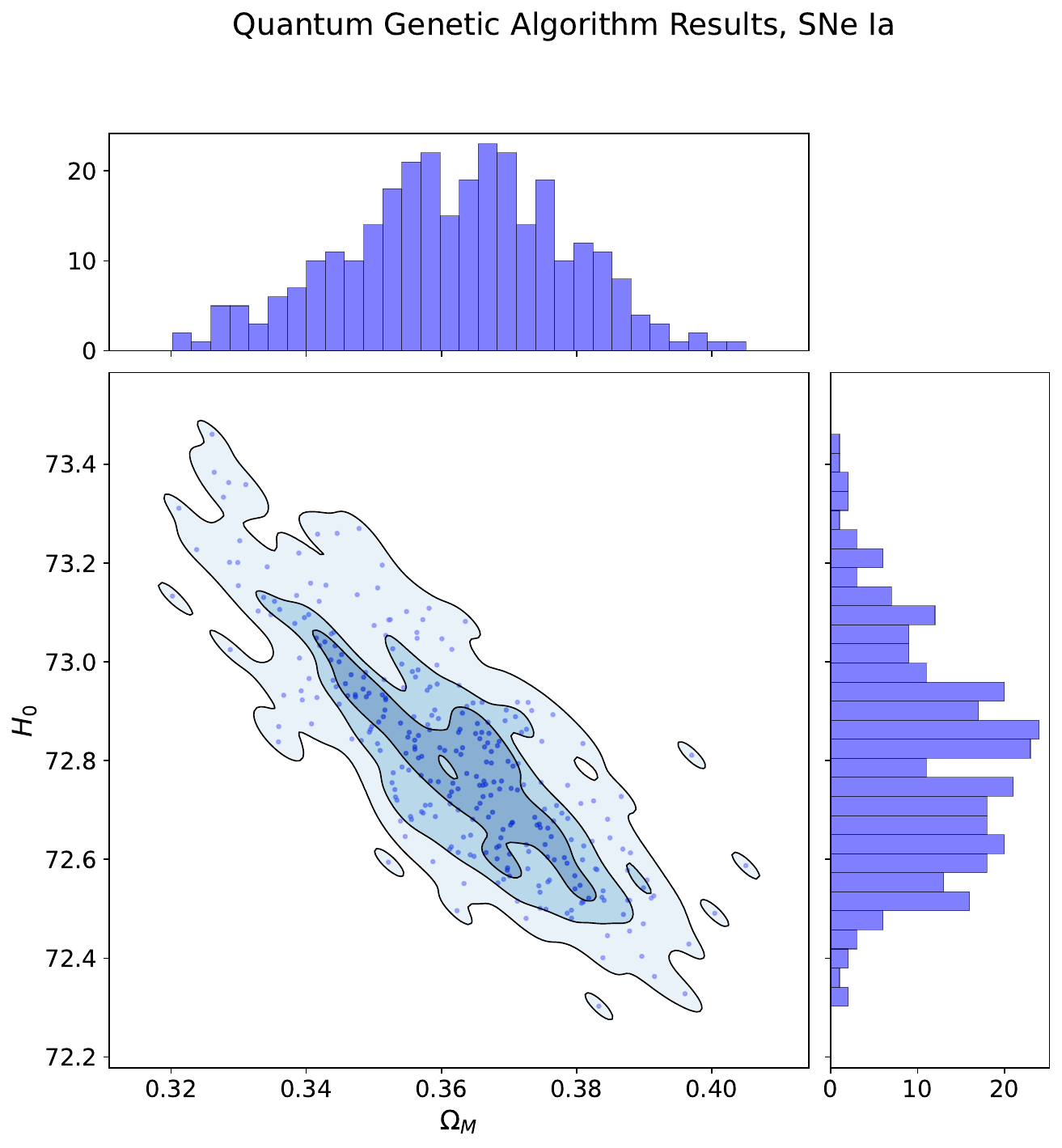}
\includegraphics[width=0.45\hsize]{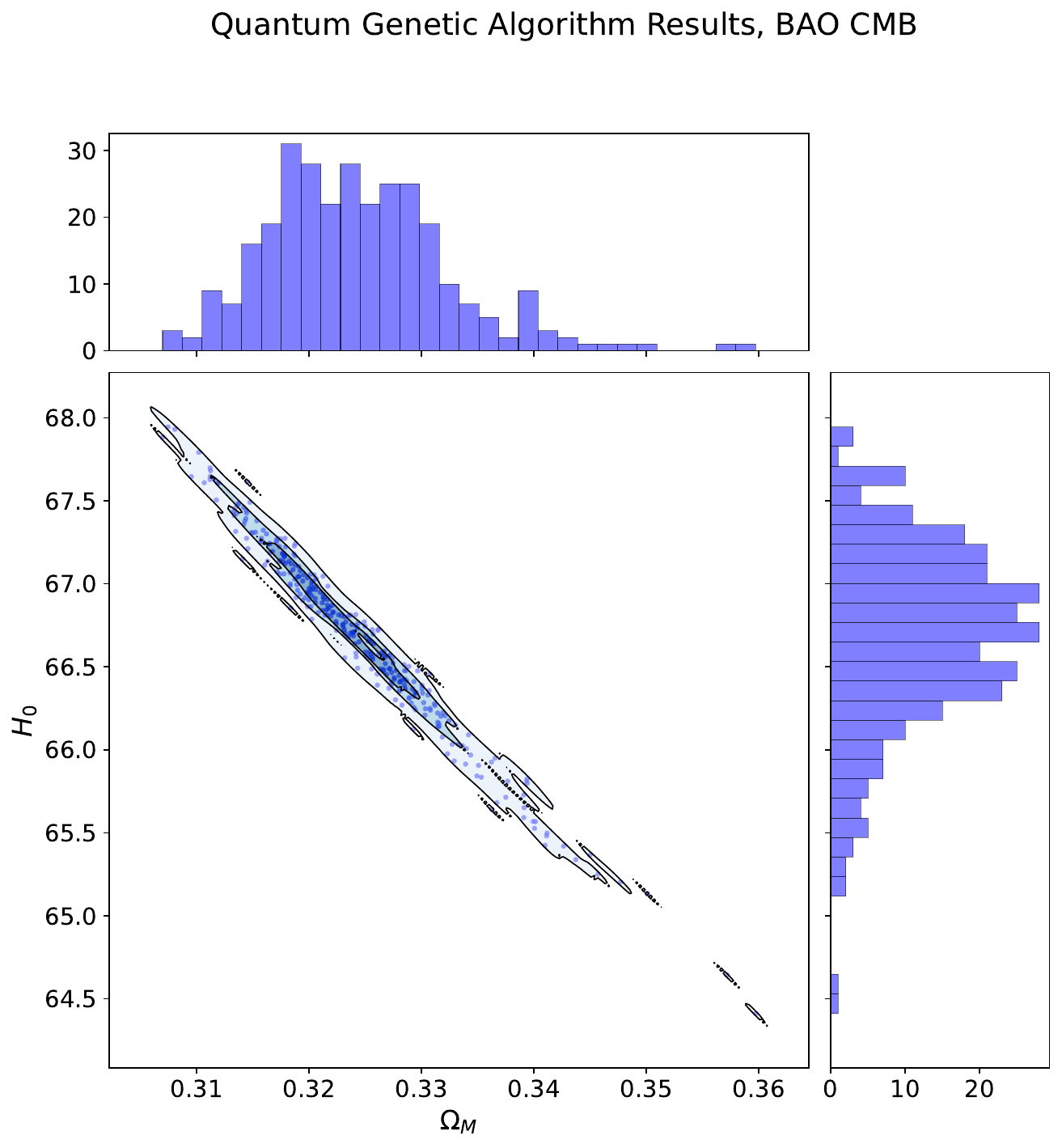}
\caption{Results of the quantum genetic algorithm. Left panel: for the SNe Ia, $n_i=300, n_g=50,$ and $n_p=32$, crossover and mutation probabilities= 0.5. Right panel: the same for the CMB\,+\,BAO sets. We recall that $n_i$ is the number of iterations, $n_g$ is the number of generations, and $n_p$ is the number of individuals inside the population.} 
\label{Fig_Quantum_results}
\end{figure*}

In Fig.~\ref{Fig_Quantum_results} we show two examples of results for our algorithm. On the left panel for SNe Ia, on the right one for CMB\,+\,BAO.
In both cases, we set the quantum crossover and mutation probabilities to 0.5, which corresponds to one of the points with the highest precision on the results (as we will show in $\S$~\ref{sub_sec_experiments_crossover_mutation}). Here, $n_p=32$ and $n_g=50$ (Tests on the influence on the results from the values of $n_p$ and $n_g$ are detailed in $\S$~\ref{sub_sec_experiments_generation_individuals}, here we show only an example run of AEQGA). We note how the results are consistent with the maps of the objective function shown in Fig.~\ref{Fig_Contour_Maps} for both cases. Indeed, the mean and the standard deviations of the 300 iterations for AEQGA are ($0.362 \pm 0.016$) for $\Omega_M$ and ($72.81 \pm 0.22$) for $H_0$ for SNe Ia, and ($0.324 \pm 0.008$) for $\Omega_M$ and ($66.68 \pm 0.56$) for $H_0$ for CMB\,+\,BAO. In both cases, we obtain results that are consistent with the true minima of the objective functions, shown in $\S$~\ref{sub_sec_overview_Experiments}, within $1 \sigma$.

We emphasize that the errors obtained by AEQGA on the results are purely statistical, computed by computing the mean and variance on the different positions found by the 300 iterations of the algorithm for each run, thus having a fundamentally different meaning from, for instance uncertainty contours that one could obtain by using Bayesian methods, which is a fundamental method widely used in cosmology. Indeed, our aim is to minimize the $\chi^2$ function, and not find posterior probabilities around the minimum.

Having proven that the minima found by AEQGA are consistent with those of the real cosmological objective functions, we now show the analysis regarding the influence of the hyperparameters of AEQGA. 

\subsection{Experiments on crossover and mutation}
\label{sub_sec_experiments_crossover_mutation}
\begin{figure*}\centering
\includegraphics[width=0.49\hsize]{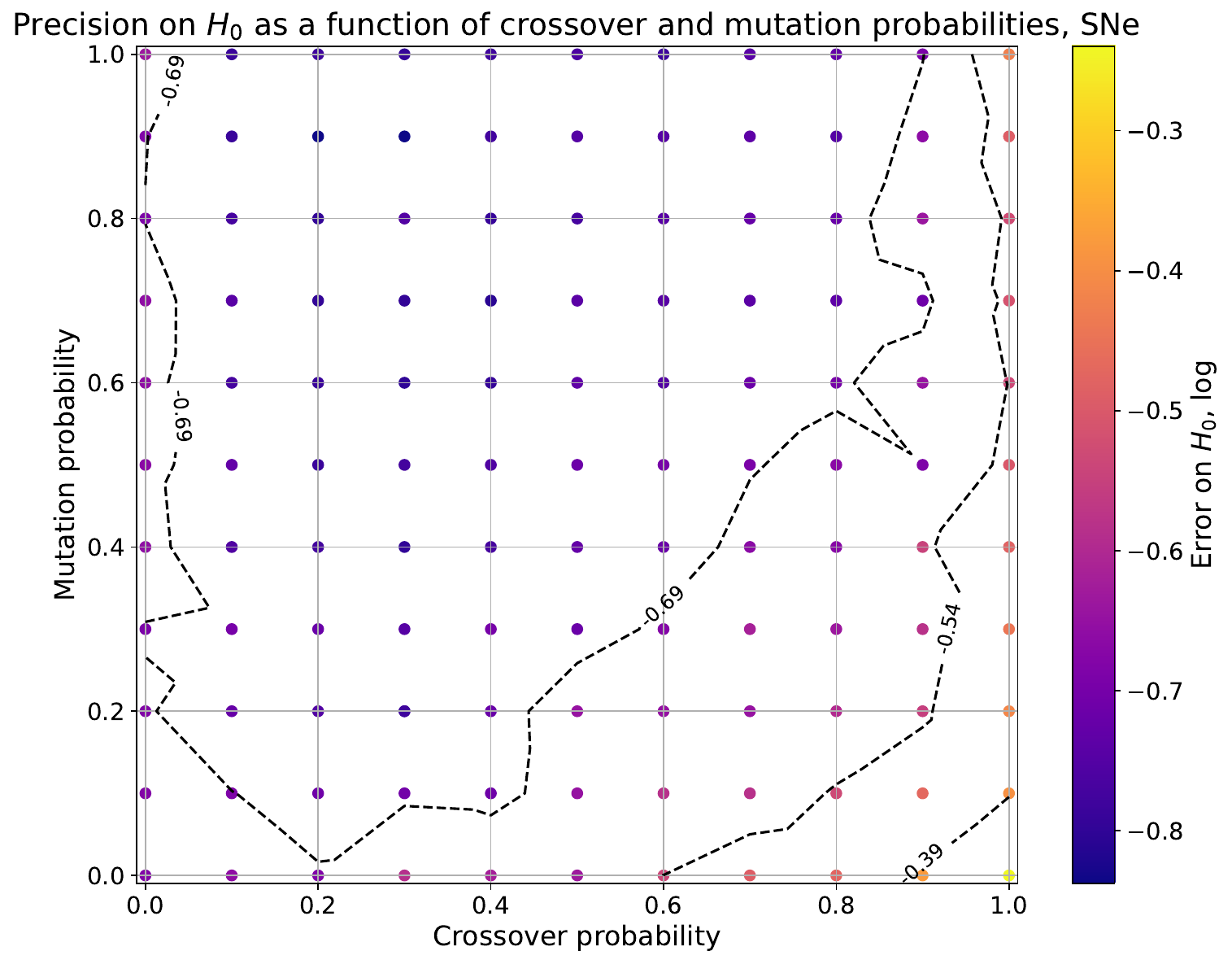}
\includegraphics[width=0.485\hsize]{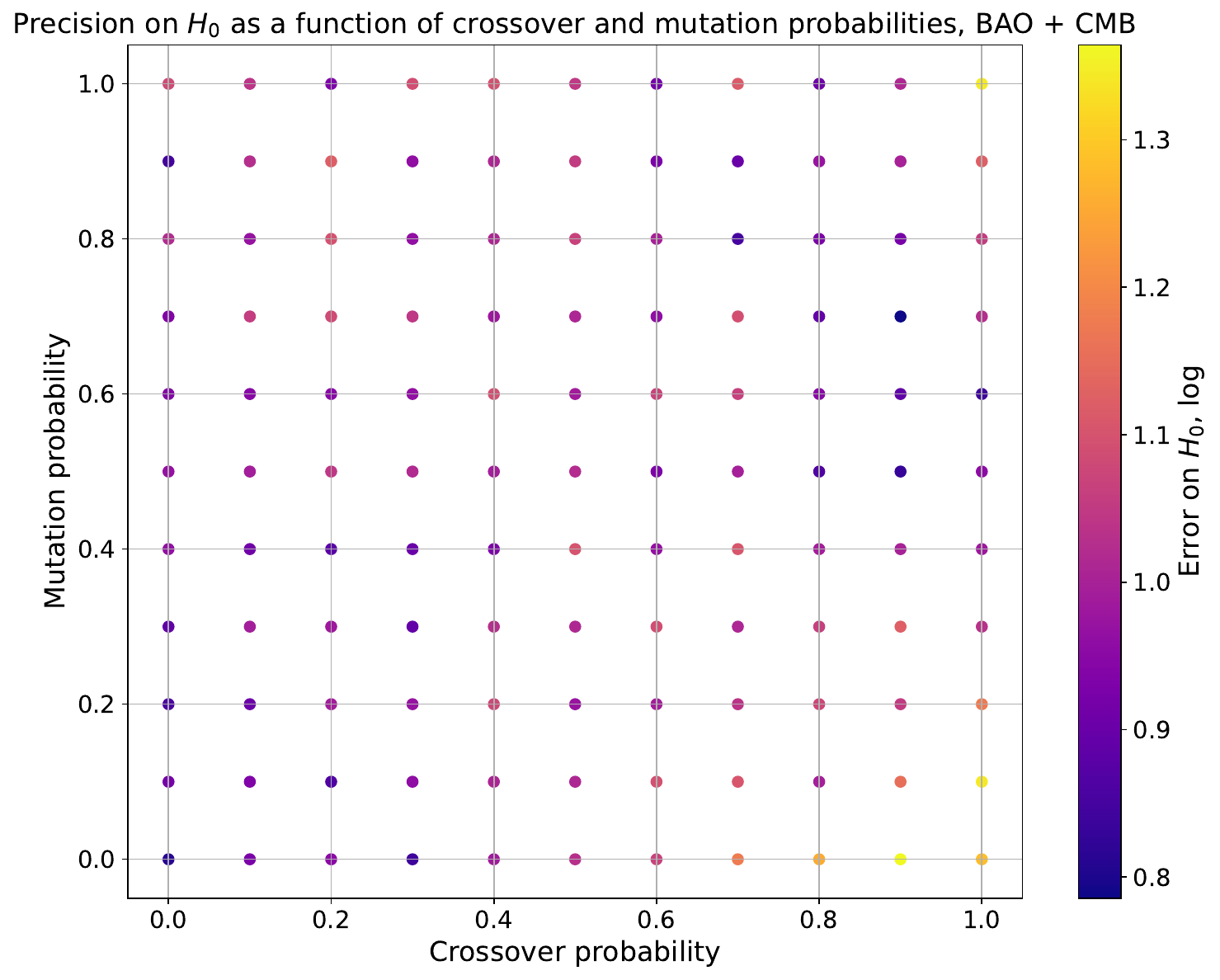}

\caption{The standard deviations on $H_0$ as a function of the crossover and mutation probabilities. The standard deviation as well as the contour levels are expressed in a logarithmic scale. Left panel: for the SNe Ia.  Right panel: for the CMB\,+\,BAO set. The contour levels are shown only for the SNe IA results because they show a clear trend, which is not the case for the  CMB\,+\,BAO ones. }
\label{Fig_Crossover_mutation_results}
\end{figure*}

We study the dependency with the probabilities associated with quantum crossover and mutation, obtaining results from AEQGA for different pairs of them, moving from probabilities 0 (0\%) to 1 (100\%) of these computations to happen inside the quantum circuit with a step of 0.1. For these tests, $n_p=16$ and $n_g=50$. The results for $H_0$ are shown in Fig. \ref{Fig_Crossover_mutation_results}. We note how for the SNe Ia the worst results in terms of precision are at the extreme values for the hyperparameters, especially for crossover probability = 1. This is consistent with what one would expect from a typical classical genetic algorithm, in which crossover and mutation probabilities are chosen to make these operations significant but do not completely disrupt the initial population. This observation indicates that our definitions of quantum crossover and mutation have positively impacted the exploration of the parameter space of the population, thereby enhancing the precision of the results for appropriate values of the related probabilities. The regular behavior of our results with the hyperparameters has enabled us to draw contour levels depending on the standard deviation of the results for the SNe Ia, as shown in the left panel of  Fig.~\ref{Fig_Crossover_mutation_results}. The minimum value for the precision on $H_0$ and $\Omega_M$ has been obtained for (0.2, 0.9), but we also note that many uncertainties on the results are close one to the other, as shown in the following.

We now discuss the results concerning the CMB\,+\,BAO set. In this case, there is no defined trend correlated with the probabilities, differently from the SNe Ia case. The main point in common between the two maps is that the results are generally worse for crossover probability 1.0, but there is no clear region where we can find results with better precision, as shown in the right panel of Fig.~\ref{Fig_Crossover_mutation_results} for $H_0$ (the same discussion applies to $\Omega_M$). 
This means that the results are more stable when we vary crossover and mutation probabilities than the SNe Ia computations. Regarding the lowest standard deviation, it has been obtained for (0.9, 0.7), followed by (0.0, 0.0) (thus without both crossover and mutation).

\begin{figure}
\centering
\includegraphics[width=0.95\hsize]{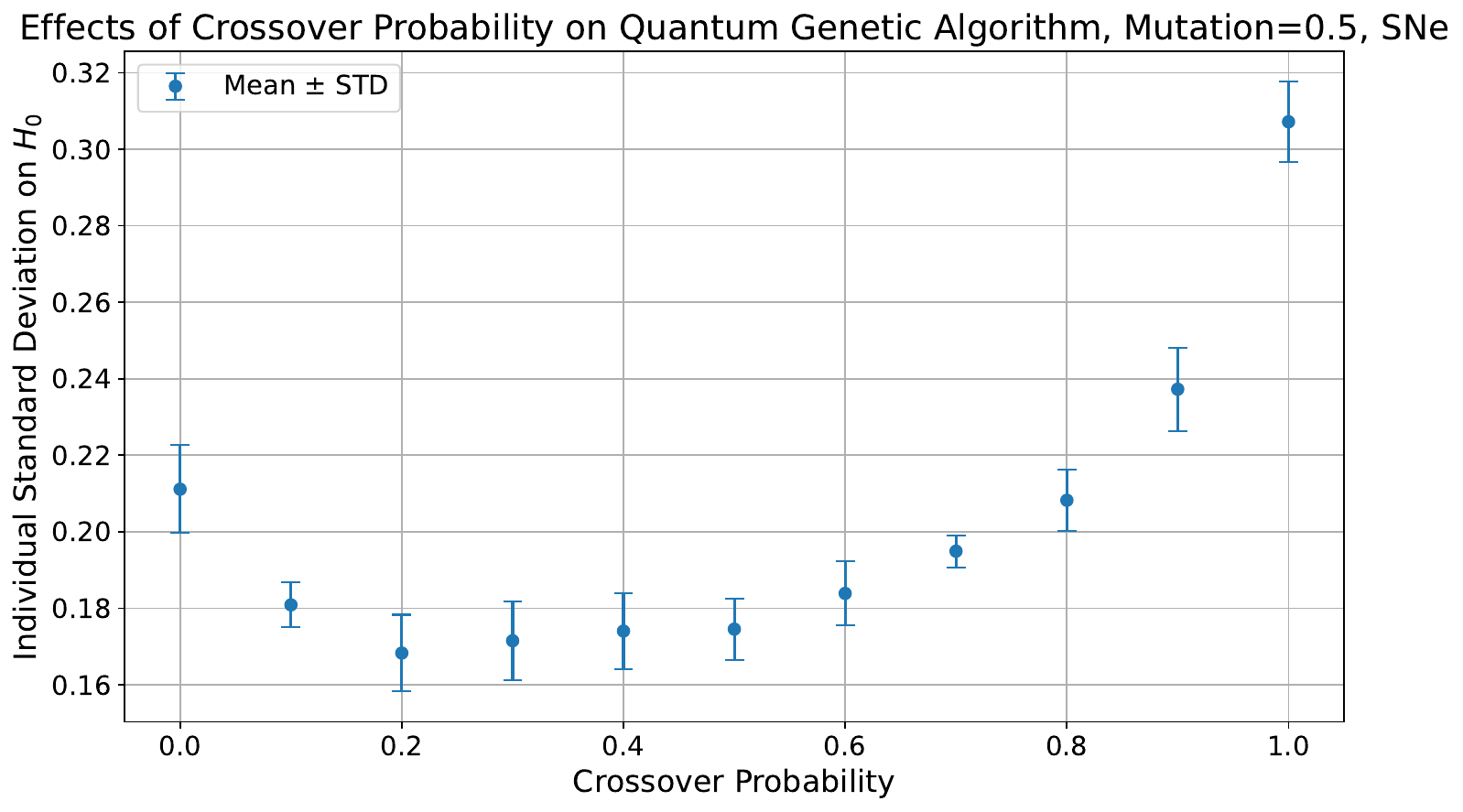}
\includegraphics[width=0.95\hsize]{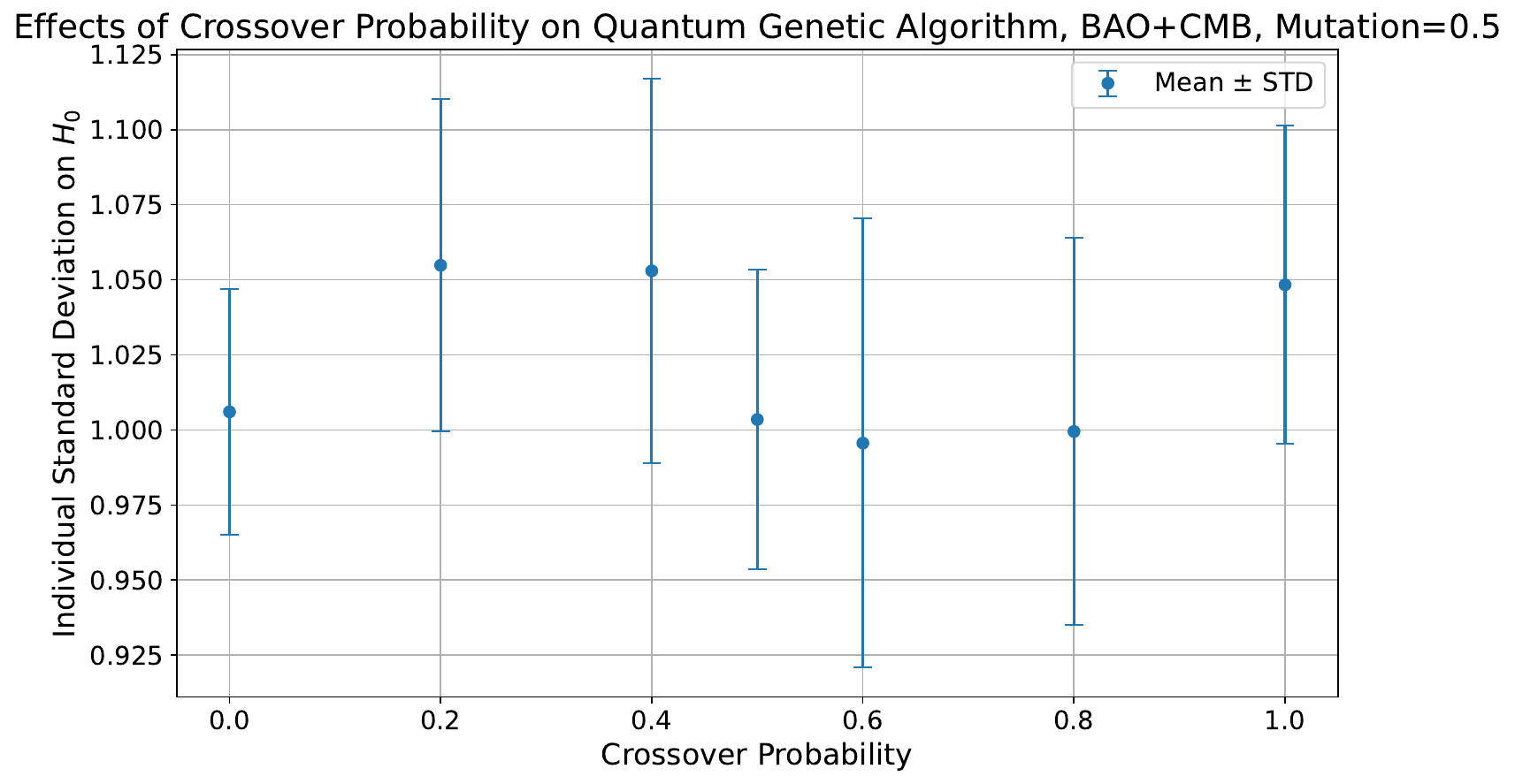}
\caption{Fixing the value of the mutation probability, the mean and standard deviation for the uncertainties on $H_0$ obtained by running AEQGA 10 times. Top panel: for the SNe Ia. Bottom panel: for the CMB\,+\,BAO set. }
\label{Fig_Std_mean}
\end{figure}

To further study the dependency of the precision of the results on the crossover and mutation probabilities, we choose some pairs of these hyperparameters and run the algorithm 10 times, to evaluate the mean and standard deviation of the errors with the goal of understanding if the differences between the points in the maps shown in Fig.~\ref{Fig_Crossover_mutation_results} are relevant or not. The results of this analysis, obtained by fixing the value of the mutation probability to 0.5, are shown in Fig.~\ref{Fig_Std_mean}, on the top panel for the SNe Ia, on the bottom for CMB\,+\,BAO. In both cases, the data shown refer to $H_0$, but the behavior for $\Omega_M$ is very similar.

SNe Ia results are characterized by a general trend in the uncertainty. A minimum is reached for intermediate values of the crossover probability, while the error increases both for very low and high values of this hyperparameter. We also note that, considering the standard deviation, the values inside the lowest contour level in the top panels of Fig.~\ref{Fig_Crossover_mutation_results} are statistically equivalent. This check confirms the discussion presented in the previous paragraph regarding the results of the SNe Ia. This analysis has also been performed on other interesting points in the parameter space; more specifically, (0.2, 0.9), where the minimum on the uncertainties of the results has been found from the map, (0, 0), (1,0), (0,1), (1,1), confirming the results obtained by the map itself, while also proving that the points (0.2, 0.9) and (0.5, 0.5) are statistically equivalent for the precision on the results. Thus, for the other tests, we keep the crossover and mutation probabilities for the SNe Ia at (0.5, 0.5), symbolizing an equal possibility of quantum crossover and mutation operations inside AEQGA.

Regarding CMB+BAO, instead, we do not see a strong variability of the results as a function of the crossover probability if we fix the mutation at 0.5, as we did in the top panel of Fig.~\ref{Fig_Std_mean}. This effect could be due to the dispersion of the results, which is generally higher than the SNe Ia case after the rescaling for the mean value of the uncertainties. It can also be the reason why there is stability on the two hyperparameters that emerge from the bottom panels of Fig.~\ref{Fig_Crossover_mutation_results}. The results directly analyzed, which are different and present the highest precision, are the ones corresponding to (0,0) and (0.9, 0.7), chosen by us because they correspond to the highest precision results found in our map. In summary, taking as a reference point the (0, 0) probabilities for crossover and mutation, we can conclude that in the SNe Ia case we find regions in which such probabilities find more precise results for probabilities different from 0. This instead cannot be concluded for the CMB+BAO tests, for the reasons previously detailed.

\subsection{Experiments on number of generations and individuals} \label{sub_sec_experiments_generation_individuals}

\begin{figure*}
\centering
\includegraphics[width=0.45\hsize]{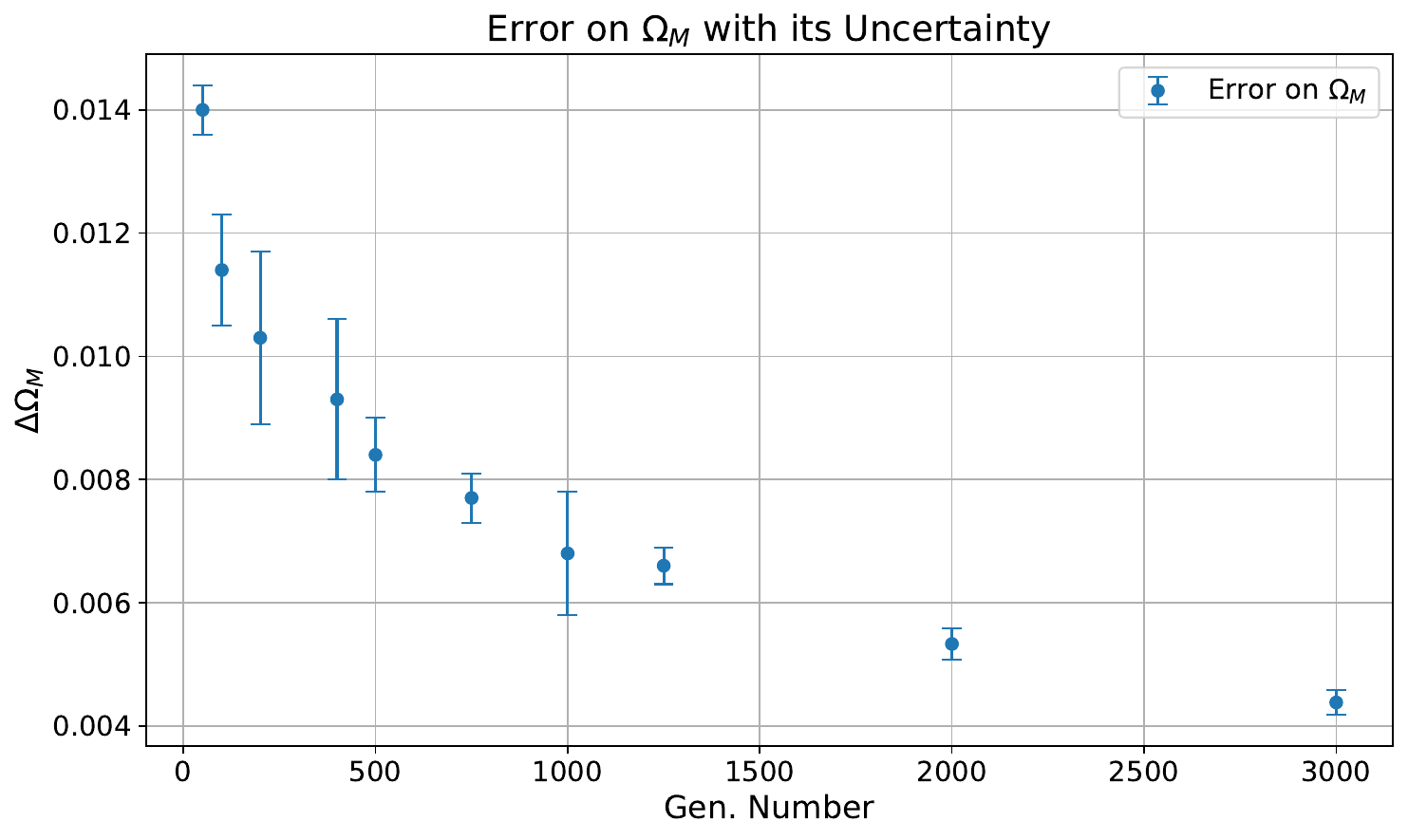}
\includegraphics[width=0.45\hsize]{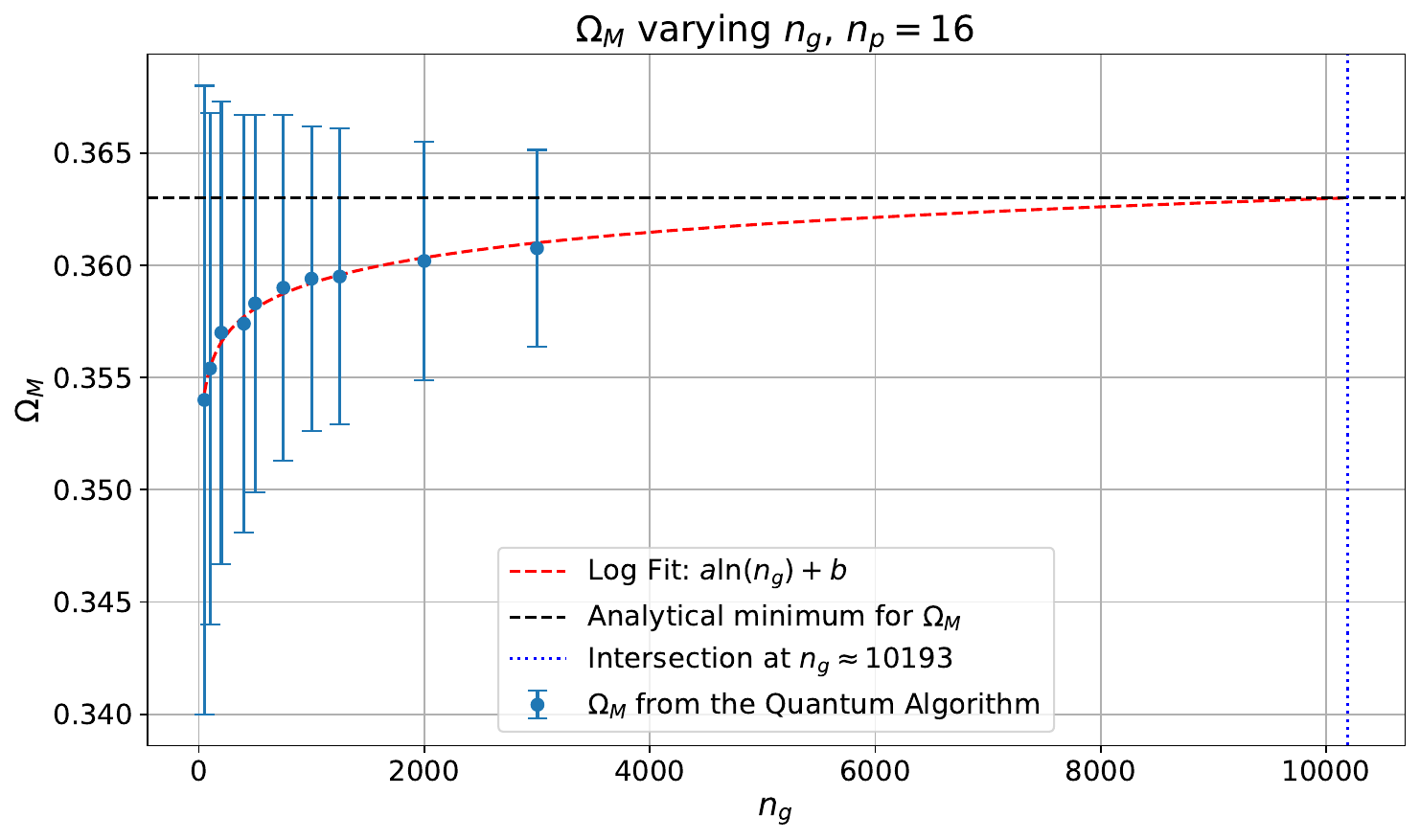}

\caption{Left panel: results on the uncertainties on $\Omega_M$ with their dispersion as a function of the number of generations. Right panel: results on $\Omega_M$ versus $n_g$, with a logarithmic fit to extrapolate the values to higher $n_g$. For both $n_p=16$.} 
\label{Fig_Quantum_generations}
\end{figure*}

Let us now study how our results change with $n_p$ and $n_g$. We focus on the SNe Ia results, fixing crossover and mutation probabilities at (0.5, 0.5). For the tests on $n_g$ we fix $n_p=16$, and the results are shown in Tab.~\ref{Tab_Gen_number}. The computations for $\Omega_M$ and $H_0$ have been performed as the mean of the various tests used for our comparison. All of them are consistent within $1 \sigma$. We also note the monotonic trend both for $\Omega_M$ and $H_0$ with $n_g$, approaching the absolute minimum of our merit function, and a general decreasing trend of the errors on both $\Omega_M$ and $H_0$ with $n_g$. This is expected given how our algorithm works. This gain becomes less significant in the runs with very high $n_g$ (for instance, the difference in the uncertainty on $\Omega_M$ between the runs with $n_g=1000$ and $n_g=50$ is $0.072$, while it is only $0.009$ between the runs with $n_g=3000$ and $n_g=2000$). This aspect also emerges in the plot shown in Fig. \ref{Fig_Quantum_generations}, where the errors on $\Omega_M$ are plotted against $n_g$.

\begin{table}
\centering 
\scriptsize
\begin{tabular}{|c|c|c|c|c|}
\hline
Gen. Number & $\Omega_M$ & $H_0$ & $\Delta \Omega_M$ & $\Delta H_0$ \\ \hline
50 & 0.3540 & 72.944 & $0.0140 \pm 0.0004$ & $0.174 \pm 0.008$ \\ \hline
100 &  0.3554 & 72.919  & $0.0114 \pm 0.0009$ & $0.135 \pm 0.013$ \\ \hline
200 &  0.3570 & 72.899  & $0.0103 \pm 0.0014$ & $0.118 \pm 0.018$\\ \hline
400 &  0.3574 & 72.894  & $0.0093 \pm 0.0013$ & $0.106 \pm 0.017$ \\ \hline
500 &  0.3583 & 72.883  & $0.0084 \pm 0.0006$ & $0.093 \pm 0.007$ \\ \hline
750 &  0.3590 & 72.877 & $0.0077 \pm 0.0004$ & $0.086 \pm 0.004$ \\ \hline
1000 &  0.3594 & 72.870  & $0.0068 \pm 0.0010$ & $0.076 \pm 0.010$ \\ \hline
1250 &  0.3595 & 72.869  & $0.0066 \pm 0.0003$ & $0.072 \pm 0.002$ \\ \hline
2000 &  0.3602 & 72.860  & $0.0053 \pm 0.0003$ & $0.058 \pm 0.003$ \\ \hline
3000 &  0.3608 & 72.856  & $0.0044 \pm 0.0002$ & $0.047 \pm 0.002$ \\ \hline
\end{tabular}
\caption{Results on $\Omega_M$ and $H_0$ with the means and standard deviations for the errors on both these quantities obtained considering the SNe Ia and varying $n_g$. Here, $n_p=16$.}
\label{Tab_Gen_number}
\end{table}

In the right panel, we show the $\Omega_M$ obtained with our algorithm with $n_g$. While all the results are consistent within $1 \sigma$ to the analytical minimum and between each other, we note a clear trend in the mean value, which tends to the analytical minimum of the objective function. In doing a simple logarithmic fit, we find that it reaches the true minimum for $n_g \approx 10193$. It is important to stress that this particular behaviour has been noted only for $n_p=16$, as we will show in the following.

For the analysis on $n_p$, we fix $n_g=50$, and the crossover and mutation probabilities to (0.5, 0.5). The results are shown in Tab.~\ref{Tab_Pop_number}. Differently from the previous case, increasing $n_p$ does not increase the precision of the results in a significant way, differently from the tests with $n_g$. This could be due to how we have defined the subset of best individuals of a given population, recalling that it is the $25\%$ of the entire sample. Increasing the sample size increases the number of chosen best individuals, which in turn increases the interval range from which the new individuals are computed from the corresponding quantum circuit, thus increasing the variability of the set itself. Nevertheless, the dispersion of the errors for both $\Omega_M$ and $H_0$ is smaller than in the case where we increase the number of generations. Again, all the results on the measurement are consistent within $1 \sigma$.

\begin{figure*}[ht]\centering
\includegraphics[width=0.44\hsize]{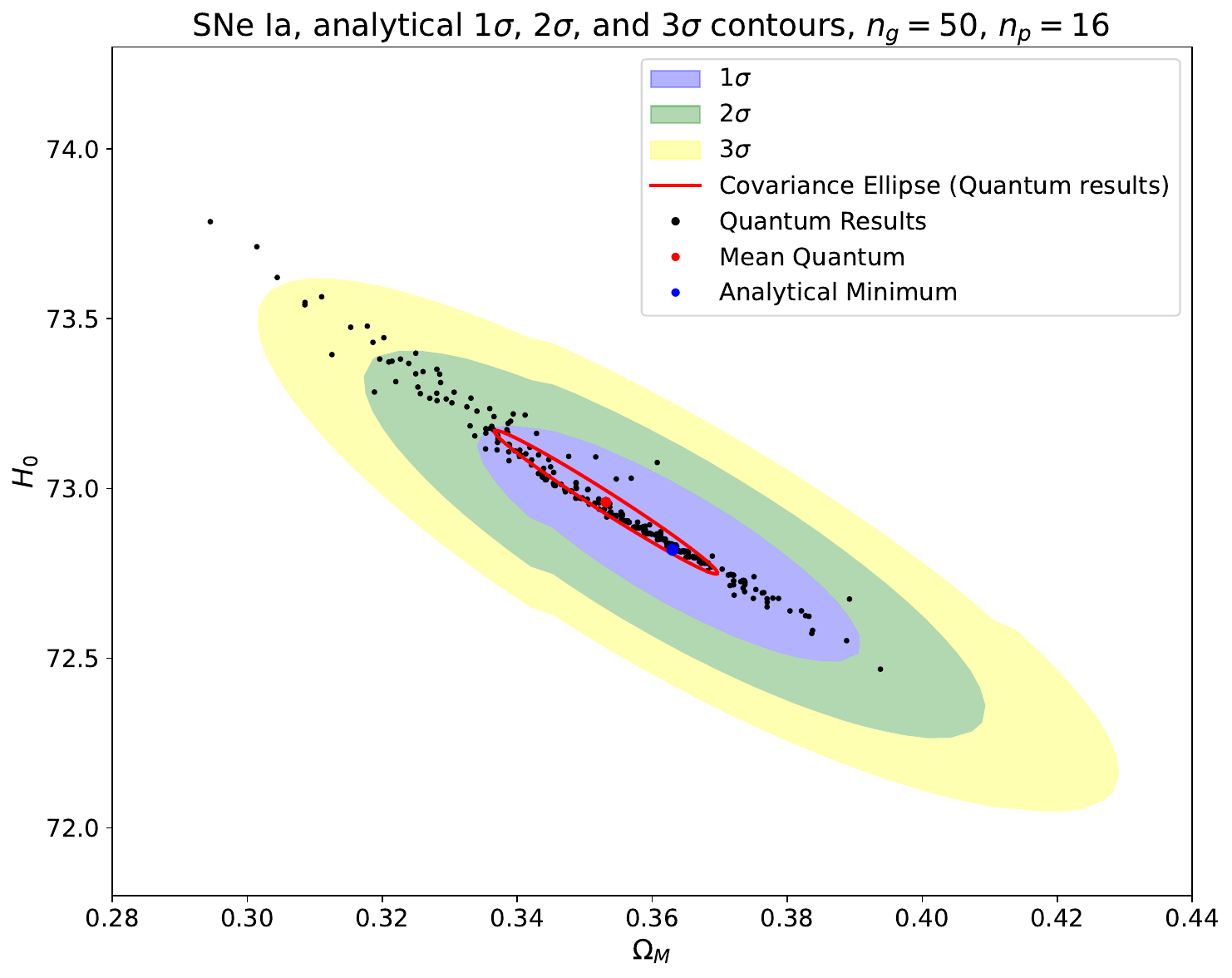}
\includegraphics[width=0.44\hsize]{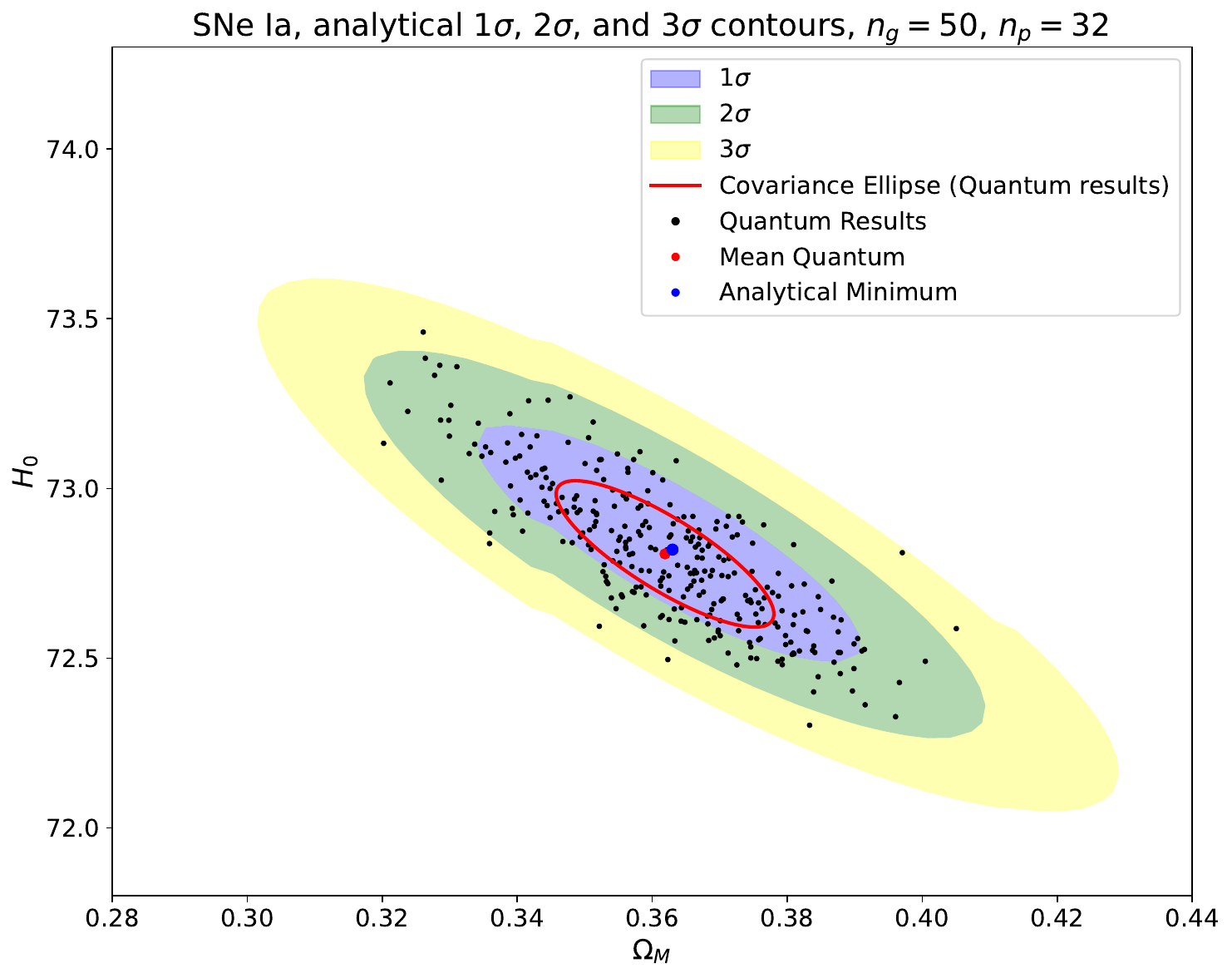}
\includegraphics[width=0.44\hsize]{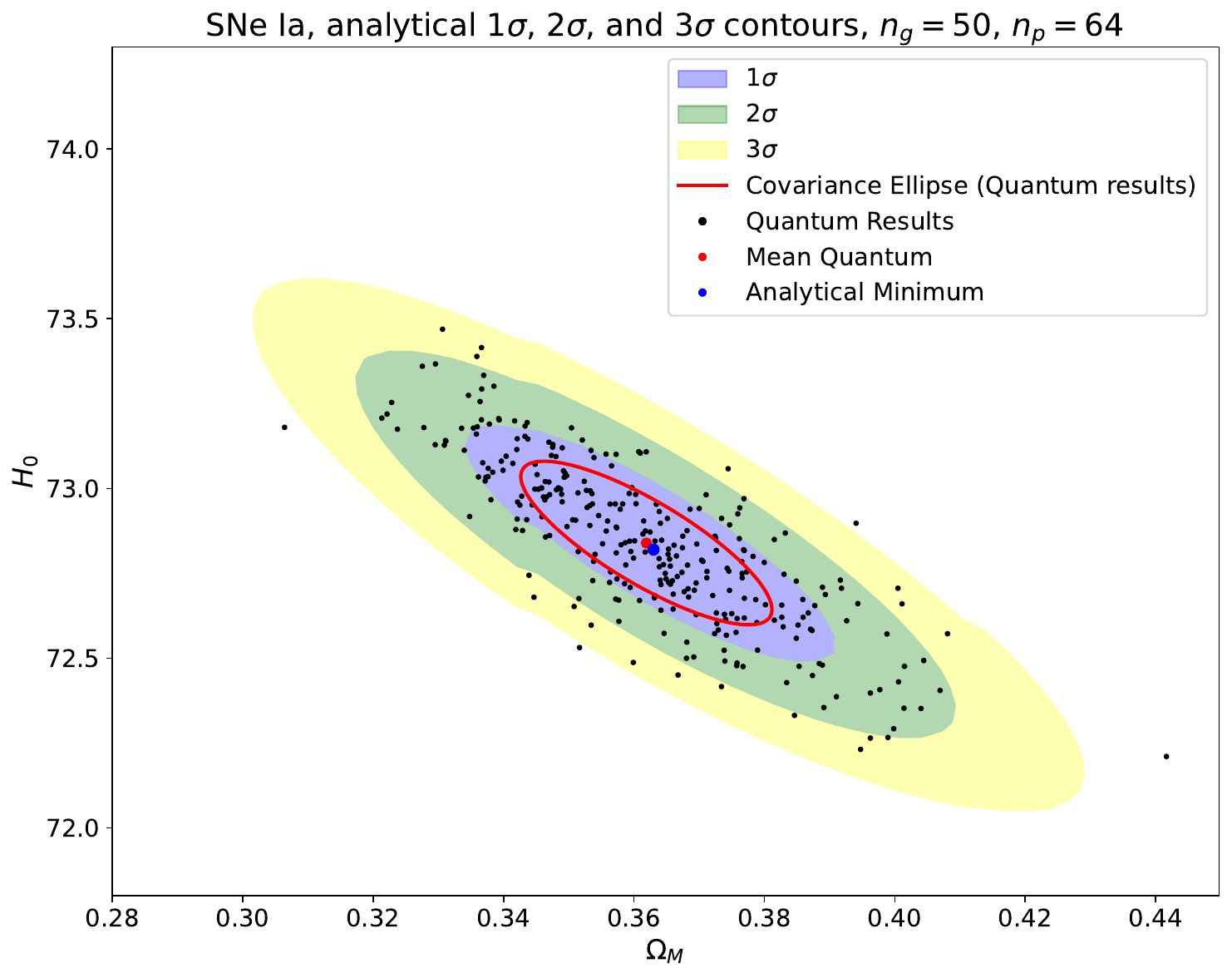}
\includegraphics[width=0.44\hsize]{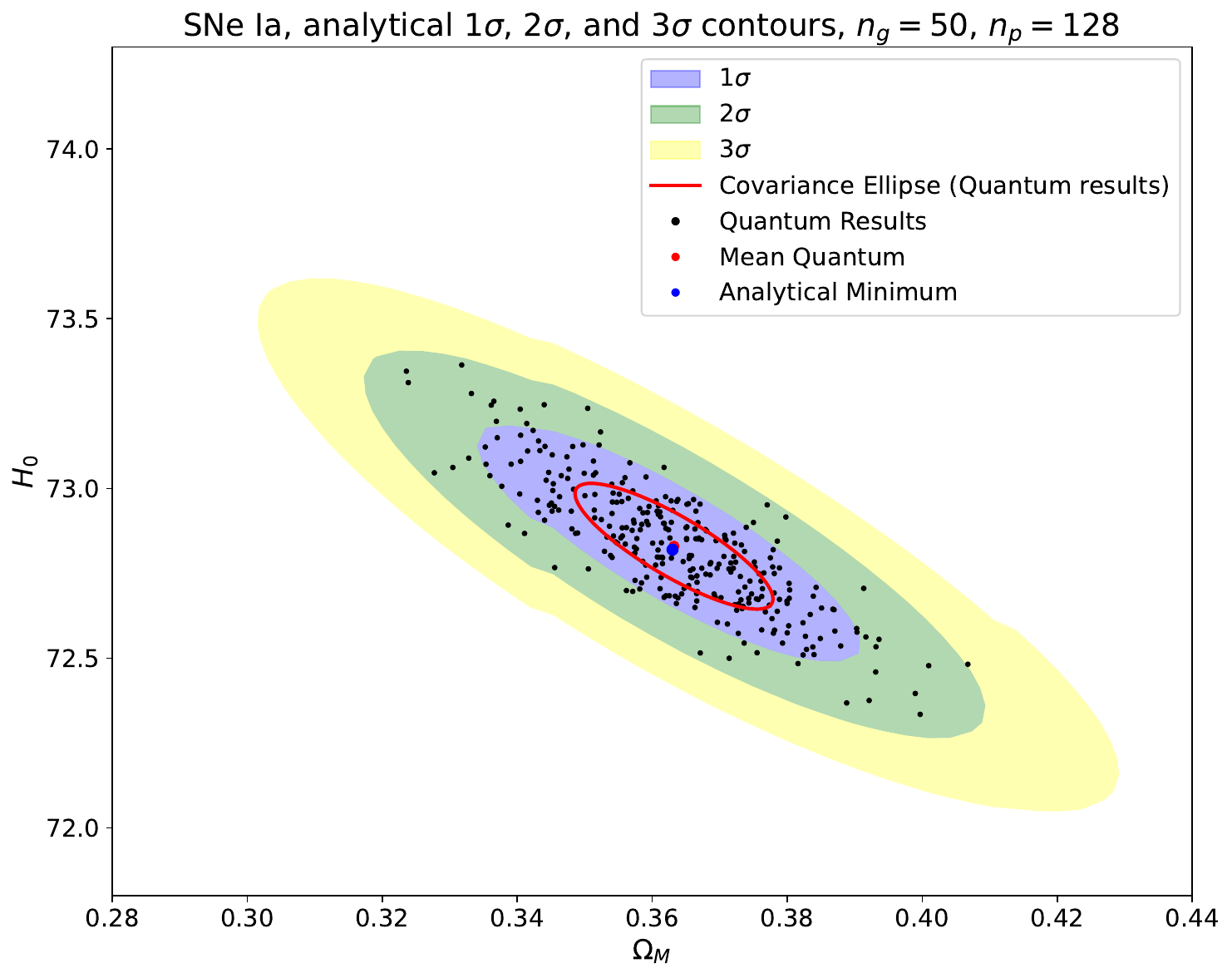}
\caption{Comparison between the results of AEQGA with the analytical contour levels of the SNe Ia merit function. Each filled color represents 1 sigma level of the analytic function, while the red ellipse details the covariance ellipse obtained by considering the quantum results.} Top left panel: with $n_p=16$. Top right panel: with $n_p=32$. Bottom left panel: with $n_p=64$. Bottom right panel: with $n_p=128$.  
\label{Fig_Population_analytic}
\end{figure*}

\begin{table}
\scriptsize
\centering 
\begin{tabular}{|c|c|c|c|c|}
\hline
Pop. Number & $\Omega_M$ & $H_0$ & $\Delta \Omega_M$ & $\Delta H_0$ \\ \hline
16 & 0.3540 & 72.944 & $0.0140 \pm 0.0004$ & $0.174 \pm 0.008$ \\ \hline
32 & 0.3618 & 72.807   & $0.0175 \pm 0.0006$ & $0.237 \pm 0.011$ \\ \hline
64 & 0.3615 & 72.841  & $0.0196 \pm 0.0007$ & $0.243 \pm 0.005$\\ \hline
128 & 0.3627 & 72.840  & $0.0147 \pm 0.0004$ & $0.183 \pm 0.004$ \\ \hline
\end{tabular}
\caption{Results on $\Omega_M$ and $H_0$ with the means and standard deviations for the errors on both these quantities obtained considering the SNe Ia and varying $n_p$. Here, $n_g=50$.}
\label{Tab_Pop_number}
\end{table}


There is an interesting behaviour we obtain by increasing $n_p$, noticeable in Fig. \ref{Fig_Population_analytic}, where we compare the results obtained by AEQGA with different values for $n_p$ with the analytical merit function of the SNe IA, in particular considering the 1, 2, and 3 $\sigma$ contours. With $n_p=16$, we obtain a stronger correlation between the individual outcomes of the genetic algorithm. This trend does not appear with a bigger $n_p$, which instead follows more closely the merit function. We stress that this is a secondary effect for our results, as we are not interested in reproducing the shape of the merit function. Nevertheless, this is an interesting effect to highlight, probably linked to the smaller number of qubits used for the populations with $n_p=16$, and this is also the reason why we show the results of a single run with $n_p=32$.  This stronger correlation in the results also explains why in Tab.~\ref{Tab_Pop_number} the results for $n_p=16$ show the smallest errors on $\Omega_M$ and $H_0$, as well as why the mean values of these quantities appear different (albeit still consistent within 1 $\sigma$) with respect to the other runs. Indeed, apart from this value, also the trend varying $n_p$ becomes closer to what we have found for $n_g$, with the most precise result being achieved with the higher value of $n_p$, even if the previous discussions on the sample size still apply.

\begin{figure*}[ht]\centering
\includegraphics[width=0.44\hsize]{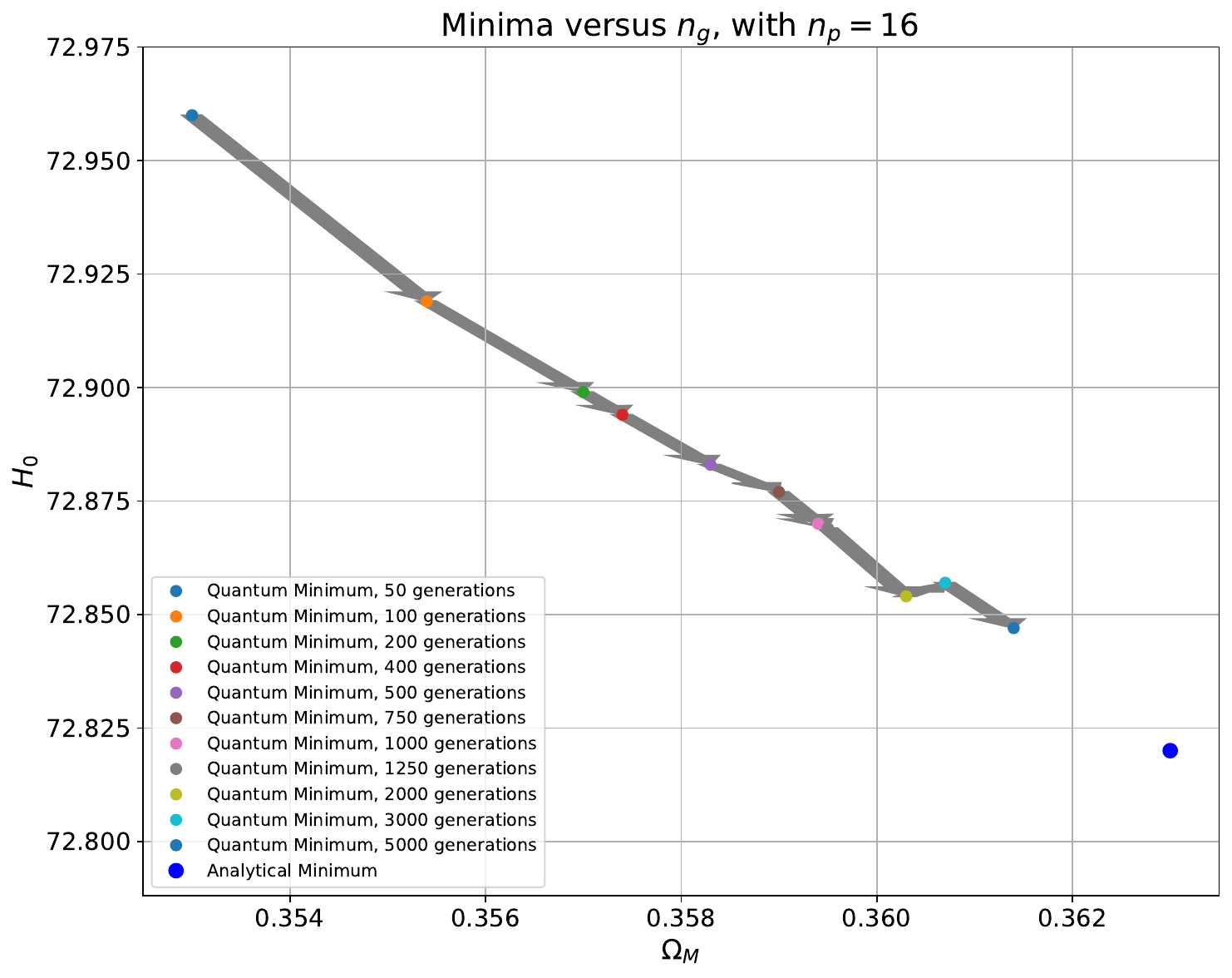}
\includegraphics[width=0.44\hsize]{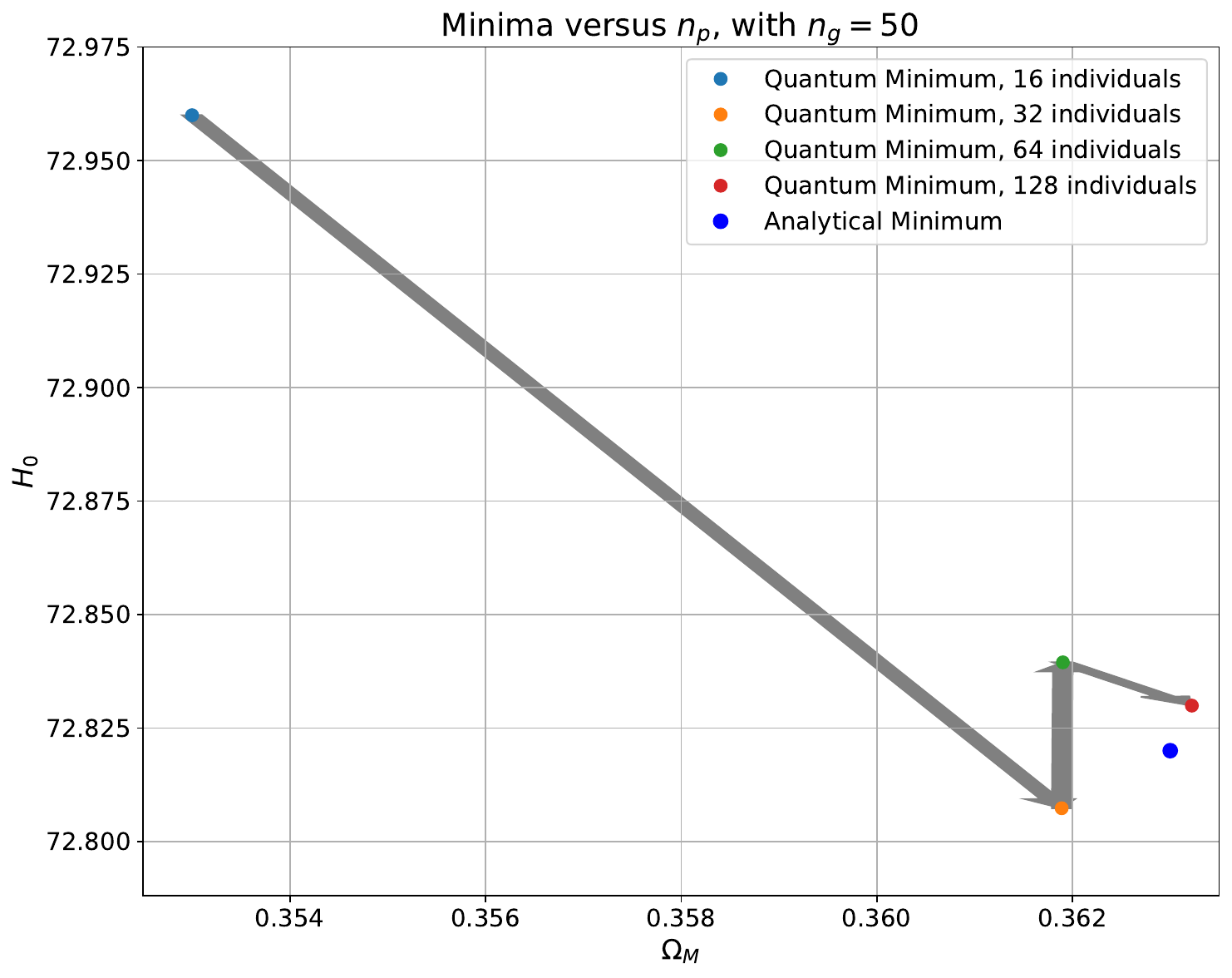}
\includegraphics[width=0.44\hsize]{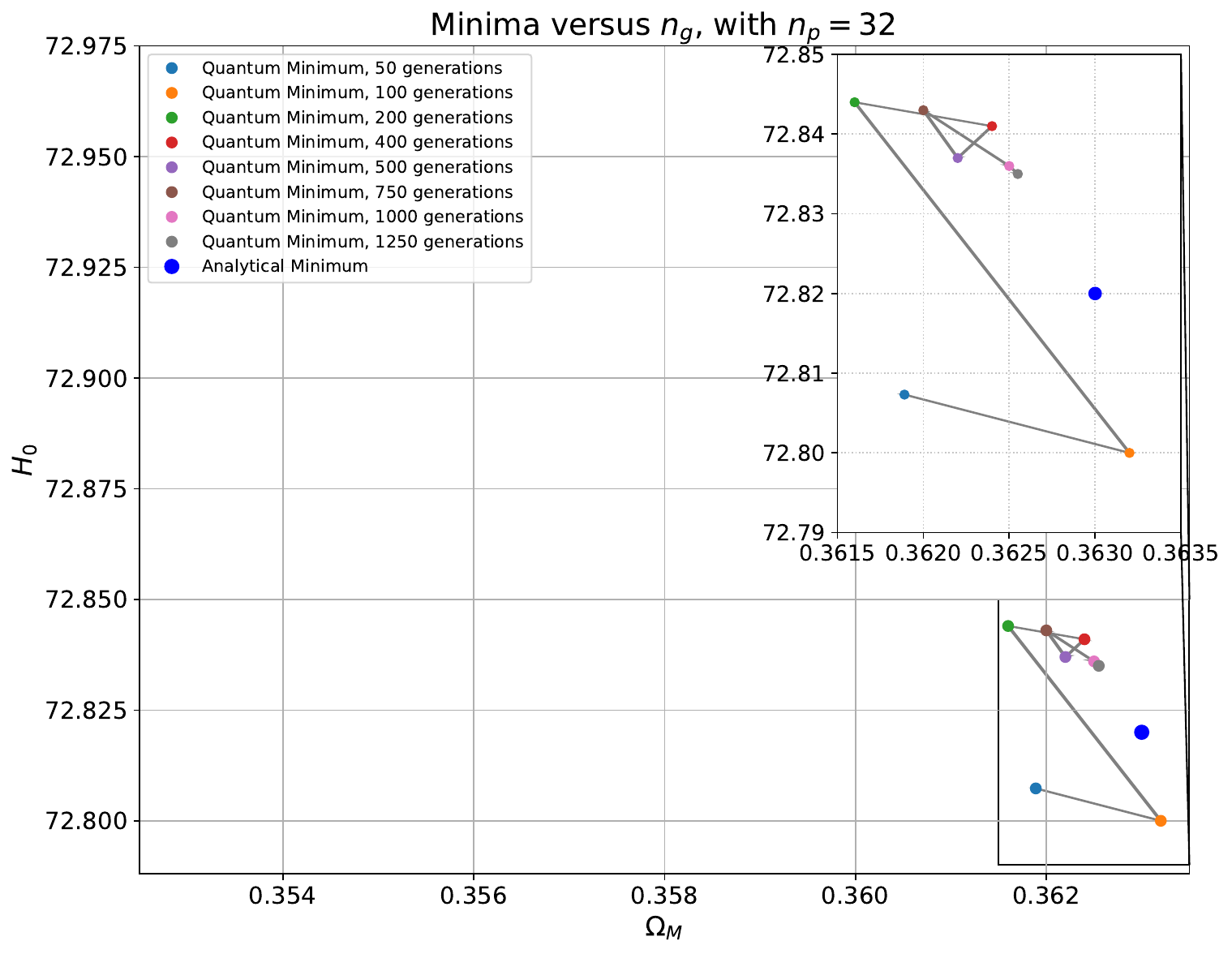}
\includegraphics[width=0.44\hsize]{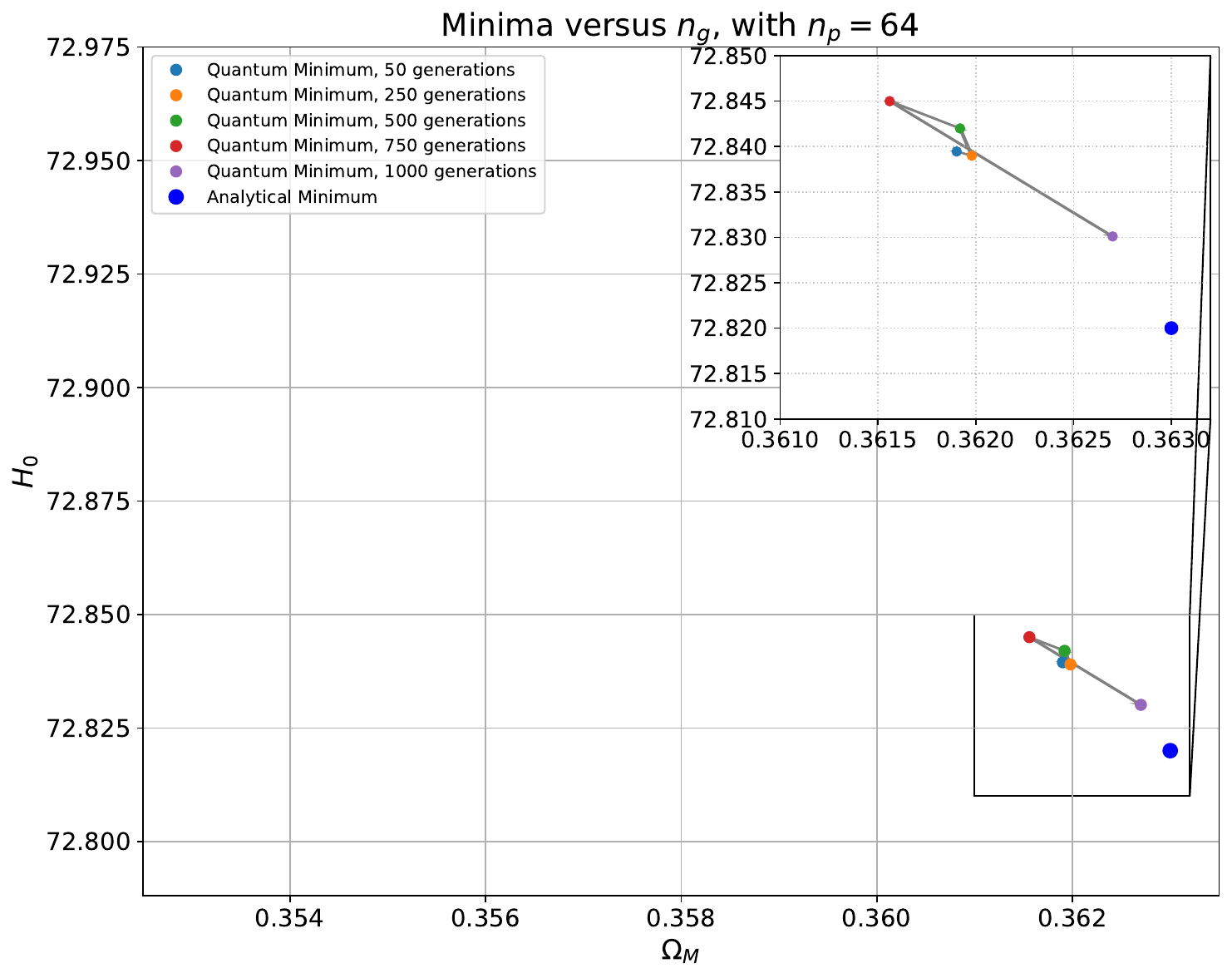}
\caption{The minima found by AEQGA, with a gray arrow linking each subsequent point to indicate the overall trend. Top left panel: varying $n_g$ and fixing $n_p=16$. Top right panel:  fixing $n_g=50$ and varying the number of individuals. Bottom left panel: varying $n_g$ and fixing $n_p=32$. Bottom right panel: the same as the bottom left panel but fixing $n_p=64$. In the bottom panels a zoom in the region around the minimum is shown.}
\label{Fig_arrow_minima}
\end{figure*}

Another test regards the analysis of the position of the minimum found by AEQGA, given by the mean of the results of all the iterations of one run. In particular, we present how, for the SNe Ia, this changes with $n_g$ (top left panel of Fig. \ref{Fig_arrow_minima}) and $n_p$ (top right panel of the same figure). In the first analysis, where $n_p=16$, we note how by increasing $n_g$ the results get closer to the absolute minimum. We also note that, as we have shown in Tab. \ref{Tab_Gen_number}, the related uncertainties decrease with $n_g$, so that all our results are consistent at the same level with the absolute minimum. In the top right panel, instead, where we vary $n_p$ fixing $n_g=50$, we see that by increasing $n_p$ from 16 to 32, we get very close results to the analytical minimum. This is still because of the stronger correlation found for $n_p=16$, which is not found for higher $n_p$. This is further confirmed by the bottom panels of Fig. \ref{Fig_arrow_minima}, where we show the same analysis fixing $n_p=32$ and $n_p=64$, respectively, and varying $n_g$. Here, the quantum results are close to the absolute minimum for every $n_g=32$. 

\section{Comparison with other Algorithms}
\label{sec_comparison}
Let us now compare the results obtained by AEQGA with the ones obtained via other methods. We implemented two classical algorithms. The first is based on recursive computations at each step, keeping the best $25 \%$ from the previous generation, while $25\%$ are randomly generated within a box around the region where the best individuals were found, and the remaining $50\%$ are randomly generated inside the entire interval range in which the initial population is defined. The second algorithm, instead, is a simple classical genetic algorithm with operations of crossover and mutations, the former defined as the mixing of the elements of different individuals, while the latter as the mutation of a given element of a random quantity. This algorithm also presents a duplication step, but it does not preserve a priori the best individuals of the population at each generation. The idea is to test how AEQGA compares with two simple classical paradigms, where in the former we keep the same preservation of the best individual at each generation, while in the latter we define crossover and mutation operations which resemble what was done in AEQGA, with the caveat that more refined classical operations are possible.

The results are shown in Tab.~\ref{Tab_Rand_Gen}. Here $n_g=50$,  while the crossover and mutation probabilities for the classical genetic algorithm are fixed at (0.5, 0.5), as previously done for the quantum genetic algorithm. Comparing these results with the ones from AEQGA, shown in the first row of Tabs.~\ref{Tab_Gen_number} and~\ref{Tab_Pop_number}, we note how the recursive algorithm is more precise for both $\Omega_M$ and $H_0$, than the other two (for instance, the uncertainties on $\Omega_M$ and $H_0$ derived from the recursive algorithm are $40\%$ and $52\%$ of the ones given by the classical one for $n_p=16$). Again, the results are consistent within $1 \sigma$.

\begin{table}
\centering 
\scriptsize
\begin{tabular}{|c|c|c|c|c|c|}
\hline
Algorithm & $n_p$& $\Omega_M$ & $H_0$ & $\Delta \Omega_M$ & $\Delta H_0$ \\ \hline
Recursive & 16 & 0.3622 & 72.840 & $0.0101 \pm 0.0003$ & $0.124 \pm 0.004$ \\ \hline
Genetic  & 16 & 0.3624 & 72.840 & $0.0248 \pm 0.0012$ & $0.315 \pm 0.019$ \\ \hline
Quantum & 16 & 0.3540 & 72.944 & $0.0140 \pm 0.0004$ & $0.174 \pm 0.008$ \\ \hline
Recursive & 32 & 0.3624 & 72.840 & $0.0078 \pm 0.0004$ & $0.094 \pm 0.004$ \\ \hline
Genetic & 32 & 0.3629 & 72.830 & $0.0158 \pm 0.0005$ & $0.192 \pm 0.007$ \\ \hline
Quantum & 32 & 0.3618 & 72.807 & $0.0175 \pm 0.0006$ & $0.183 \pm 0.004$ \\ \hline

\end{tabular}
\caption{The means and standard deviations for the errors on $\Omega_M$ and $H_0$ obtained considering the SNe Ia for the two classical algorithms. Here $n_g=50$. For the sake of direct comparison, we also repeat the results obtained with AEQGA in the same conditions.}
\label{Tab_Rand_Gen}
\end{table}

This result is due to how the recursive algorithm explores the parameter space, being completely random, especially in the box defined around the best results. But, on the other hand, it is also more sensitive to the size of the box itself, while, for the decoding strategy we employed, where we recall we have a non-linear multiplication of the probabilities to better explore the central regions of the box, AEQGA is less sensitive to this parameter. The classical genetic algorithm is not as precise because of its very simple formulation and because it does not preserve the best individuals a priori at each generation. We have still kept this result because it is interesting to see if AEQGA compares favorably to a classical one, even if the latter can be improved in its formulation. These considerations are especially true for $n_p=16$, where we also recall the stronger correlation between the two parameters for the quantum results explained in the previous section. For $n_p=32$, instead, the precision on the results between the classic and quantum algorithms becomes more similar, while the recursive is still more precise (quantitatively, the uncertainties on $\Omega_M$ and $H_0$ derived from the recursive algorithm are both $51\%$ of the ones given by the classical one). We note that for $n_p=32$ the mean values for $\Omega_M$ and $H_0$ for AEQGA become closer to the classical computations. As an example, the percentage difference between the classical and AEQGA results for $\Omega_M$ goes from $2.3\%$ for $n_p=16$ to $0.3\%$ for $n_p=32$.

\begin{figure*}
\centering
\includegraphics[width=0.49\hsize]{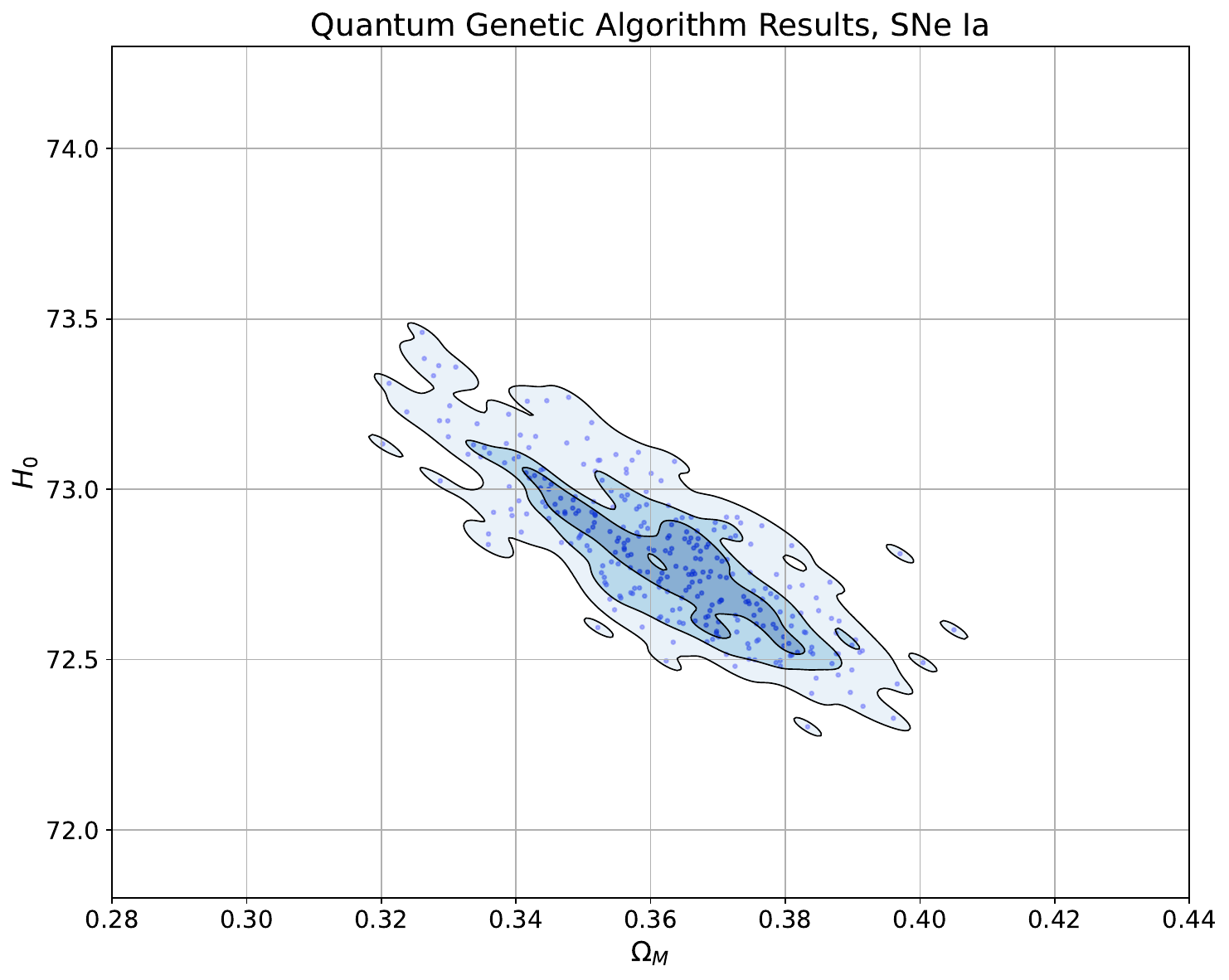}
\includegraphics[width=0.49\hsize]{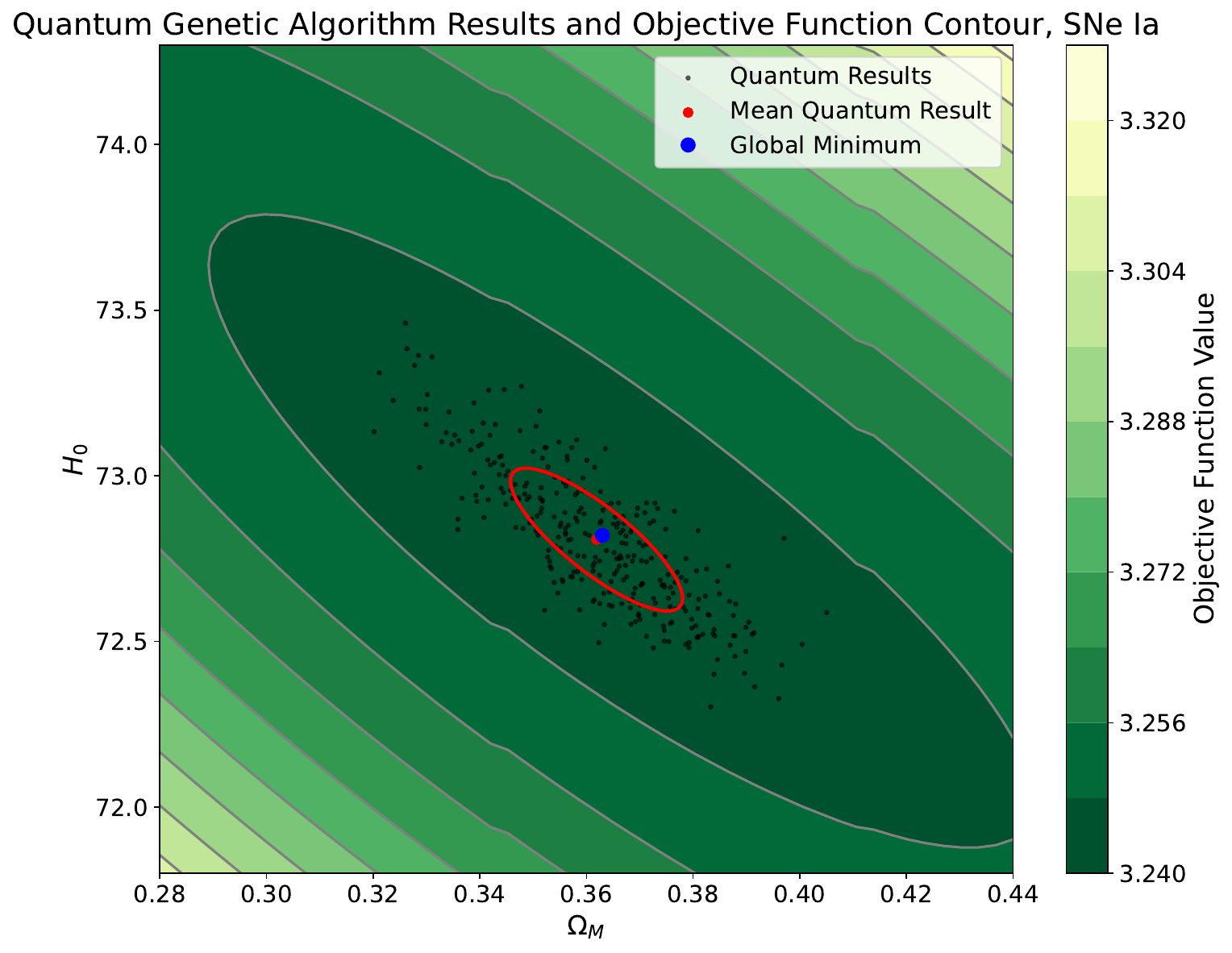}
\includegraphics[width=0.49\hsize]{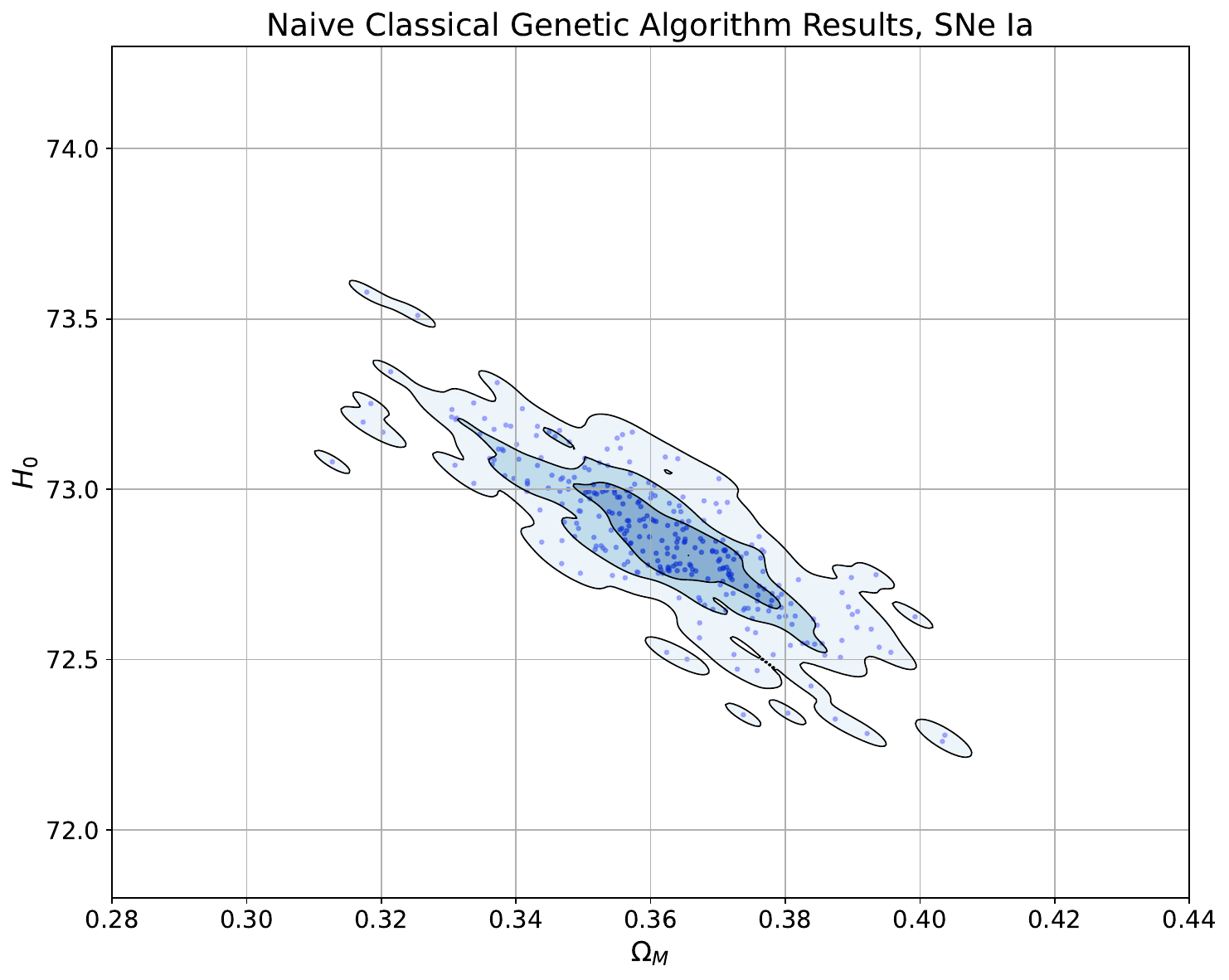}
\includegraphics[width=0.49\hsize]{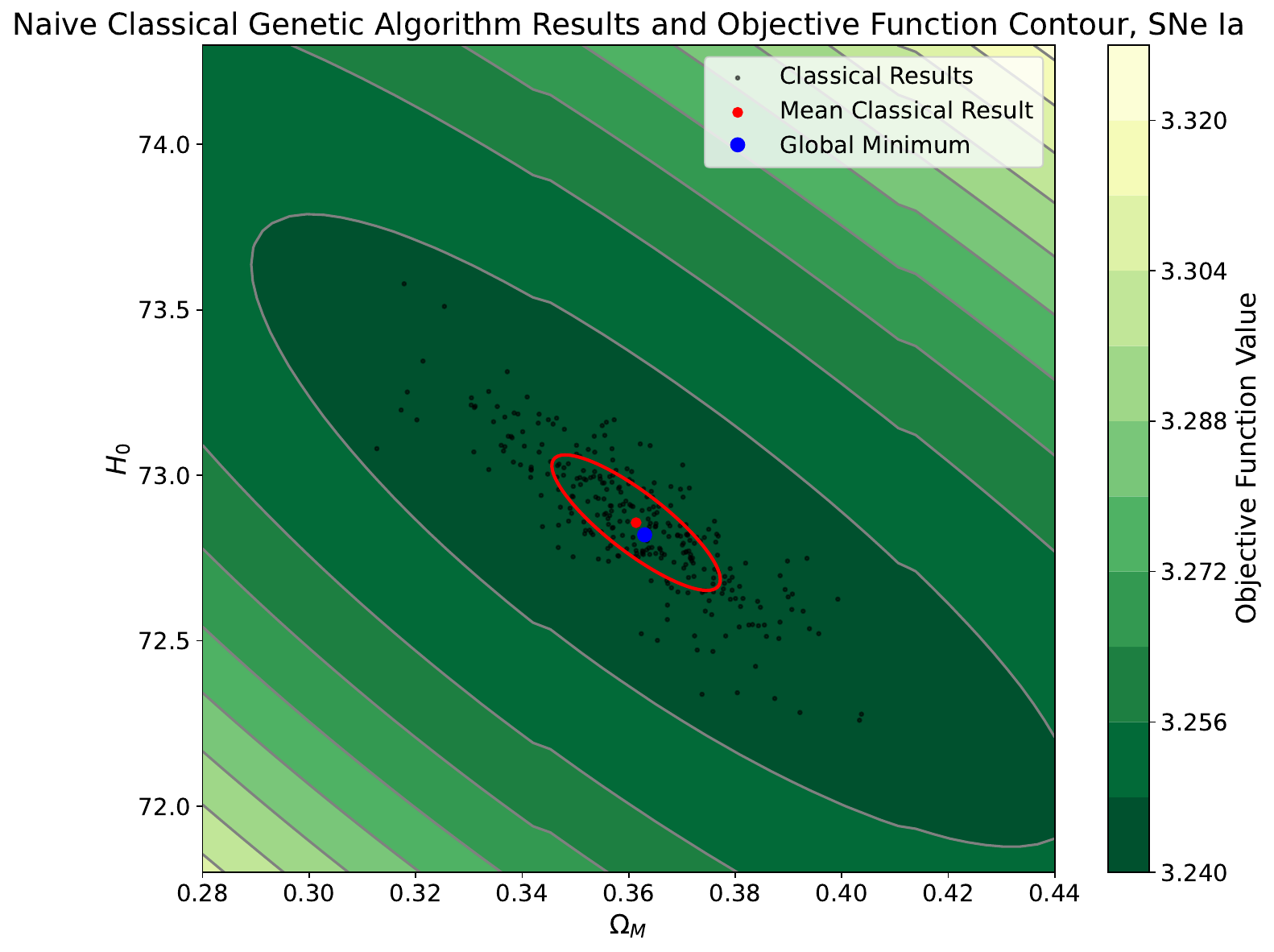}
\includegraphics[width=0.49\hsize]{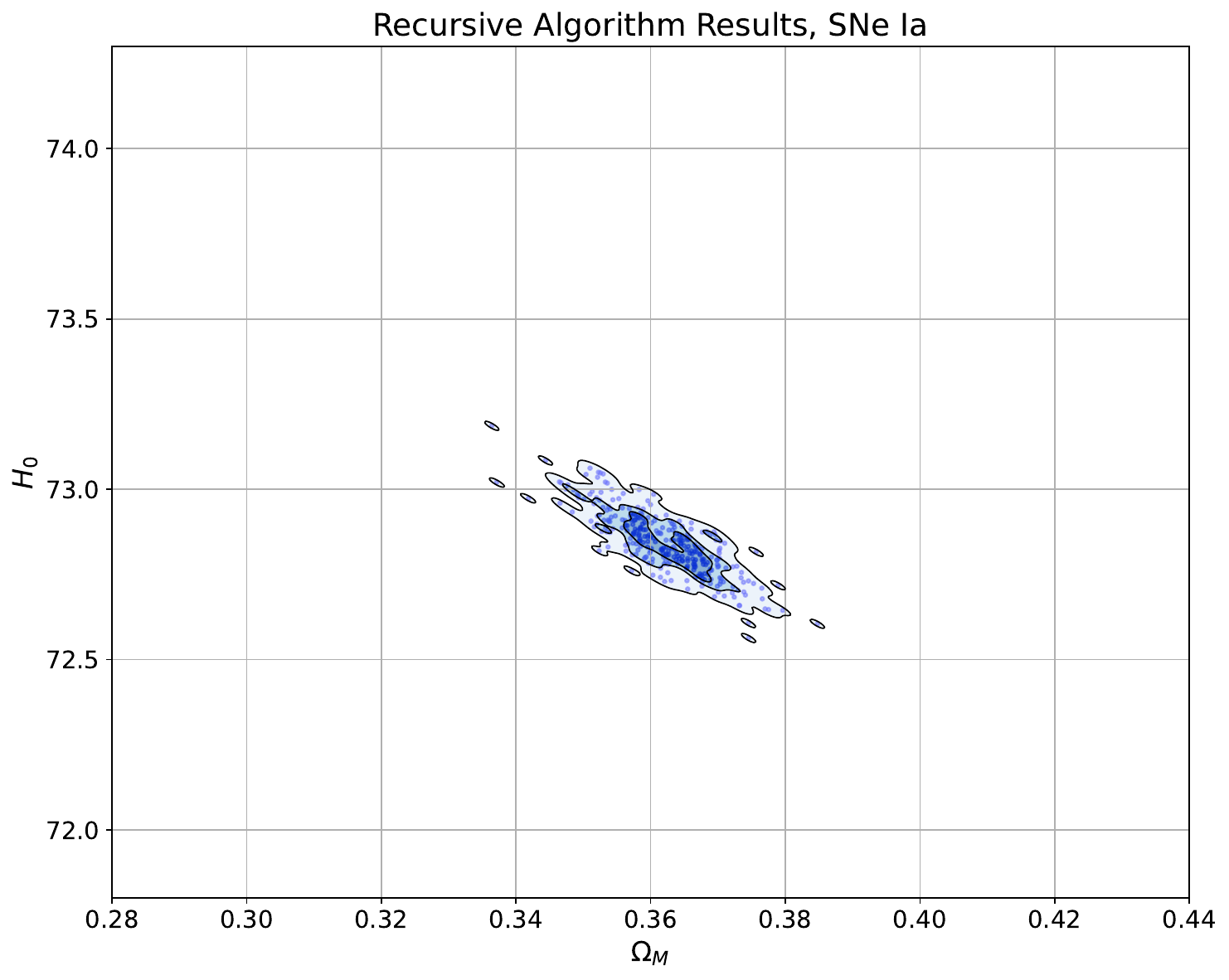}
\includegraphics[width=0.49\hsize]{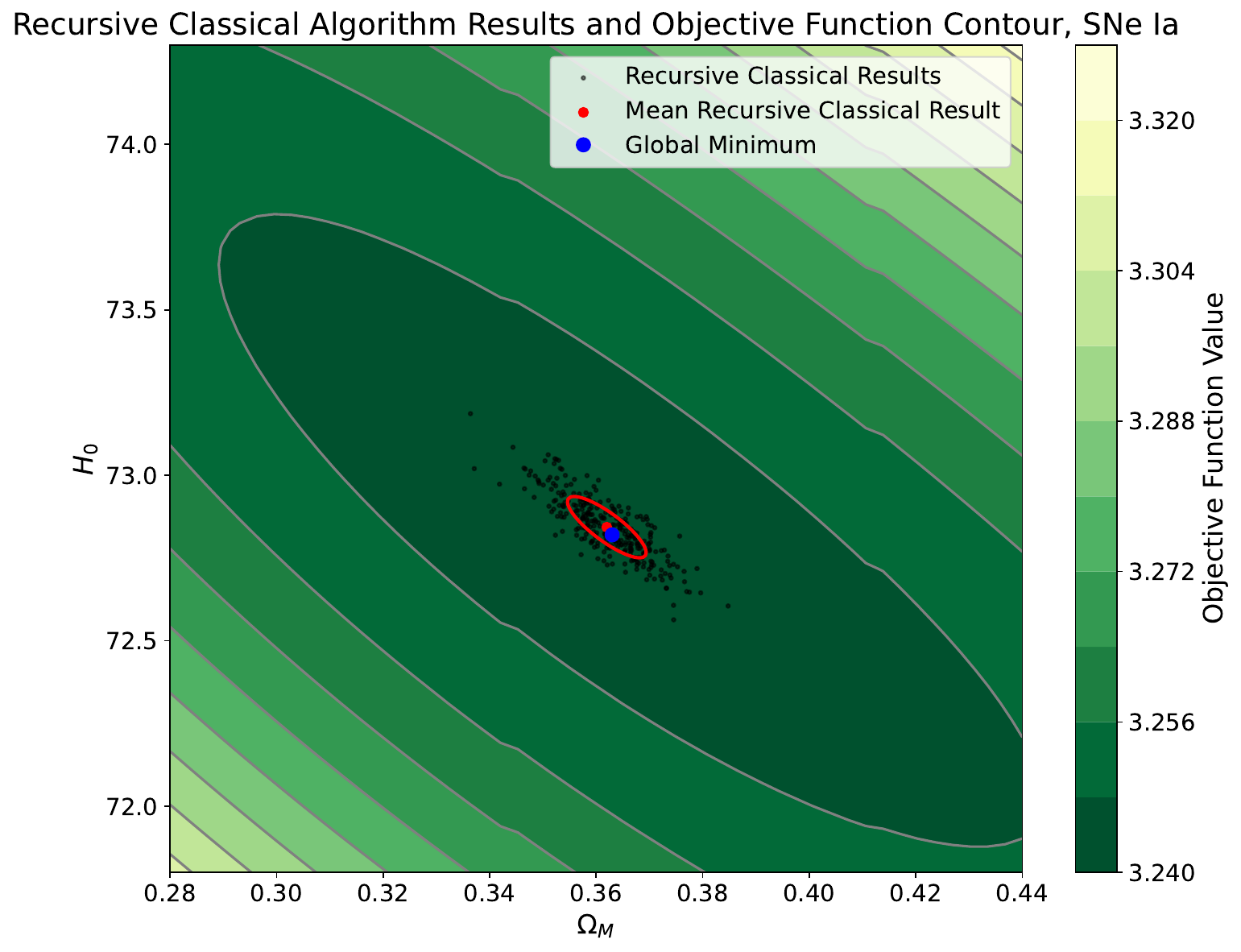}
\caption{Top left panel: the same results shown in the Top panel of Fig.~\ref{Fig_Quantum_results}. Top right panel: comparison of the quantum results with the analytical merit function, highlighting the mean and $1 \sigma$ ellipse derived from AEQGA in red, which we recall has been obtained using the different results of the algorithm as we did in Fig. ~\ref{Fig_Population_analytic}. Middle left panel: the results for the classical genetic algorithm, using the same hyperparameters for the population and generation size. Middle right panel: the comparison with the analytical function. Bottom left panel: the same for the recursive algorithm. Bottom right panel: the corresponding comparison with the analytical merit function. }
\label{Fig_Comparison_Classical}
\end{figure*}

The results for a single computation with $n_p=32$ are displayed in Fig.~ \ref{Fig_Comparison_Classical}. In this case, the overall dispersion of the quantum results is very similar to the classical one. This becomes clearer in comparison with the analytical objective function shown in the right panels of this figure, where we also show the covariance ellipse computed for our results. Again, all our results are consistent within $1 \sigma$ with the absolute analytical minimum. Regarding the scalability and computational costs, both classical algorithms scale linearly with $n_p, n_g,$ and $n_i$, as AEQGA, because of the similarities between the three algorithms. AEQGA is not faster than the two classical algorithms because of the emulation costs, the number of shots necessary to obtain results from the quantum circuit, and, more importantly, because the evaluation of the objective function remains classical. Nevertheless, both the accuracy and the computational costs of these three algorithms remain in the same order of magnitude, and scale in the same way with the hyperparameters. 

As a second test, we considered another quantum genetic algorithm already implemented in the literature and compared the results. In particular, we used, for the SNe Ia, the HQGA \citep{Acampora_2021} on which we give more insights in $\S$~\ref{sub_sec_QGA_lit}. We here just note that the philosophy of the two algorithms is intrinsically different, HQGA being based on binary encoding, discretization of the objective function, elitism for the evaluation of the fitness (i.e. evaluating the fitness function not for all the individuals of the population but selecting the best ones before the computation), and a different definition of the quantum crossover operation, that in HQGA is always performed.

All these differences translate into different dependencies on the hyperparameters and computational costs. Even so, the comparison is still interesting, to understand which approach can be more efficient for the particular case under analysis, as well as to compare AEQGA with another one proven to be reliable for various benchmarking functions \citep{Acampora_2021}. 

\begin{figure}
\centering
\includegraphics[width=0.85\hsize]{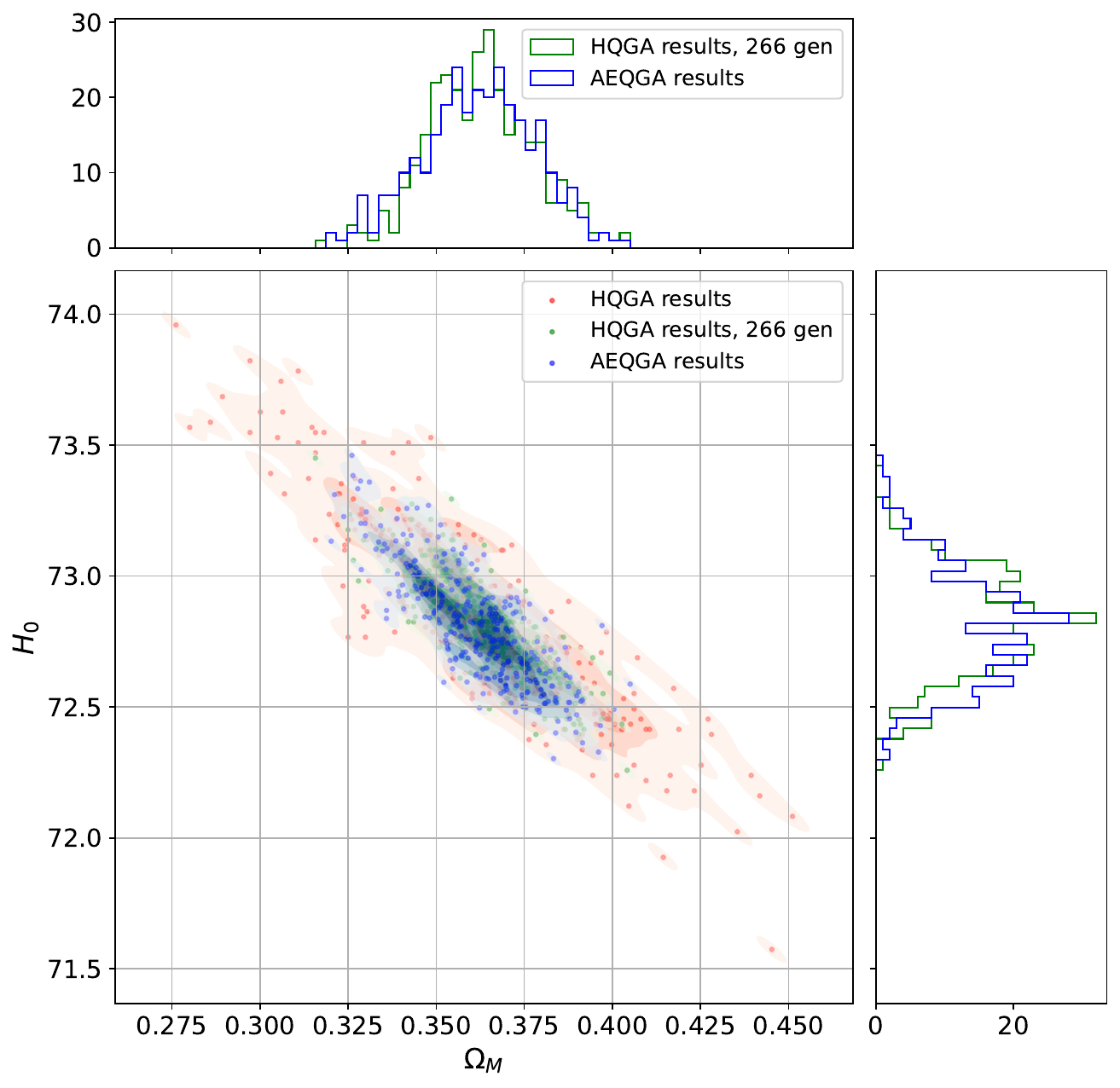}
\caption{The results obtained with two different tests of HQGA compared with AEQGA. The histograms compare the distributions of $\Omega_M$ and $H_0$ obtained by AEQGA and HQGA with 266 generations. The standard deviations for $\Omega_M$ and $H_0$ obtained for the single runs shown here for AEQGA and HQGA with 266 generations are $0.016$ and $0.015$ for $\Omega_M$, and $0.22$ and $0.19$ for $H_0$, respectively.}
\label{Fig_HQGA_comparison}
\end{figure}

Again, we focus on the SNe Ia computations, whose merit function has been introduced in the HQGA case studies. The results are shown in Fig.~\ref{Fig_HQGA_comparison}, where we compare the results derived by AEQGA ($n_g=50$, $n_p=32$, $n_i=300$, crossover and mutation probabilities= 0.5) with two tests performed using HQGA.

For the tests shown in this plot, the hyperparameters used for the HQGA are itemized as follows: 
\begin{itemize}
    \item binary string=10 (which indicates how many data points are used to discretize the function inside our interval range);
    \item number of individuals for which the merit function is computed at each step=3;
    \item mutation probability=0.5 (the crossover is always set to 1 in HQGA);
    \item angle for the rotation gates used as initialization of the algorithm $\delta= \pi/8$;
    \item Quantum Elitism (QE), because HQGA allows choosing between three different approaches to select the individuals inside the population to compute the merit function: QE, Deterministic Elitism (DE), and Quantum with Reinforcement Elitism (RE);
    \item finally, 300 iterations, to derive means and uncertainties as we did for AEQGA.
\end{itemize} The difference in the two HQGA tests is in the number of generations, which is 100 for the results displayed in red, and 266 for the ones in green (chosen so that the total number of fitness evaluations needed by HQGA matches those needed by AEQGA with $n_p=16$, i.e. our result with the smallest number of evaluations of the objective function). Finally, we have tested the stability of the HQGA algorithm by running it 10 times with QE, 100 generations, and mutation probability=0.5, finding for the errors $\Delta \Omega_M=0.027 \pm 0.001$ and $\Delta H_0=0.33 \pm 0.02$.

The algorithms present consistent results, thus confirming the validity of AEQGA. In comparing the precisions of the results shown in \ref{Fig_HQGA_comparison}, AEQGA is $41\%$ and $20\%$ more precise with respect to HQGA with 100 generations for $\Omega_M$ and $H_0$, respectively, while it is $6\%$ and $13\%$ less precise for HQGA with 266 generations.

\begin{table*} \label{Tab_HQGA_Results}
\centering
\begin{tabular}{|c|c|c|c|c|c|c|c|}\hline
Elitism & Mut. prob. & Num. Gen. & $\delta$ & $\Omega_M$ &  $H_0$ & $\Delta \Omega_M$ & $\Delta H_0$ \\ \hline
QE & 0.5 & 100 & 0 & 0.365& 72.77& 0.027 & 0.32 \\ \hline
QE & 1 & 100 & 0 & 0.363 & 72.83 & 0.030 & 0.36 \\ \hline
QE & 0 & 100 & 0 & 0.363 & 72.85 & 0.038 & 0.50 \\ \hline
RE & 0.5 & 100 & $\pi/16$ & 0.354 & 72.98 & 0.042 & 0.56 \\ \hline
RE & 0.5 & 100 & $\pi/8$ & 0.351 & 73.00 & 0.045 & 0.61 \\ \hline
RE & 0.5 & 100 & $\pi/4$ & 0.347 & 73.06 & 0.044 & 0.60 \\ \hline
RE & 0.5 & 100 & $\pi/2$ & 0.349 & 73.04 & 0.048 & 0.65 \\ \hline
DE & 0.5 & 100 & 0  & 0.346 & 73.08 &  0.045 & 0.62 \\ \hline
QE & 0.5 & 266 & 0 & 0.363 & 72.82 & 0.015 & 0.19 \\ \hline
RE & 0.5 & 266 & $\pi/16$ & 0.356 & 72.94  & 0.025 & 0.34 \\ \hline
DE & 0.5 & 266 & 0 &  0.356 & 72.93  &  0.024 & 0.31 \\ \hline
QE & 0.5 & 500 & 0 & 0.363 & 72.83  &  0.012 & 0.15 \\ \hline
QE & 0.5 & 1000 & 0 & 0.363 & 72.83  & 0.009 & 0.11 \\ \hline
QE & 0.5 & 1500 &  0 & 0.362  &  72.84 &  0.008 & 0.09 \\ \hline
\end{tabular}
\caption{Results for the SNe Ia using HQGA as a function of the Hyperparameters. We here recall that QE stands for Quantum Elitism, DE for Deterministic Elitism, and RE for Quantum Elitism with Reinforcement.}
\end{table*}

The tests we performed with HQGA are reported in Tab.~\ref{Tab_HQGA_Results}. Here, $\delta$ is the angle associated with the further step provided by RE. 

The results obtained with QE have smaller uncertainties on $\Omega_M$ and $H_0$, while the best mutation probability and $\delta$ are $0.5$ and $\pi/16$, respectively. Again, the results for $\Omega_M$ and $H_0$ are consistent with all tests and with AEQGA within $1 \sigma$ ($\S$~\ref{Sec_Results}). The improvement in the precision of the results obtained using 500 generations closely resembles what we have obtained in Tab.~\ref{Tab_Gen_number} from AEQGA when comparing the number of times the merit function has been evaluated by the two algorithms (1500 here for HQGA vs 1600 in the second row of that table for AEQGA). Indeed, looking also at the results with a higher number of generations, the overall accuracy improves with the number of generations in a similar fashion to what we have seen for AEQGA and $n_g$.

Regarding the scalability and the computational costs, because HQGA uses binary encoding, for multidimensional functions the total length of the binary string corresponding to the qubits also depends on the number of parameters defining the objective function,  which makes it hard to scale with more complex cosmological functions with respect to AEQGA, given that more qubits would be required. The computational time also depends on this parameter, alongside the number of individuals for which the objective function is evaluated.

In summary. while AEQGA may be better suited for the application of cosmological functions than HQGA, it is no faster than classical computations because of the previously mentioned bottlenecks. Although a gain in speed was not our main aim for this first application, we note that a fair comparison should be made by counting the number of merit evaluations needed to find the best fit. Moreover, it is still to be investigated how this latter scales with the number of parameters. It is, however, worth wondering how the present bottleneck can be circumvented. Some possible solutions could be:
\begin{itemize}
     \item implementing the merit evaluation in the quantum circuit;
     \item following HQGA, inserting some form of elitism in AEQGA;
     \item decreasing the number of generations and individuals to acquire a specific precision on the results.
 \end{itemize}

\section{Conclusions}
\label{Sec_Conclusions}
The impressive increase in the quality and quantity of data has turned cosmology from the {\it `` search for two numbers"} (in the words of Allan Sandage in the mid-70s) into a challenging exploration of many-dimensional parameter space. It has then become imperative to look for algorithms capable of performing the challenging task of finding the best-fit values, minimizing the merit function in a huge number of dimensions, especially because these are usually associated with huge collections of datasets which could also be cross-checked, increasing even more the computational complexity. Motivated by this necessity, we presented the first application of the QC techniques to the realm of cosmological parameter determination. We started from the relatively simple case with two parameters only, namely $(\Omega_M, H_0)$, and two objective functions based on different datasets, i.e., SNe IA and CMB\,+\,BAO. Our aim here was to investigate whether QC methods can retrieve similar or compatible results to classical ones.

We have therefore built a hybrid genetic algorithm in which the merit evaluation is performed classically, while the crossover and mutation operations are built in a quantum circuit once the classical populations are translated into quantum information via amplitude encoding. 

AEQGA finds results consistent with the objective function maps for both the SNe Ia and CMB\,+\,BAO sets. We have thus analyzed its behavior as a function of the main hyperparameters of our procedure, namely the crossover and mutation probabilities, the population size $n_p$, and the number of generations $n_g$ for each iteration. The precision of the results increases with $n_g$, but so does the computation time, which scales linearly with $n_g$, so one has to choose whether more precise results or a faster computation are needed. Instead, the precision does not monotonically improve with $n_p$, but it gets closer to the analytical minimum, especially in the jump between 16 and 32 individuals (see Fig. \ref{Fig_arrow_minima}). The number of iterations $n_i$ has been set to 300 as a compromise between computation time and statistical significance, but this can also be modified. Finally, an in-depth analysis has been performed on the effects of mutation and crossover probabilities, finding interesting differences between the two studied sets.

We also compared our results with two different classical genetic algorithms and a different hybrid quantum genetic algorithm, HQGA, whose philosophy is significantly different from AEQGA. These comparisons show that AEQGA derives results whose precision is consistent with the ones obtained with other methods.

It is important to stress that this is the first approach of a novel computational paradigm, QC, to a very relevant problem for the cosmological community (i.e. optimization of cosmological parameters and functions). This approach could pave the way for the applications of QC to more complex cosmological optimization and inference problems, which will be studied in future works. Here, we simply note that, as it is now, increasing the dimensions of the parameter space (for instance, to $6$ dimensions, which is the minimal number of parameters to describe the $\Lambda$CDM model) would require a bigger number of quantum circuits (in this case, 12 circuits), and thus, the computational time due to the emulation also increases. Not only that, but it could also be necessary to use more individuals per population, thus more qubits for each circuit. All these considerations mean that to solve realistic cosmological optimization problems, a more elaborate strategy could be necessary, especially for even more complex parameter spaces necessary for cosmological computations involving probes such as weak lensing. Indeed, for the latter, one not only has to derive the fundamental cosmological parameters that one wishes to study, but also two parameters for the intrinsic alignment along with a certain number of nuisance parameters (e.g., $N$ shear multiplicative bias parameters + $N$ mean redshift shift parameters, where $N$ is the number of redshift bins \cite{Laureijs_2011}). This means that if one wishes to study this problem with AEQGA, a very large number of quantum circuits would be necessary. Not only that, but even keeping ourselves with the probes used in this analysis, if one were to consider the proper CMB likelihood with all the nuisance parameters involved in these observations, the parameter space expands to more than 20 dimensions \citep{Planck2020}. And this can also be true with other cosmological probes. Nevertheless, this could be worth the investigation in future works, especially in conjunction with Bayesian inference method, finding the region where the minimum of the objective function is and starting from that for the posterior inference, possibly saving a lot of computational time in the exploration of the parameter space.

As an example of future work, also in conjunction with Bayesian inference applied to cosmological functions, in \citet{Sarracino_2025}, a hybrid Quantum Markov Chain Monte Carlo (QMCMC) has been developed to find Bayesian contours of cosmological functions using quantum computing to define the exploration steps into the parameter space. While the scope of AEQGA is not finding the shape of the objective functions (even if in $\S$~\ref{sub_sec_reiability_tests} we show how it can be reconstructed a posteriori by the algorithm results), but only finding the best-fit values, this algorithm could be used in conjunction with QMCMC (or any other classical Bayesian sampler) to help the inference of the Bayesian contours, by locating a priori the regions in the parameters space to explore finding the minima. Or, in the other sense, one could run a coarse-grained QMCMC to find approximately the regions in which AEQGA should be used.

Another very interesting test to do in a future analysis would be to test this algorithm on a real Quantum Machine.  These kinds of tests are important because, given that we are currently in the Noisy intermediate-scale quantum (NISQ) era, noise propagation in the circuit is an important phenomenon that could potentially destroy the actual quantum coherence in the circuit, thus losing the quantum properties. Actual error propagation plays a fundamental role as well. Not only that, but in the current quantum circuit, not all the Qubits are directly linked to each other. Because of all these caveats, understanding if AEQGA can be properly executed by proper quantum circuits while maintaining its precision becomes an interesting test worth performing to understand its reliability.  

As a starting point of this analysis, we have run our AEQGA on simulated backends of real quantum hardware offered by Qiskit, which are built to mimic the behaviors of IBM Quantum systems using system snapshots containing the fundamental properties of these hardware \footnote{https://quantum.cloud.ibm.com/docs/en/api/qiskit-ibm-runtime/fake-provider}. More specifically, we have considered a sample run with 5 generations and tried to run it using all the V2 fake backends present here. Out of these, on 25 we have managed to obtain an estimate on the QPU runtime, considering also the multiplication by the number of shots. In the vast majority of cases, on a run of approximately 8 seconds on the classical device, the estimated quantum runtime is less than 1 second, with the best result obtained by the FakeHanoi backend, with 0.22. This estimate does not take into account other time losses one would encounter in using quantum hardware, like the queue time or the submission time. Nevertheless, this also proves that, ideally, using real hardware would be beneficial for our computations with respect to the classical simulations, albeit one has to note that the runtime would still be dominated by the evaluation of the objective function. In this sense, a possible way to find the quantum advantage that will be worth exploring could be in changing the aim of AEQGA, not fixing the hyperparameters a priori, but deriving the number of evaluations of the objective function a posteriori, giving as a convergence check for the algorithm a specific precision on the results.

It is important to conclude here that this work is only a part of the research line looking for possible investigations that are currently being tested for the applications of Quantum Computing in Astrophysics and Cosmology. Apart from the QMCMC, other ideas involve quantum machine learning for signal detection \citep{farsian_2025}, Quantum Fourier Transform for CMB  \citep{farsian_2025_b}, and the numerical integration of differential equations \citep{Cappelli_2025}. Other possible applications will also be investigated, like for instance data compression and analysis/classification of large datasets, a task of fundamental importance for astrophysical and cosmological datasets.

\section*{Acknowledgements}
This work is supported by the ICSC - Centro Nazionale di Ricerca in HPC, Big Data e Quantum Computing, CN00000013, spoke 10 ``quantum computing", CUP C53C22000350006. We wish to thank the group of prof. Acampora for their insight and discussions related to HQGA. For our computations, we acknowledge the use of the DAME machine at the Osservatorio Astronomico di Capodimonte, Naples. We also wish to thank the two anonymous referees for their useful insights.

\bibliographystyle{elsarticle-harv} 
\bibliography{main}
\appendix

\section{Brief Introduction to Quantum Computing} \label{sub_sec_brief_introd}
\begin{figure}
    \centering
\includegraphics[scale=0.5]{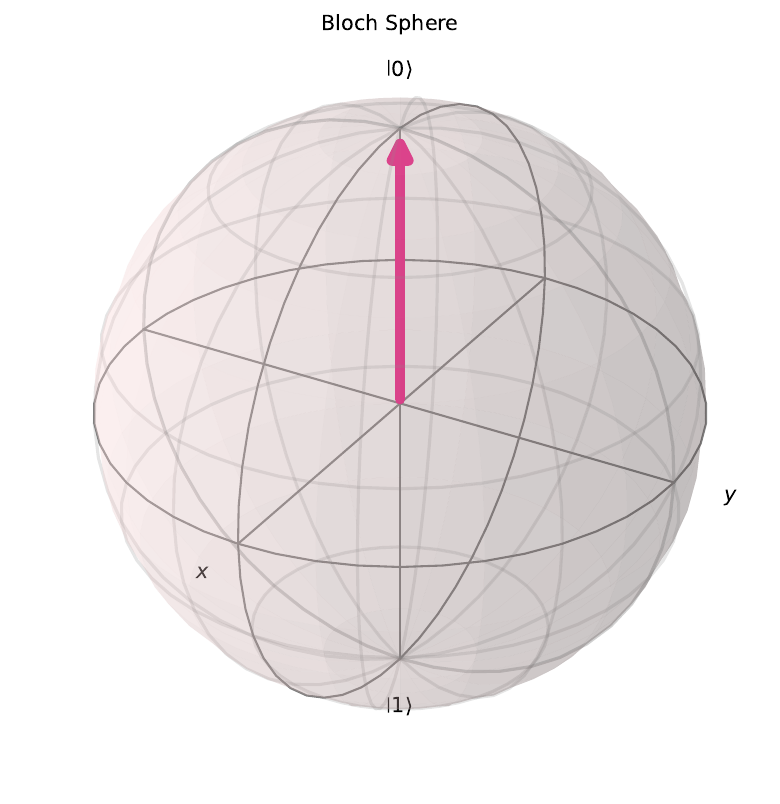}
\caption{Geometrical representation of a qubit in the $\ket{0}$ state on the Bloch Sphere.}
\label{Fig_Bloch_Sphere}
\end{figure}

    The main difference between QC and Classical Computing is in the fundamental information unit: indeed, while the latter is the classical bit which can either be 0 or 1, the former is the qubit, which in the simplest case is a unit vector in a 2-dimensional space whose basis can be written as $[\ket{0},\ket{1}]$ \citep{Rieffel_2000, Chae_2024}. A single qubit can be in any superposition
    \begin{equation} \label{qubit}
        \psi=\alpha\ket{0}+\beta\ket{1},
    \end{equation}
    with the constraint
    \begin{equation} \label{normalization}
        |\alpha|^2+|\beta|^2=1.
    \end{equation}
    Geometrically, a qubit can be represented as a unitary vector lying in the so-called Bloch Sphere, which is a unit 2-sphere, whose north and south poles are usually chosen to represent the $\ket{0}$ and $\ket{1}$ states, respectively. A  representation of the $\ket{0}$ on the Bloch Sphere is shown in Fig.~\ref{Fig_Bloch_Sphere}.

    This novel formulation for the fundamental information unit allows to use concepts of Quantum Mechanics in the context of Computer Science, namely superposition and entanglement. The former is expressed as in Eq.~(\ref{qubit})  and generalizes straightforwardly to an $n$-qubit register (i.e, a collection of qubits from which $2^n$ states can be defined via superposition). Entanglement, on the other hand, gives rise to multi-qubit states that cannot be expressed as tensor products of individual qubit states. An example is the Bell state for two qubits, expressed as follows:
    \begin{equation} \label{Bell State}
        \phi=\frac{1}{\sqrt{2}}\ket{00}+\frac{1}{\sqrt{2}}\ket{11},
    \end{equation}
    where the notations $\ket{00}$ and $\ket{11}$ describe the tensor products $\ket{0} \otimes \ket{0}$ and $\ket{1} \otimes \ket{1}$ of the states of the single qubits, respectively. The existence of such states allows for quantum phenomena that would be impossible in a classical setting, such as instantly knowing the state of a qubit by measuring the one entangled to it independently of the distance between the two qubits. These correlations appear instantaneous but obey the no-signaling principle: they cannot be used for faster-than-light communication and are compatible with relativistic causality. Superposition and entanglement are pivotal concepts for QC, allowing to express $2^n$ possible states using only $n$ different qubits. These state spaces can not be efficiently explored using classical resources, thus requiring the proper usage of quantum resources and hardware.

    We focus on the gate-based QC framework, which has been used in our algorithm. Gates are unitary matrices that can act on a small number of qubits, and represent the quantum operations that can be performed on them. As an example, a very famous 1-qubit gate is the Hadamard, schematized as 
    \begin{equation} \label{Hadamard}
        H = \frac{1}{\sqrt{2}}
            \begin{bmatrix}
                    1 & 1 \\
                    1 & -1 \\
            \end{bmatrix}.
    \end{equation}
    After the application of this gate to a qubit whose initial state is $\ket{0}$ and measuring classically the new state, we find an equal probability of measuring $\ket{0}$ and $\ket{1}$. In general, any gate applied on a single qubit may be schematized as a rotation around three different axes (from now on, we refer to the Bloch sphere and its axes when we describe rotations)
    \begin{equation} \label{rx}
        R_x(\theta) = \begin{bmatrix}
\cos(\frac{\theta}{2}) & -i\sin(\frac{\theta}{2}) \\
-i\sin(\frac{\theta}{2}) & \cos(\frac{\theta}{2})
\end{bmatrix},
    \end{equation}
    \begin{equation} \label{ry}
        R_y(\theta) = \begin{bmatrix}
\cos(\theta/2) & -\sin(\theta/2) \\
\sin(\theta/2) & \cos(\theta/2)
\end{bmatrix},
    \end{equation}
    \begin{equation} \label{rz}
        R_z(\theta) = \begin{bmatrix}
e^{-i\theta/2} & 0 \\
0 & e^{i\theta/2}.
\end{bmatrix}
    \end{equation}
    The most general rotation around a given axis may be expressed as the following Unitary matrix:
    \begin{equation} \label{rotation}
        U = e^{i \gamma} \begin{bmatrix}
e^{-i(\beta+\delta)/2} \cos(\theta/2) & -e^{-i(\beta-\delta)/2} \sin(\theta/2) \\
e^{i(\beta-\delta)/2} \sin(\theta/2) & e^{i(\beta+\delta)/2} \cos(\theta/2)
\end{bmatrix},
    \end{equation}
    where $\gamma$ is a global phase, $\beta$ and 
$\delta$ are the phases corresponding to rotations about the 
$z$-axis before and after the rotation 
$\theta$ around the $y$-axis. We note how the rotation around the $z$-axis is a phase shift that does not change the probability associated with a quantum state. As we described in $\S$~\ref{sub_sec_Quantum_Crossover_Mutation}, this is important for the implementation of our quantum genetic algorithm.

Regarding the gates acting on more than one qubit, arguably, the most important quantum gate of this type is the controlled-NOT operation (CNOT), represented by the following matrix 
\begin{equation} \label{CNOT}
    \text{CNOT} = \begin{bmatrix}
1 & 0 & 0 & 0 \\
0 & 1 & 0 & 0 \\
0 & 0 & 0 & 1 \\
0 & 0 & 1 & 0
\end{bmatrix}.
\end{equation}
This operation uses two qubits, called control and target, and acts as follows:
\begin{itemize}
    \item If the control qubit is in state $\ket{0}$, the state of the target qubit remains unchanged.
    \item If the control qubit is in state $\ket{1}$, the state of the target qubit is flipped.
\end{itemize}
The operations implemented in the quantum part of our algorithm are compositions of rotational gates acting on one qubit and CNOT gates linking two qubits, as shown in $\S$~\ref{Sec_our_QGA}. It is also possible to define directly gates acting as controlled rotations, i.e. compositions of unitary gates and CNOT. Generally, if one applies three rotation gates $U_1, U_2, U_3$ then the composed gate $C$ is defined as
\begin{equation} \label{composed_gate}
    C = U_3 \, U_2 \, U_1.
\end{equation}

These gates are used in quantum circuits, that is, a series of operations acting on the qubits. These are concluded with the classical measurements, which collapse the quantum states into classical information.

\section{Genetic algorithms, cosmological applications, and an overview on Quantum Genetic Algorithms}
    \label{Sec_Introduction_QGA}

   In this section, we introduce genetic algorithms. First, we describe classical genetic algorithms and various cosmological applications. Then, we describe the quantum genetic algorithms already available in the literature, the idea behind their implementation, and their use cases.

\subsection{Classical genetic algorithms and cosmological applications} \label{sub_sec_CGA_lit}

  Classical genetic algorithms \citep{Goldberg_1989} have now become a stable technique for a variety of problems, such as parameter estimation and, in general, optimization of different kinds of merit functions. Given an initial population of candidate solutions, genetic algorithms perform various operations, such as crossover and mutation among the individuals of the population, and then evaluate them via a merit function that assesses how good the individuals are as the solution of a particular problem. The best individuals are selected in relation to the problem at hand and kept, while the rest of the population is re-populated according to a possible plethora of ways, taking into consideration the operations performed at the previous generation. This process is repeated for different generations until a stopping criterion is reached. As previously mentioned, we aim to reproduce this idea with the means of QC, for the parameter estimation of cosmological functions. 

 These kinds of algorithms are called genetic algorithms because they are inspired by natural evolutionary processes \citep{Bagavathi_2019}. For a more detailed discussion on how they generally behave and what kind of operations can be performed by imitating biological computing, especially in the selection, crossover, and mutation operations, see \citet{Medel_Esquivel_2023}. Here, we mention that classical genetic algorithms have been used both for combinational and continuous optimization problems. For a simple example of the former, following \citet{Mitchell_1996}, one uses a bit representation of the genes (similar to the quantum binary encoding, as we will show in $\S$~\ref{sub_sec_QGA_lit}). The algorithm goes as follows:
 \begin{itemize}
     \item Start with a randomly generated population of chromosomes represented by l bits.
     \item Calculate the fitness evaluation for each of these chromosomes.
     \item Repeat the operations of crossover and mutation with given probabilities, the former defined as cross over the pair at a randomly chosen point of the string to form two offspring. If no crossover takes place, form two offspring that are exact copies of their respective parents. The latter, instead, is a mutation in that given point. These operations are performed on two "parents" chromosomes, selected with a given probability dependent on the previous evaluation of the merit function.
     \item Replace the current population with the new population.
     \item Repeat these operations until a given generation number is reached.
     \item Repeat the entire run to obtain different results on the genetic algorithm to perform statistics on the results.
 \end{itemize}
 This framework can be used in any problem in which a binary representation is suitable \citep{Eiben_Smith_2015}, thus, it can also be extended to continuous functions, as it has already been done in the literature.


Genetic algorithms have been used in a cosmological setting, as a way to solve optimization problems. For instance, in \citet{Medel_Esquivel_2023,Gomez_Vargas_2023, Gomez_Vargas_2024} there are applications of classical genetic algorithms for the optimization of cosmological parameters. Here, it is argued that optimization and sampling algorithms typically used in a cosmological setting (like the Bayesian posterior inference) are two different tasks that can be used in conjunction. It is important to stress that in this context, optimization means to find the best-fitting values for the cosmological parameters, while sampling means to explore the posterior distributions, thus allowing for a more complete idea of the cosmological parameter space. For instance, the optimization from the genetic algorithm allows one to find the region around the global minimum to sample, thus limiting the issues of relative minima in sampling bimodal distributions \citep{Medel_Esquivel_2023}. It has also been shown that genetic algorithms can be used together with other machine learning methods for cosmological applications \citep{Gomez_Vargas_2023, Gomez_Vargas_2024}. 

 Genetic algorithms have also been used for the reconstruction of cosmological functional forms, going beyond parameter inference. Indeed, in \citet{Arjona_2020} genetic algorithms have been used to reconstruct the luminosity distance equation from a sample of SNe Ia. Related to this point, in \citet{Alestas_2022}, a comprehensive analysis using cosmological data of different kinds is shown, for which genetic algorithms are used to reconstruct functions showing possible deviations from the $\Lambda$CDM model and General Relativity.

 Instead, in \citet{Orjuela-Quintana_2023}, machine learning methods, and in particular genetic algorithms, have been used in relation to the matter transfer function $T(k)$ related to the matter power spectrum $P(k,z)$ for the large-scale structures. Other analyses of this kind can be found in \citet{Kamerkar_2023}, where genetic algorithms have been used to reconstruct the inflationary potential via cosmological data, and in \citet{Lodha_2024}, where genetic algorithms have been used to explore features of the CMB spectra in a model-independent way. For other works of this kind, see also \citet{Arjona_2020B, Arjona_2021}.

\subsection{Quantum Genetic algorithms: an overview} \label{sub_sec_QGA_lit}
Having presented some applications of quantum computing to cosmological problems, we now show some examples of quantum genetic algorithms, and more general optimization routines involving QC, already implemented in the literature. For a more general introduction on QC, we refer to $\S$~\ref{sub_sec_brief_introd}. Here, we will show some details relevant to our algorithm.

The vast majority of optimization problems in gate-based QC can be divided into two main categories \citep{Abbas_2024}: algorithms implemented to solve combinatorial problems, and those whose main goal is to find the ground state of a given Hamiltonian. 

For the former, the most implemented is the  Quantum Approximate Optimization Algorithm (QAOA, \citealt{farhi_2014}), which is a hybrid iterative method successfully applied to combinatorial problems, among which the max-cut problem: i.e, partition nodes of a graph in a way that maximizes the number of edges between nodes in differing groups. QAOA are one of the most interesting class of quantum algorithms from a computational complexity point of view, giving promising results for a variety of applications. For more details see \citet{Wang_2018, Blekos_2024}

In the second group, the most popular class of algorithms is known as the Variational Quantum Algorithms (VQA, \citealt{Cerezo_2021}). These are hybrid algorithms that use classical optimization routines to train parametrized quantum circuits, i.e. circuits composed of rotational gates (like the ones shown in Eq. \ref{rotation} in $\S$~\ref{sub_sec_brief_introd}) depending on parameters $\theta$ to be found by the routine itself as the best solution for the problem under study. These can also be linked to quantum machine learning algorithms \citep{Benedetti_2019, farsian_2025}.

We stress that neither of the aforementioned classes of algorithms applies directly to our case. 

 We now go more into detail in the literature regarding quantum genetic algorithms. The interjection of quantum computing and evolutionary algorithms has been studied before the birth of proper quantum hardware  \citep{Spector_1998, Han_2000, Spector_2004, Sofge_2008}. Here, the first concepts of evolutionary algorithms applied to quantum computing have been theoretically analyzed, finding applications for combinational optimization problems as well as Deutsch’s
 Early Promise Problem for the nature of the qubits \citep{Deutsch_1985}.

 In \citet{Ibarrondo_2022, Ibarrondo_2023} a quantum genetic algorithm has been implemented with the goal of minimizing a Hamiltonian as in the VQA formalism, but with the key difference of not having a classical optimization routine complementing the algorithm nor a parameterized quantum circuit.  More specifically, in these algorithms, the individuals are quantum registers, each represented by $c$ qubits, i.e.\ a state $|\psi_i\rangle$ defined in the Hilbert space $\mathcal{H}_{\text{reg}_i} \cong (\mathbb{C}^2)^{\otimes c}$. A population of $n$ individuals then spans the composite space $\mathcal{H}_{\text{pop}} = \bigotimes_{i=1}^n \mathcal{H}_{\text{reg}_i}$. In other words, each individual in the population is a small quantum register, and the full population is represented as a tensor product of these registers. The goal is to minimize a target Hamiltonian $H_p$ by evolving the population of registers with genetic-inspired operations. This is done by ranking the registers based on energy measurements with respect to $H_p$, and then applying operations of quantum crossovers (i.e, swapping the qubits of given registers) or mutations (mutating a qubit with a given probability).

 Other quantum genetic algorithms have been designed more as search algorithms, i.e. finding the specific state corresponding to the minimum of a function whose form is already known. The idea is to start from one of the most successful quantum algorithms ever implemented, the Grover algorithm \citep{Grover_1998}, which has already proven to be computationally faster than the classical counterpart. Indeed, while classical selection algorithms are of computational cost $\mathcal{O}(n)$, where $n$ is the number of array elements in which the selection is taken place, Grover's quantum algorithm for selection is proven to be of cost $\mathcal{O}(\sqrt{n})$  if the oracle (i.e., the black box component of the quantum circuit which ``knows" what particular state to select as the solution of the problem) of the algorithm is already built. 

 Following this idea, an implementation of a quantum genetic algorithm \citep{Lahoz-Beltra_2016} discretizes the values of the function to be minimized, then binary encodes these discrete values in quantum states using their binary representation as follows
\begin{equation} \label{binary_encoding}
    x_k \longrightarrow |x_k\rangle = |b_1 b_2 \dots b_n\rangle, \quad b_i \in \{0,1\}.
\end{equation}

 Assuming that the position of the minimum is already known, a quantum oracle is built to mark the state corresponding to the minimum by applying a phase flip to this state 
\begin{equation} \label{Oracle}
\mathcal{O}_{f} |x\rangle =
\begin{cases}
-|x\rangle & \text{if } x = x_{\min} \\
\;\;\;|x\rangle & \text{otherwise}
\end{cases}
\end{equation}

 This oracle is then used for Grover's algorithm to increase the probability that the correct quantum state is selected after the measurement in a few iterations.

Another quantum genetic algorithm based on the binary encoding of continuous functions has been implemented in \cite{Acampora_2021, Acampora_2022, Acampora_2023, Acampora_2023b}, called Hybrid Quantum Genetic Algorithm (HQGA). In this case, the use of Grover's algorithm is avoided, following a different scheme for the quantum selection that does not assume knowing the position of the minimum beforehand. Also, another, different, definition of quantum crossover and mutation is implemented, the former by using conditional operations considering ancillary qubits, while the latter as rotations on the main qubits. Finally, the classical merit operation is computed for the selected state. This concept has been implemented for different benchmark functions and can also be extended to external functions. Indeed, as shown in $\S$~\ref{sec_comparison}, we implemented this algorithm for our cosmological case, comparing the results with AEQGA.

 From this summary of the quantum genetic algorithms in the literature, one notices that many different implementations have been proposed, giving different meanings to the population, crossover, and mutation concepts, while keeping the core idea of an evolutionary algorithm. Our AEQGA follows this scheme.

\section{Reliability Tests}
\label{App_Reiability_tests}
In this appendix, we present some reliability tests we have performed to test the validity of our approximations on the $\chi ^2$ functions to be minimized as well as the performance of AEQGA with different functions.

\subsection{Testing the approximations on the objective functions} \label{sub_sec_approximations}
\begin{figure*}
\centering
\includegraphics[width=0.4\hsize]{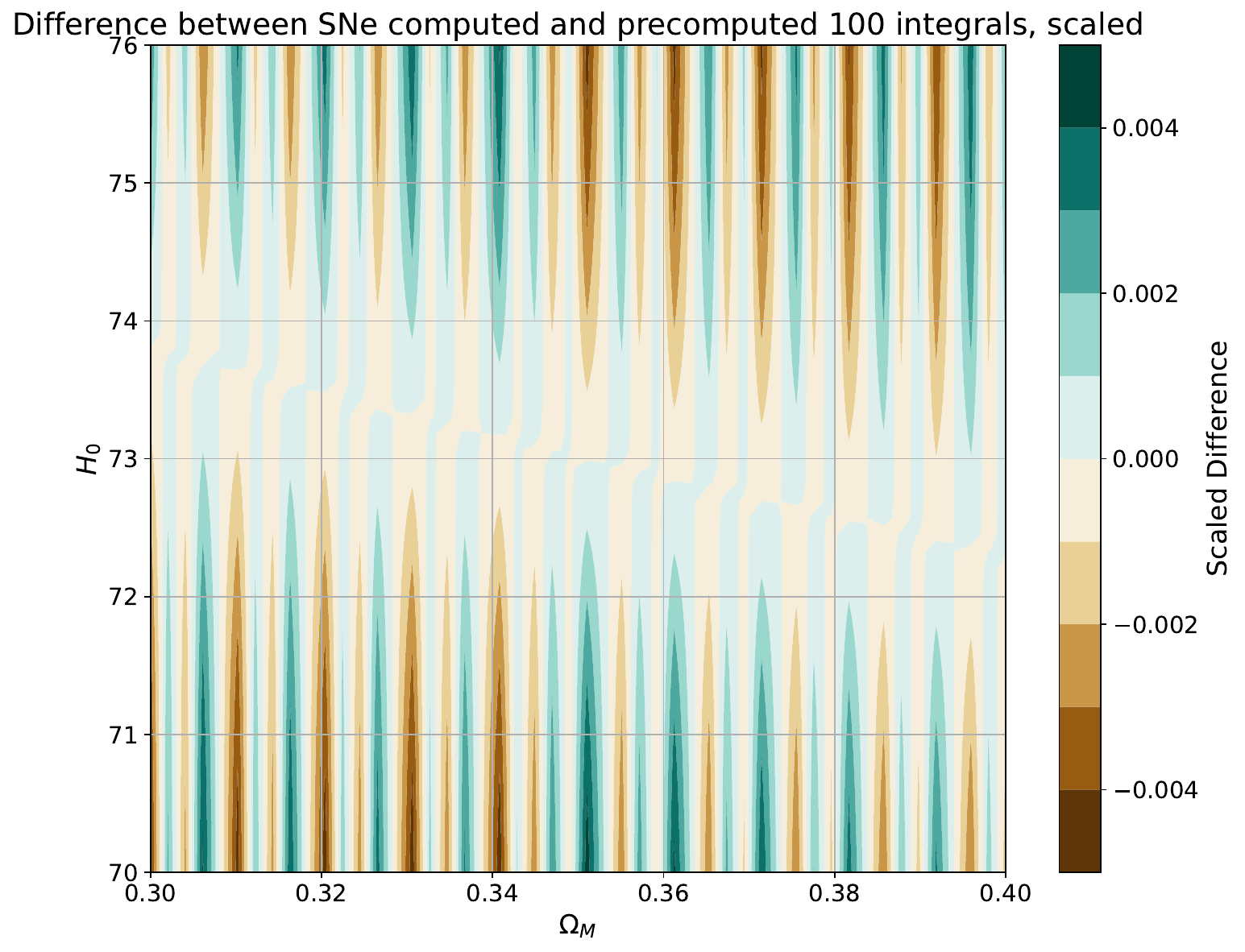}
\includegraphics[width=0.4\hsize]{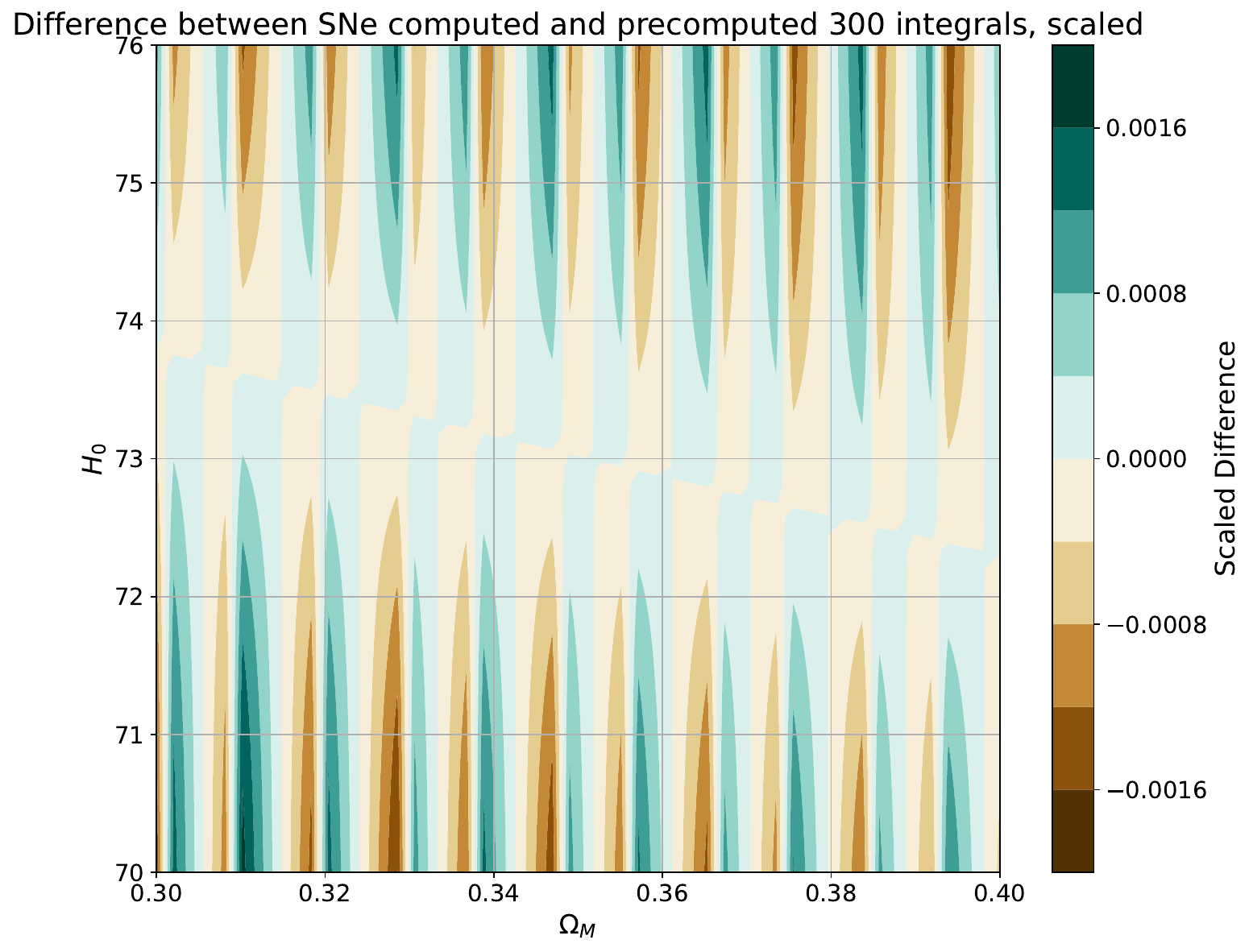}
\caption{The fractional differences between the map for the objective function derived by pre-computing the integrals defining the luminosity distance for different values of $\Omega_M$ between 0 and 0.5. Left panel: for 100 values. Right panel: for 300 values. These have been computed as the difference between the two maps divided by the pre-computed results. }
\label{Fig_Contour_Maps_difference}
\end{figure*}

Firstly, we investigate whether having precomputed the luminosity distance entering the SNe Ia data comparison impacts the estimate of the objective function. To this end, we produce two maps varying the number of steps in $\Omega_M$ over the range (0.0, 0.5) where we pre-compute the integrals. We then rescale the difference maps with respect to the fiducial one (i.e., the top panel of Fig.~\ref{Fig_Contour_Maps}, computed analytically) shown in Fig.~\ref{Fig_Contour_Maps_difference}. Note that $\Omega_M$ is the only parameter entering the integral in Eq.~(\ref{E(z)}). 

\begin{figure*}[ht!]\centering
\includegraphics[width=0.4\hsize]{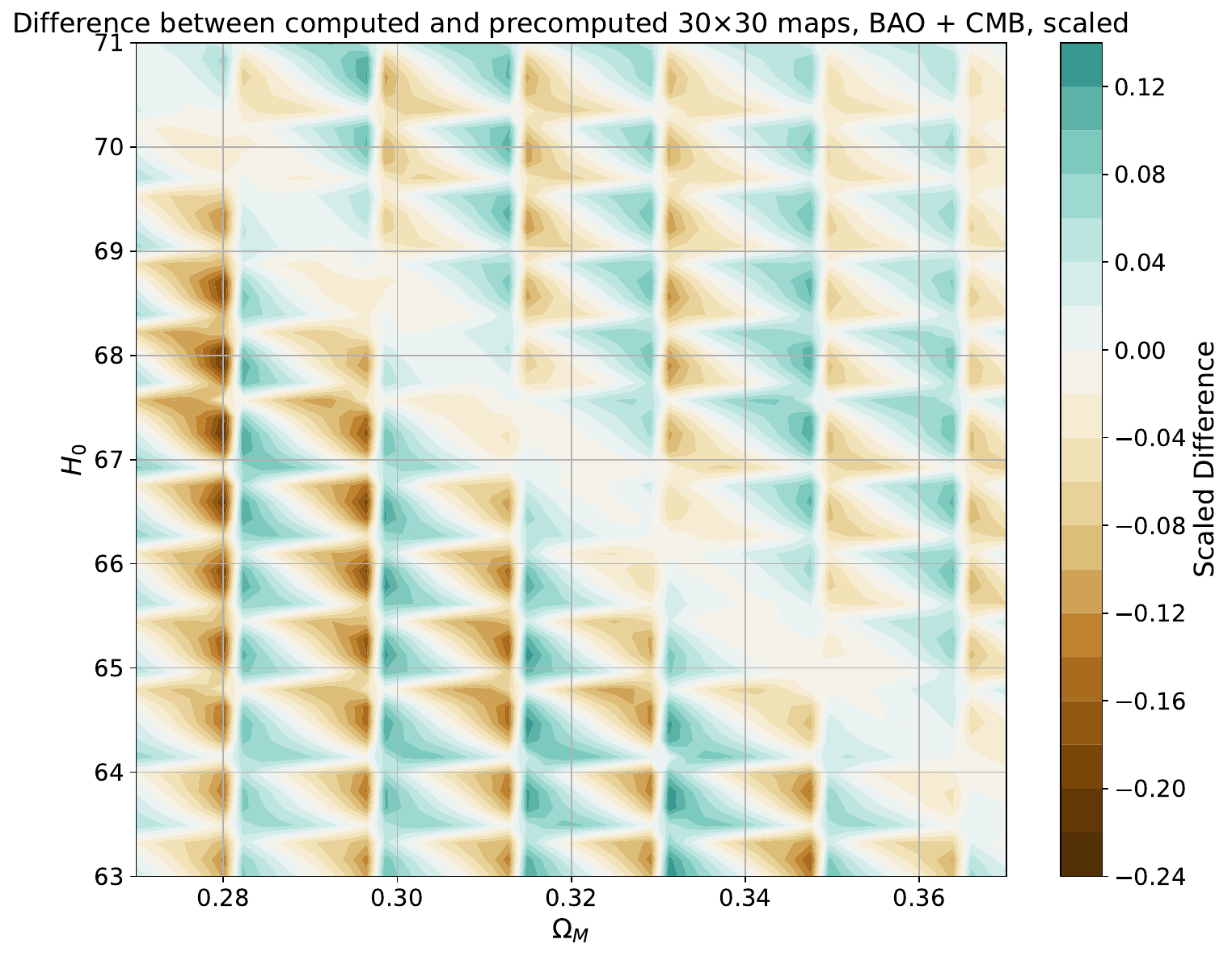}
\includegraphics[width=0.4\hsize]{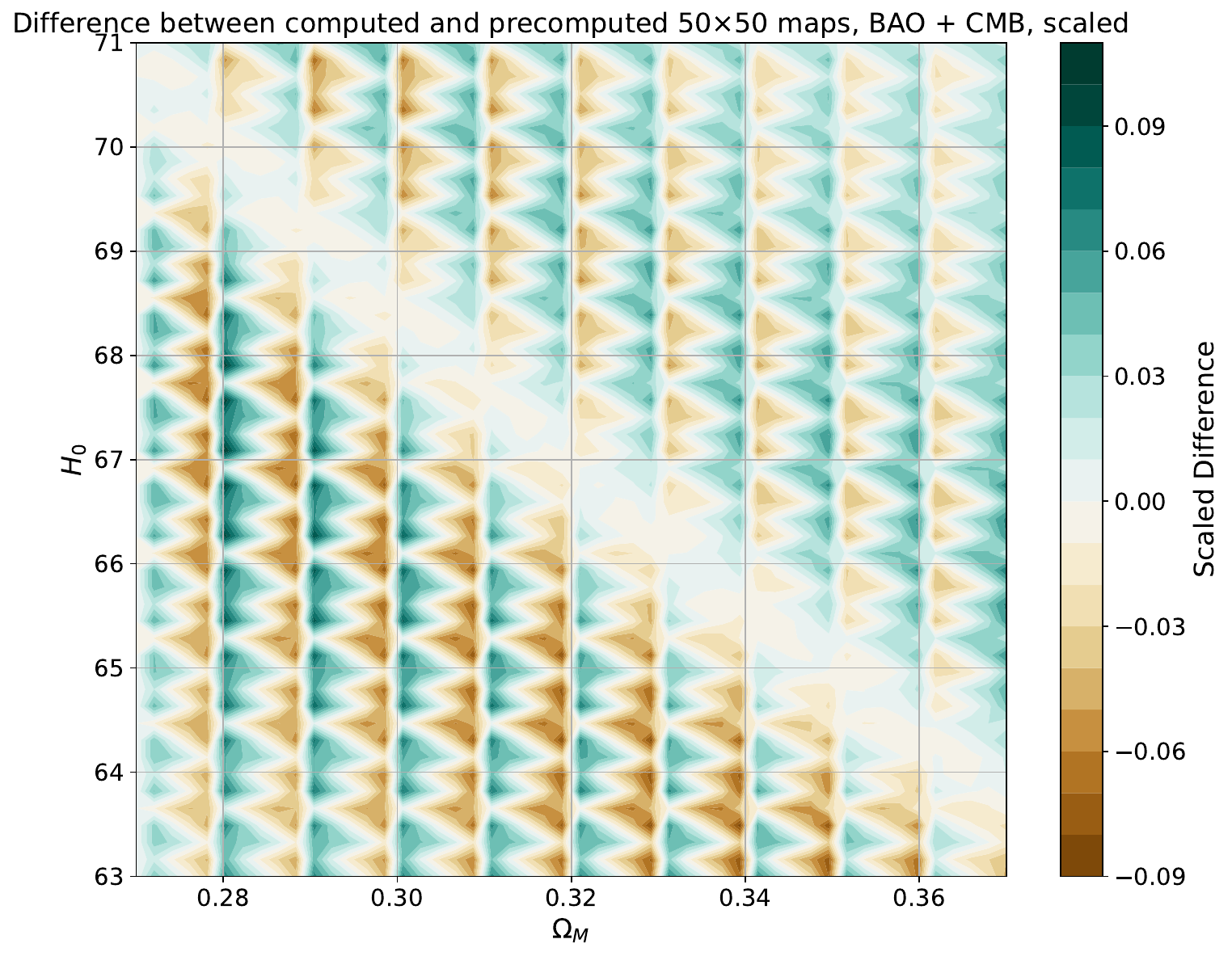}
\includegraphics[width=0.4\hsize]{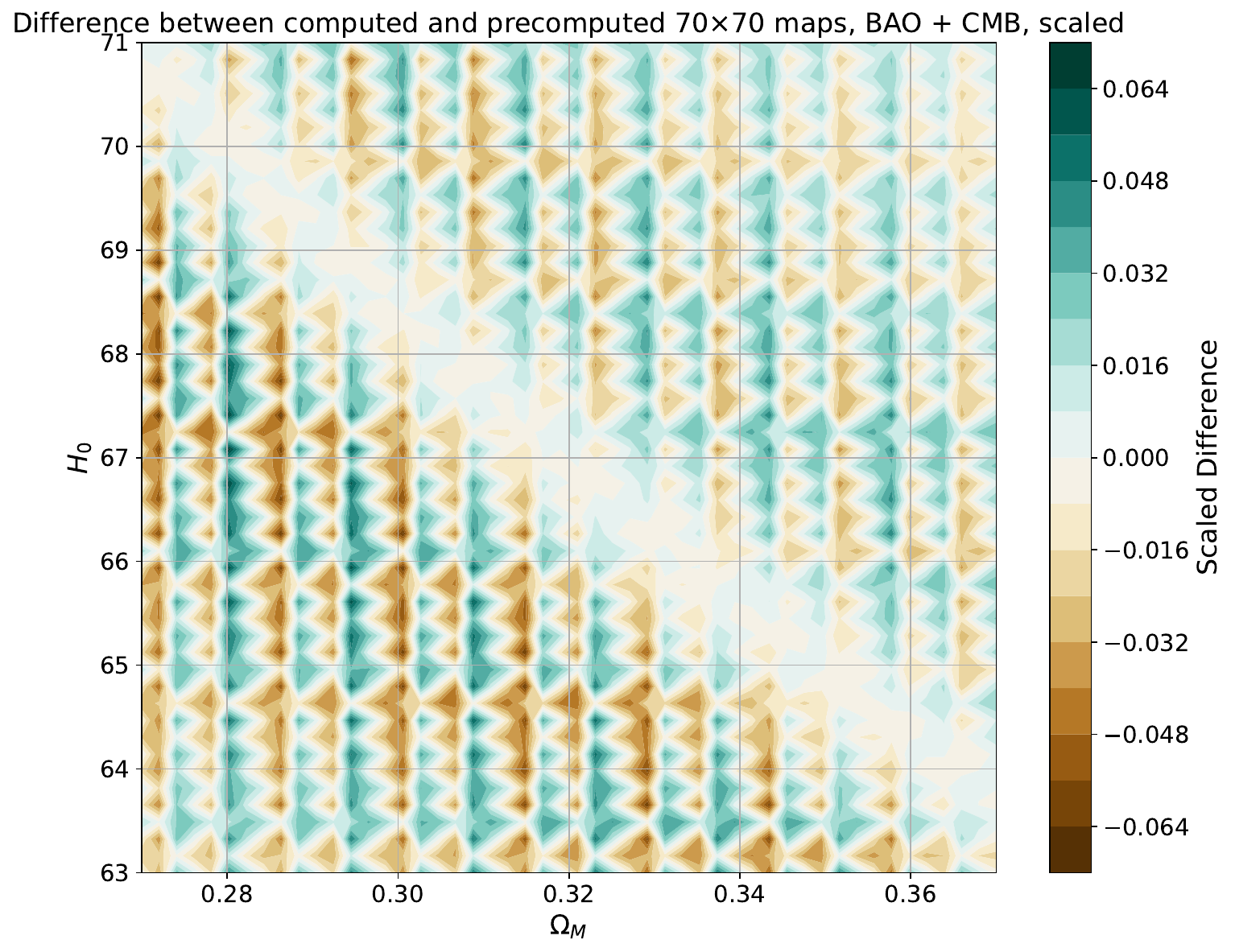}
\includegraphics[width=0.4\hsize]{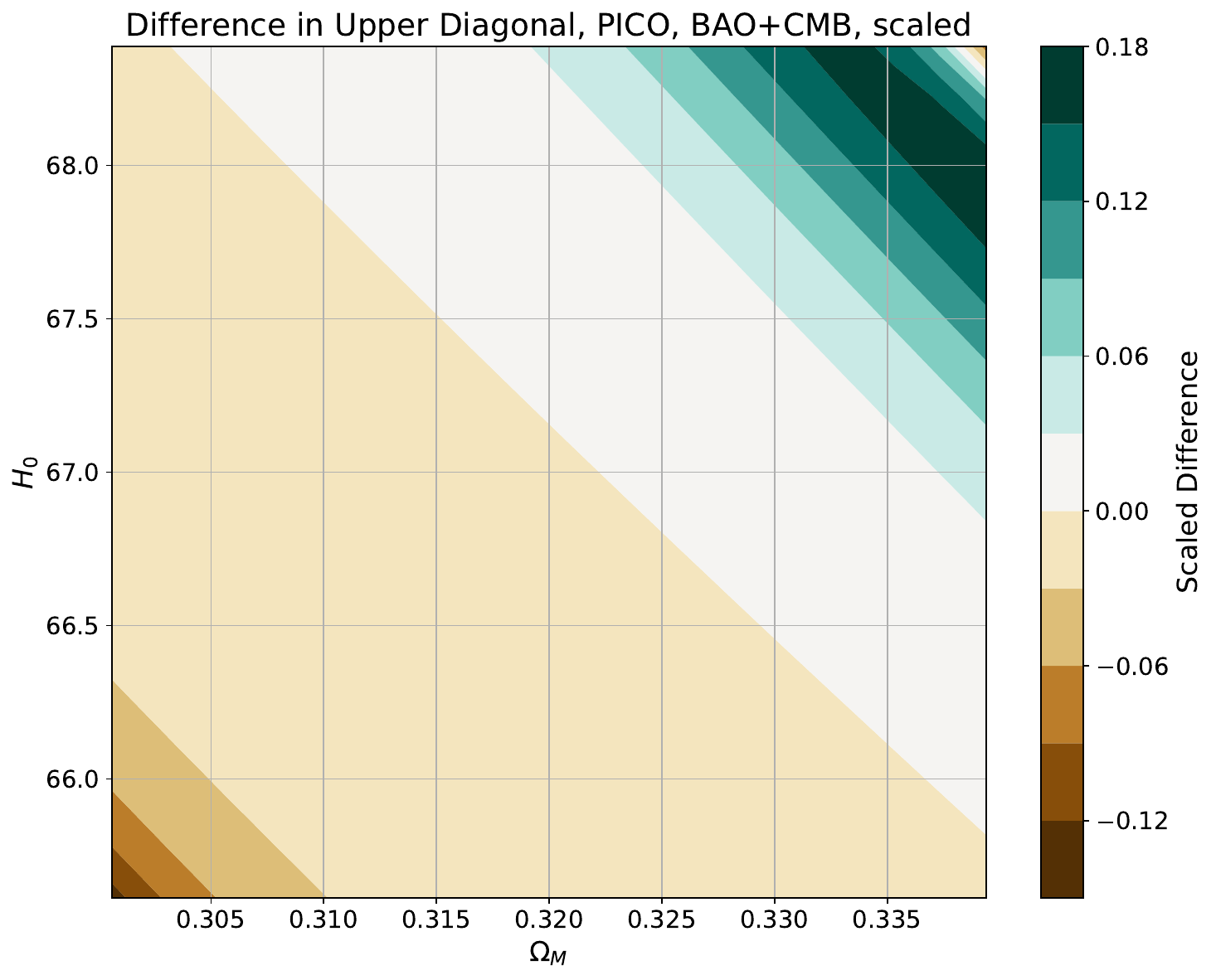}
\caption{The fractional difference, as computed in Fig. \ref{Fig_Contour_Maps_difference}, between the map for the objective function derived by pre-computing the TT spectrum for the CMB for different couples of values of $\Omega_M$ and $H_0$ between 0 and 0.5 and 60 and 80, respectively. Top left panel: for 30 couples.  Top right panel: for 50 couples. Bottom left panel: for 70 couples. Bottom right panel: the same fractional difference but using \texttt{PICO}, zoomed in the region around the absolute minimum.  }
\label{Fig_Contour_Maps_difference_BAO_CMB}
\end{figure*}

We note the very small difference between the pre-computed maps and the original one, especially in the region with the smallest value for the objective function, which is also the region we wish to study in our analysis. This test has demonstrated that we can use both strategies for the pre-computation. Indeed, we used the 100 pre-computed integrals to infer general statistical information from the results, and the 300 pre-computed integrals to draw the plots we show in our analysis.

A similar check has been done for the TT spectra of the CMB, comparing the results by \texttt{CAMB} with the ones derived from the pre-computed spectra and \texttt{PICO}. For the pre-computation, we have chosen three grids: (30, 30), (50, 50), and (70, 70) different values for $\Omega_M$ and $H_0$, selected in the same intervals considered for the SNe Ia. The paucity of grid points compared with the SNe Ia test is due to the more intensive and time-consuming computations involved in the evaluation of the different spectra.

The comparisons are shown in Fig.~\ref{Fig_Contour_Maps_difference_BAO_CMB}. Increasing the number of pre-computed spectra increases the precision of the results, as expected. We also note how, in the region around the absolute minimum \texttt{PICO} shows homogeneity and smoothness in its contours, thus does not artificially create local minima in the objective function, but its results are worse than the pre-computed spectra in regions far from the absolute minimum. Thus, we used this emulator for our experiments involving the CMB\,+\,BAO dataset.

\subsection{Tests on simulated datasets} \label{sub_sec_reiability_tests}

To understand if our results are consistent with a cosmological model used to create mock data, we simulate a dataset of luminosity distances logarithmically equispaced in the redshift interval [0.01, 2]. These have been created in a flat $\Lambda$CDM model with $\Omega_M= 0.3$ and $H_0=70$, adding error bars on the data and a Gaussian dispersion around the curve given by the input model, thus having data spread around it rather than perfectly aligned on it, as one expects from a real dataset. 

\begin{figure*}\centering
\includegraphics[width=0.49\hsize]{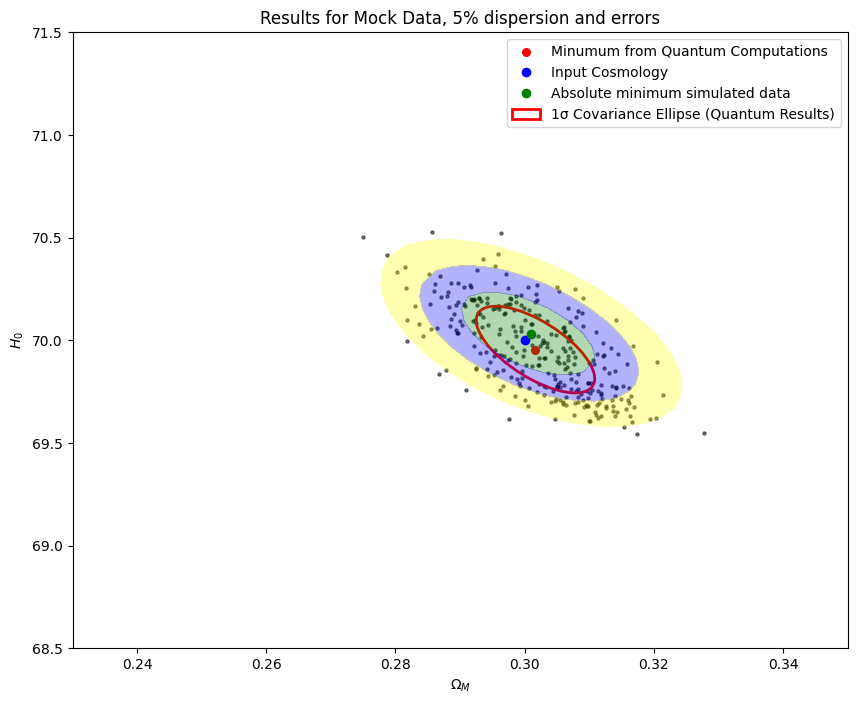}
\includegraphics[width=0.49\hsize]{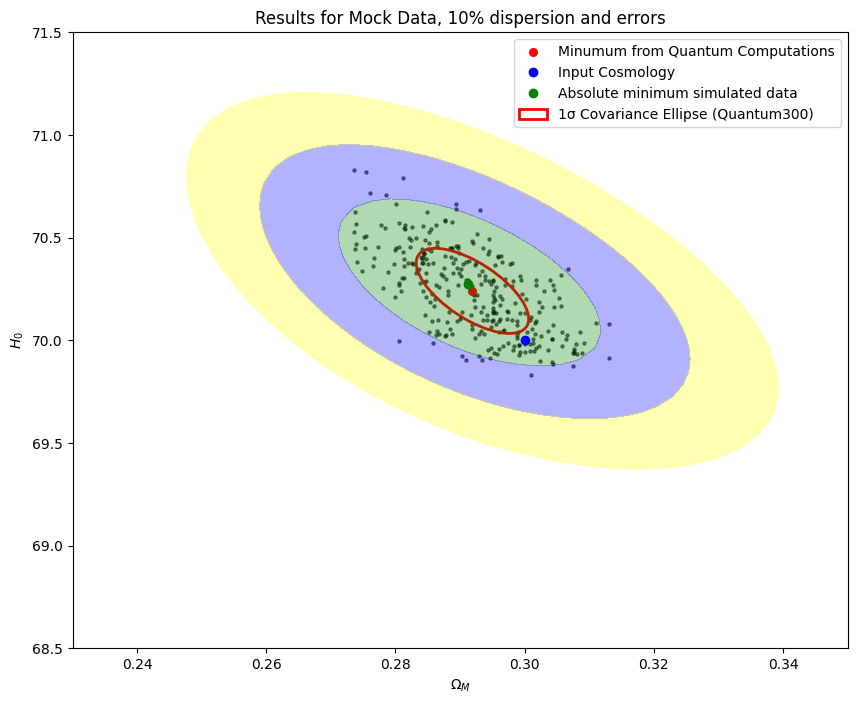}
\includegraphics[width=0.49\hsize]{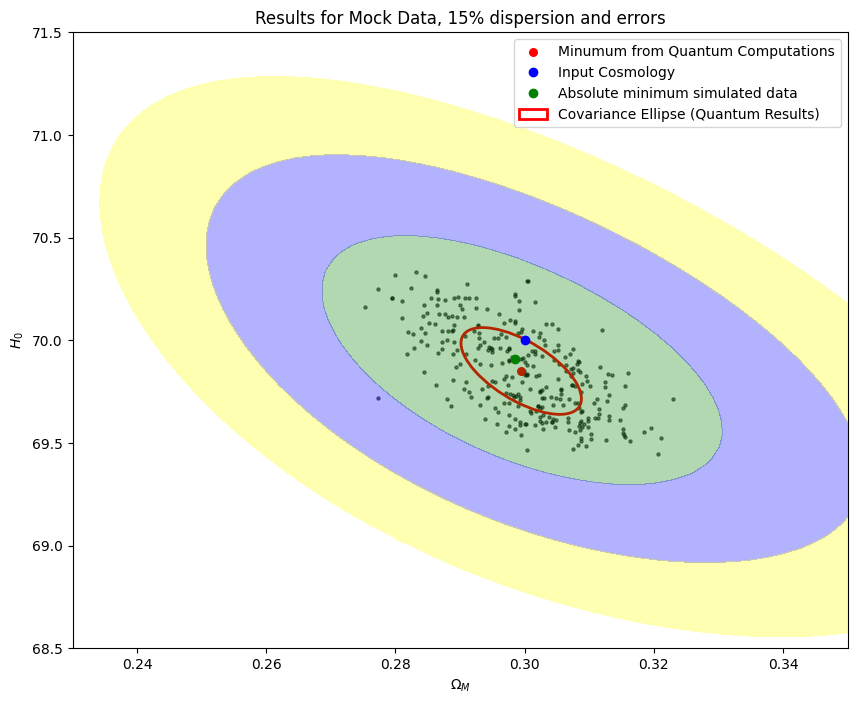}
\includegraphics[width=0.49\hsize]{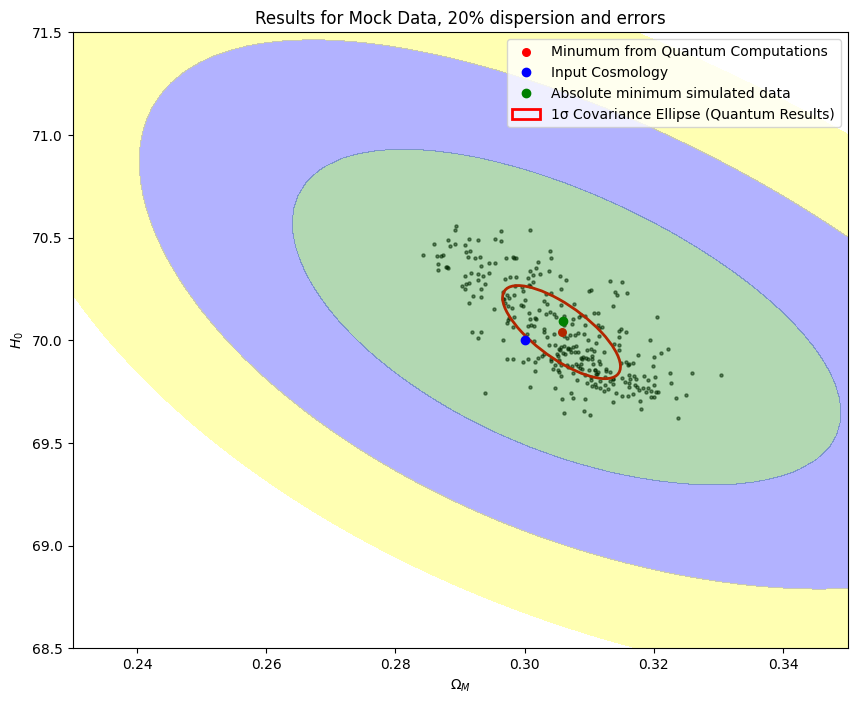}
\includegraphics[width=0.49\hsize]{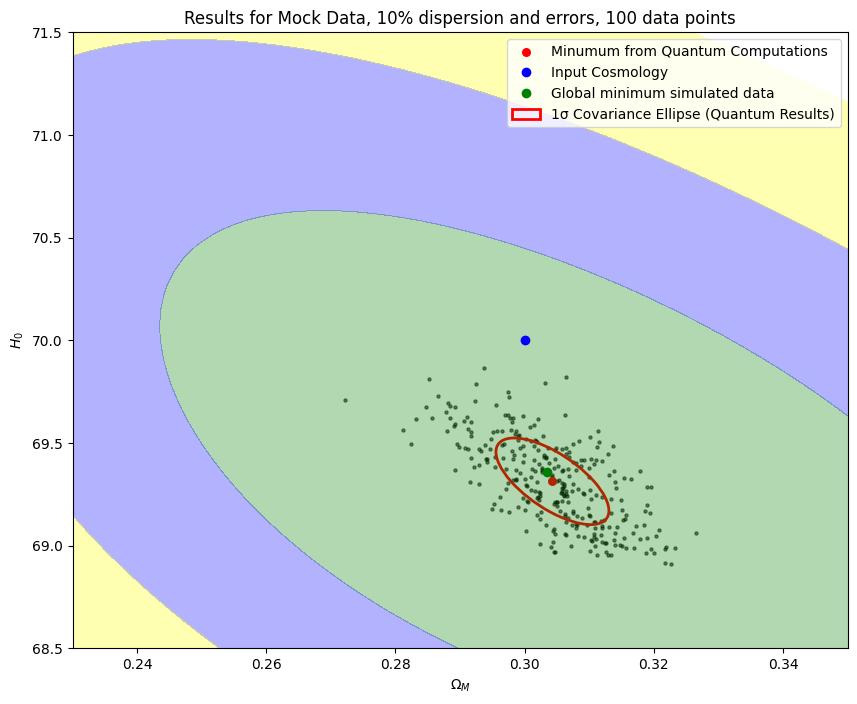}
\includegraphics[width=0.49\hsize]{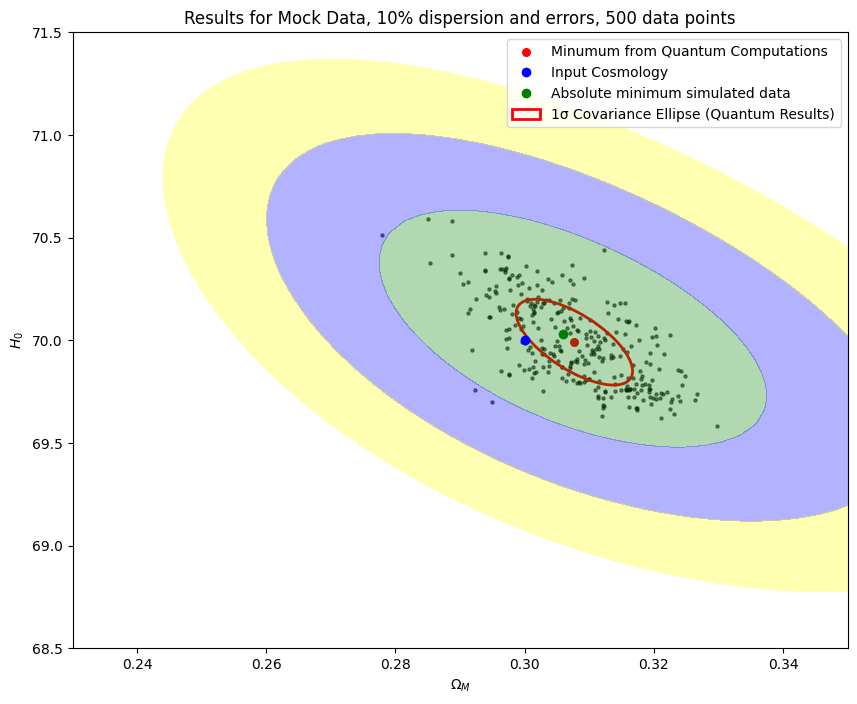}
\caption{The results for AEQGA with the mock dataset. Top panels: with 1000 simulated data and $5\%$ (left) and $10 \%$ (right) error and dispersion. Middle panels: for $15\%$ (left) and $20 \%$ (right) error and dispersion. Bottom panels: fixing error and dispersion to $10 \%$ and using 100 (left) and 500 (right) simulated luminosity distances.  }
\label{Fig_Mock}
\end{figure*}

We performed six tests. In the first four, we simulated 1000 luminosity distances, then we added an error equal to $5\%, 10\%, 15 \%,$ and $20 \%$ of the simulated luminosity distance and a Gaussian dispersion with a standard deviation equal to the errors themselves. Then, we fixed the errors and the dispersion to the $10\%$ and used smaller sets of 100 and 500 simulated luminosity distances, respectively. All the results are shown in Fig. \ref{Fig_Mock}, with: $n_g=50$, $n_i=300$, $n_p=32$, and crossover and mutation probabilities= 0.5, following the previously mentioned general scheme. For the study of these hyperparameters and their effects on the results derived from the real data, see $\S$~\ref{Sec_Results}.

\begin{figure*}
\centering
\includegraphics[width=0.7\hsize]{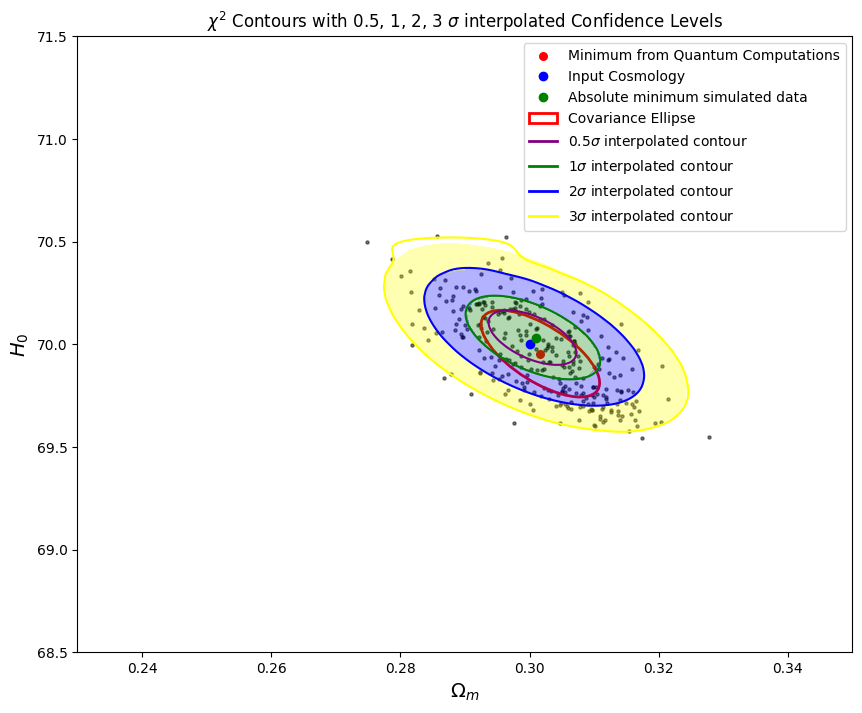}
\caption{Interpolation of the contour levels from the results of AEQGA, compared with the analytical contour levels and the covariance matrix. The results for the mock dataset with $5\%$ error and dispersion have been used here.}
\label{Fig_Interpolated_Mock}
\end{figure*}

In these panels, the black dots are the results of AEQGA, for which the means and covariance ellipses have been derived, while the underlying contours are the analytical $1,2,$ and $3 \sigma$ contours of the objective function computed from the mock datasets. We conclude that AEQGA is always consistent with the minimum of the objective function (which differs from the input cosmology given the dispersion we have simulated) within $1 \sigma$. The ellipse of the quantum results follows roughly the same shape and orientation as the ones given by the analytical objective function, while its size is independent of the underlying analytical dispersion, meaning that is mainly dependent on the hyperparameters of the algorithm itself than the shape of the underlying objective function.  
To link the results of AEQGA with the shape of the function, we interpolated the contour levels of the $\chi^2$ function for the mock data, starting from the results of AEQGA. The idea is to see if we can recover the analytical contours, while showing the difference with the covariance uncertainty on the minimum. The results are shown in Fig.~\ref{Fig_Interpolated_Mock}, where we focused on the $\chi^2$ function for the $5\%$ error and dispersion. We recovered the analytical $\sigma$ contours from our results, and we note the difference with the uncertainty covariance matrix. Given that the experiments performed in $\S$~\ref{Sec_Results} concern merit functions we can analytically evaluate, and the scope of AEQGA, in our analysis we focused on the statistical uncertainties. Still, the possibility of recovering the contour levels from our results is worth mentioning.

As a further test for AEQGA, we considered a 2D negative peak Gaussian function, with the absolute minimum in (0,0) and variable standard deviation. From this, we took into account two cases, one with a wide Gaussian ($\sigma=0.5$), and the other with a narrow dispersion ($\sigma=0.005$). The idea is see if, under the same hyperparameters, the dispersion of the results depends on the underlying width of the function. What we found is that, with $n_g=50$, AEQGA does not always converge in the region around the minimum for $\sigma=0.005$, but when it does, it is more precise than for $\sigma=0.5$, thus showing a difference due the shape of the function. For $n_g=500$, instead, we obtain precise results in both cases.

\begin{figure*}
\centering
\includegraphics[width=0.45\hsize]{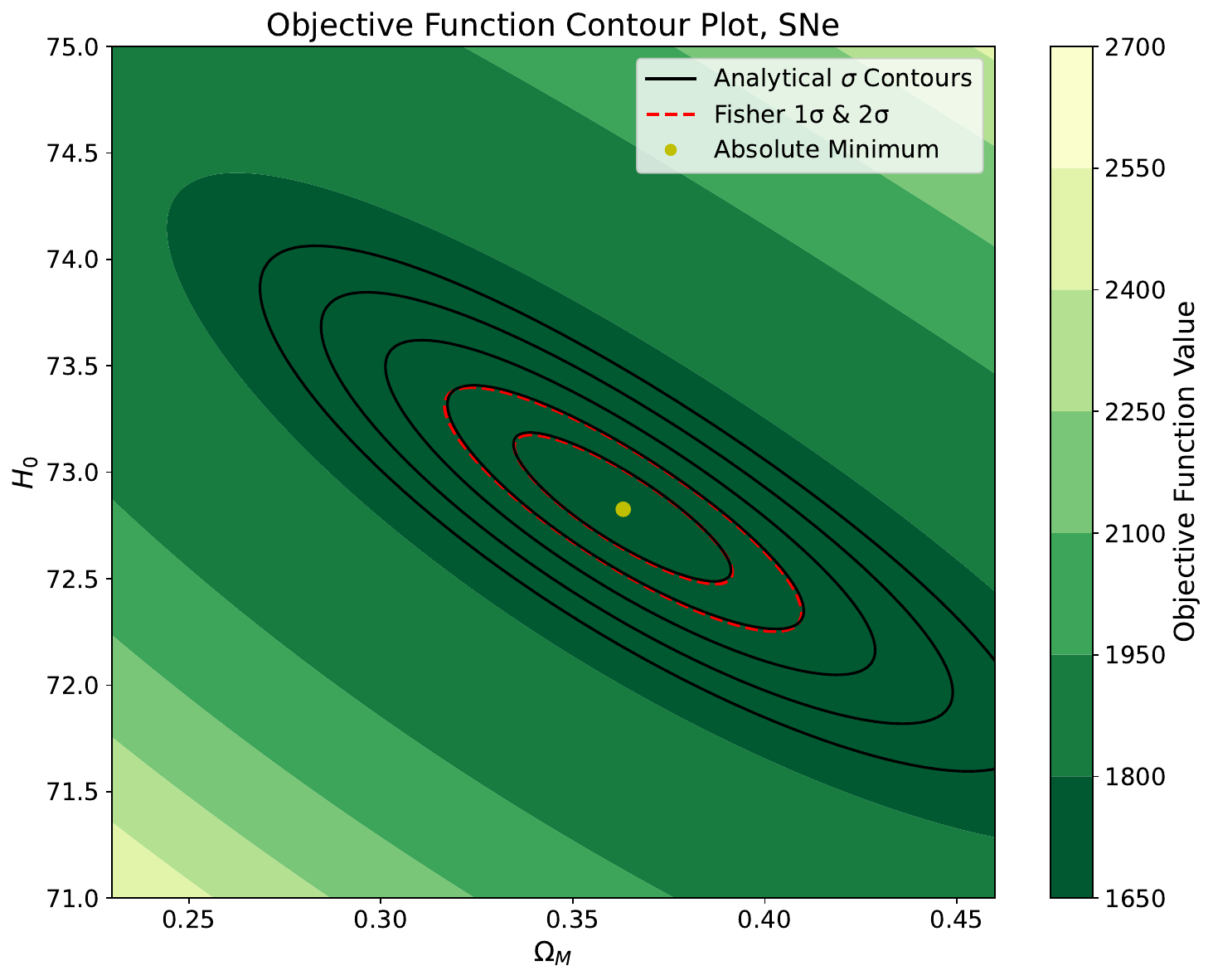}
\includegraphics[width=0.45\hsize]{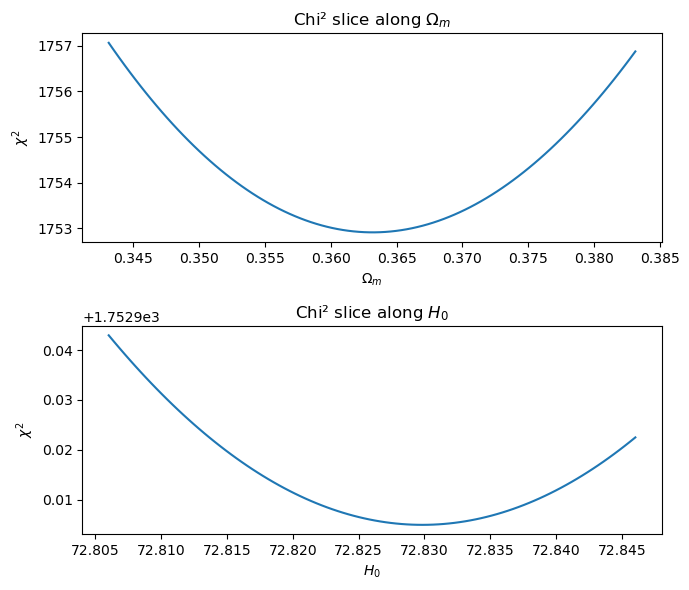}
\caption{Left panel: $1$ and $2$ $\sigma$ contours found by the Fisher approximation methodology compared with the analytical contours of the $\chi^2$ SNe Ia function. Right panel: Symmetry checks in the directions of $\Omega_M$ and $H_0$ for the region around the minimum of the $\chi^2$ SNe Ia function.}
\label{Fig_Fisher_Contours}
\end{figure*}

Related to this point, as a further way to trace contours from the results of AEQGA, we have tested a way to infer the contours of the objective functions used for our analysis via an approximation related to the Fisher matrices around the minima found by the AEQGA. More specifically, we approximate the local curvature around the best–fit point (i. e., the point with the lowest value for the objective function found by the AEQGA) $\boldsymbol{\theta}_\star$ by numerically evaluating its Hessian with central finite differences. In general,  for $n$ parameters $\boldsymbol{\theta}=(\theta_1,\ldots,\theta_n)$ and a small step $\varepsilon$, the mixed second derivatives can be approximated as
\begin{multline} \label{Hessian_approximation}
H_{ij} \equiv 
\left.\frac{\partial^2 \chi^2}{\partial \theta_i \partial \theta_j}\right|_{\boldsymbol{\theta}_\star}
\approx \\
\frac{
\chi^2(\boldsymbol{\theta}_\star+\mathbf{e}_i\varepsilon+\mathbf{e}_j\varepsilon)
-\chi^2(\boldsymbol{\theta}_\star+\mathbf{e}_i\varepsilon-\mathbf{e}_j\varepsilon) }{4\,\varepsilon^2} \\
-\frac{
\chi^2(\boldsymbol{\theta}_\star-\mathbf{e}_i\varepsilon+\mathbf{e}_j\varepsilon)
-\chi^2(\boldsymbol{\theta}_\star-\mathbf{e}_i\varepsilon-\mathbf{e}_j\varepsilon)}
{4\,\varepsilon^2}\,,
\end{multline}
and, for the diagonal terms $(i=j)$,
\begin{equation} \label{diagonal_Hessian_Approximation}
H_{ii} \;\approx\; \frac{\chi^2(\boldsymbol{\theta}_\star+\mathbf{e}_i\varepsilon)
-2\,\chi^2(\boldsymbol{\theta}_\star)
+\chi^2(\boldsymbol{\theta}_\star-\mathbf{e}_i\varepsilon)}
{\varepsilon^2}\,
\end{equation}

where $\mathbf{e}_i$ is the unit vector along $\theta_i$. We then take the Fisher information matrix as $F \equiv2 H$, where the $2$ comes from the fact that we are treating a $\chi^2$ function and not the log-likelihood. Under the Gaussian–likelihood (quadratic) approximation, the parameter covariance is $C =F^{-1}$ and the 1-$\sigma$ marginalized uncertainties are
\begin{equation}
\sigma_i \;=\; \sqrt{C_{ii}}\,.
\end{equation}
In applying this computation to the merit function of the SNe Ia and comparing the contours obtained in this way with the analytical ones, we get the results shown in the left panel of Fig. \ref{Fig_Fisher_Contours}. Here, we note how this methodology recovers almost exactly the analytical contours of the $\chi^2$ SNe Ia function. The small difference could be due to the fact that our $\chi^2$ function is not exactly Gaussian, but presents an asymmetry in the direction of  $H_0$, as can be seen in the right panel of Fig.  \ref{Fig_Fisher_Contours}. This allows us to conclude that, even if the main focus of AEQGA is minimization, it is still possible to infer from its results an estimate of the contours of the objective function. The caveat here is that this method requires the direct computation of the objective function around the minimum, which could be troublesome for more complex cosmological $\chi^2$ especially when considering high-dimensional multi-parameter spaces.  

\end{document}